\documentclass[a4paper,11pt, svgnames]{article}
\usepackage{a4wide}
\usepackage{amsmath,amssymb,amsthm,mathtools,array,booktabs}
\usepackage{graphicx,hyperref}
\usepackage{xcolor}
\usepackage{cancel}
\usepackage{tikz}
\usepackage{amscd}
\usepackage{latexsym,cite}
\usepackage{subcaption}
\usepackage{ytableau}
\usepackage{scalerel}
\usepackage{calc}
\usepackage{authblk}
\usepackage{longtable}
\usepackage{booktabs}
\usepackage{array}
\theoremstyle{plain}

\theoremstyle{definition}
\newtheorem{remark}{Remark}[section]

\input xy
\xyoption{all}

\hypersetup{
	unicode,
	bookmarksnumbered,
	linktoc = all,
	pdfborderstyle = {/S/U/W 0.5}
}

\def\mC3{\mathbb{C}^3}

\def\cM{\mathcal{M}}

\def\p{\partial}

\numberwithin{equation}{section}

\usepackage{preambles_Nafiz}
\usepackage{preambles_Saebyeok}

\ytableausetup{boxsize=.5em}

\title{Refined Invariants and Quantum Curves from Supersymmetric Localization}

\date{}

\author[1]{Sibasish Banerjee}
\author[2,3]{Nafiz Ishtiaque}
\author[4]{Saebyeok Jeong}

\affil[1]{Institut des Hautes \'Etudes Scientifiques (IHES), Bures-sur-Yvette 91440, France}
\affil[2]{Center for Mathematics and Interdisciplinary Sciences, Fudan University, Shanghai
	200433, China}
\affil[3]{Shanghai Institute for Mathematics and Interdisciplinary Sciences (SIMIS), Shanghai
	200433, China}
\affil[4]{Center for Geometry and Physics, Institute for Basic Science (IBS), Pohang 37673, Korea}

\begin{document}

	\maketitle

	\begin{abstract}
		\noindent
		We study an Aganagic-Vafa brane supported on a special Lagrangian submanifold $\CalL$ in a non-compact toric Calabi-Yau threefold $\CalX$. From the perspective of geometric engineering, the Aganagic-Vafa branes give rise to a special class of half-BPS codimension-two defects in 5d $\CalN=1$ supersymmetric field theories in the presence of $\O$-background. We propose that the defect partition functions give generating functions of refined, non-negative, integral open BPS invariants of the pair $(\CalX,\CalL)$, across different K\"{a}hler moduli chambers they are expanded in. In the Nekrasov-Shatashvili limit, the partition function provides a partially resummed solution to a $q$-difference equation that quantizes the mirror curve of $\CalX$ in an unambiguous fashion, in a polarization determined by the discrete labels of the Aganagic-Vafa brane. We demonstrate our method at examples of $\BC^3$, resolved conifold, resolved $A_1$-singularity, local $F_0$, and local $F_1$.
	\end{abstract}
	\newpage
	\setcounter{tocdepth}{3}
	\tableofcontents
	\setcounter{footnote}{0}

	\section{Introduction}

Enumerative invariants of a Calabi-Yau threefold $\CalX$ can be packaged in several closely related curve-counting theories, including the map-based approach of Gromov-Witten theory and the
sheaf-theoretic approach of Donaldson-Thomas theory. A natural open analogue arises when one studies a pair $(\CalX,\CalL)$, where $\CalL\subset \CalX$ is a Lagrangian submanifold (equipped
with auxiliary brane data). In this setting, one is led to the counting of open curves where the boundary is mapped to $\CalL$. Compared to the closed sector, the non-compactness of the open moduli space introduces additional subtleties: the framing of the Lagrangian and the stability parameters. The aim of this work is to establish a field-theoretic method to compute these invariants as BPS degeneracies of the associated coupled 3d-5d system. 

A physical incarnation of these invariants arises from string/M-theory in the background $\CalX$, where they encode protected quantities capturing the spectrum of BPS states. In the
closed sector, the A-model partition function on $\CalX$ (refined by their spin content in the $\Omega$-background) admits a universal BPS expansion: the invariants can be reorganized in terms of non-negative, integral BPS multiplicities associated with M2-branes wrapping holomorphic curves in $\CalX$ \cite{Gopakumar:1998ii,Gopakumar:1998jq}. In the presence of an M5-brane supported on $\CalL$, an analogous open sector is obtained from M2-branes whose boundary lies on $\CalL$ (which descend to open strings ending on the A-brane) \cite{Ooguri:1999bv,Labastida:2000zp,Aganagic:2003db}. Their protected degeneracies are expected to give refined, non-negative, integral, open BPS invariants.

When $\CalX$ is non-compact, the 5d supergravity decouples from remaining 5d $\CalN=1$ field theory, geometrically engineered by $\CalX$ \cite{Katz:1996fh}. The remaining BPS invariants can be interpreted intrinsically within this  field theory. When there is a K\"{a}hler chamber where the engineered theory has a weakly coupled IR gauge theory description (the large-base chamber), the BPS Hilbert space is realized as the equivariant cohomology of the moduli space of framed instantons on $\mathbb C^2$. The closed BPS partition function is then given by the equivariant Euler characteristics (i.e., the K-theoretic Nekrasov partition function \cite{Nekrasov:2002qd,Nekrasov:2003rj}), a series in the gauge coupling with coefficients being exact rational functions in the equivariant parameters. Identifying the exponentiated K\"ahler moduli with the gauge coupling and the equivariant parameters, the single-particle BPS multiplicities are then obtained by taking the plethystic logarithm of the partition function, expanded in all the K\"{a}hler moduli. In this step the K\"{a}hler moduli chamber is fully specified; different expansions yield the refined closed BPS invariants in the respective chambers.

From the viewpoint of geometric engineering, an M5-brane supported on $\CalL$ produces a half-BPS codimension-two defect in the 5d $\CalN=1$ theory engineered by $\CalX$. In this work, we restrict to a distinguished class of such branes of Aganagic-Vafa type, for which $\CalL$ is a special Lagrangian submanifold with topology $\BR^2\times S^1$ \cite{Aganagic:2000gs, Aganagic:2001nx,MR666108}. An Aganagic-Vafa brane gives rise to a \textit{canonical} codimension-two defect described by a 3d $\CalN=2$ $\U(1)$ gauge theory whose flavor symmetry is partially gauged by the restriction of the 5d gauge field to the defect locus. It is then natural to expect that the vacuum expectation value of this defect $-$ equivalently, the
partition function of the coupled 3d-5d system in the $\Omega$-background $-$ computes the generating function of refined open BPS invariants of the pair $(\CalX,\CalL)$, with the newly introduced discrete and continuous open moduli translating into the 3d defect parameters. This field-theoretic approach to open BPS invariants has appeared in a number of previous works \cite{Dimofte:2010tz,Kimura:2020qns,Cheng:2021nex,Alim:2022oll,Grassi:2022zuk}; however, most explicit analyses have been restricted to settings in which the 5d gauge dynamics is absent or effectively decoupled, due to the lack of a general localization formula for the defect expectation value in the presence of a fully dynamical 5d bulk.

The main goal of the present work is to overcome this limitation by coupling the codimension-two
defect to a fully dynamical 5d $\CalN=1$ gauge theory and computing its exact expectation value by
localization. The canonical defect admits two complementary descriptions, which we refer to as
the Coulomb and Higgs phases, corresponding to different choices of parameters to be turned on in the 3d $\CalN=2$ theory (a complex scalar $X \in \BC^\times$ in the twisted chiral multiplet versus a complexified FI parameter $Z \in \BC^\times$).

In the Coulomb phase, the defect partition function receives no non-perturbative 3d contributions, and it reduces to a one-loop determinant depending on the twisted masses of the 3d chiral multiplets. When the defect is coupled to the 5d bulk, these twisted masses must be promoted to the scalar in the dynamical 5d vector multiplet restricted to
the defect. Equivariantly, this promotion is implemented by replacing the corresponding flavor bundle with the universal instanton sheaf on the instanton moduli space, thereby producing a codimension-two observable that we call the $Q$-observable (see the 4d construction in \cite{Jeong:2024onv} and its 5d uplift in \cite{Jeong:2025yys}).

In the Higgs phase, the 3d path integral receives contributions from non-perturbative BPS vortex configurations.
Following \cite{Jeong:2025yys}, we exploit the representation of the K-theoretic vortex partition function as a $q$-lattice sum whose summand is precisely the Coulomb-phase one-loop factor. This makes it possible to couple the vortex sum to the bulk instanton background term-by-term: we
promote each summand to the corresponding $Q$-observable and thereby obtain the Higgs phase
codimension-two insertion, which we refer to as the $H$-observable.

The vacuum expectation values of these codimension-two observables are then computed by equivariant localization on the moduli space of framed instantons, i.e., by summing over its isolated fixed points. Finally, taking the plethystic logarithm of the resulting defect partition function and expanding in a chosen K\"{a}hler chamber, we extract the refined open BPS invariants as
the coefficients of the series, and verify their expected integrality and non-negativity.

An Aganagic-Vafa brane is characterized by discrete and continuous data, which translate
naturally into parameters of the canonical codimension-two defect. The discrete framing $f\in\BZ$ controls the choice of open-string polarization (equivalently, how the open modulus is parametrized on the mirror curve of $\CalX$) and is identified on the defect side with the
Chern-Simons level. The continuous open modulus is mapped to a parameter of the 3d $\CalN=2$ theory: depending on the phase of the defect, it is identified either with the Coulomb parameter $X$ or with the complexified FI parameter $Z$. The choice of this phase is correlated with another discrete label of the AV brane: the choice of toric leg on which the AV brane ends.

We confirm this dictionary most explicitly via the \emph{quantum mirror curve}: in the
Nekrasov-Shatashvili limit the defect expectation value is annihilated by a $q$-difference operator, which provides a quantization of the classical mirror curve. In particular, the discrete AV brane data fix the polarization of the quantization, and we obtain an
unambiguous quantum mirror curve operator compatible with the defect construction.

The paper is organized as follows. In Section~\ref{sec:bpsgeo}, we first review the refined closed BPS invariants of $\CalX$ and their realization in the geometrically engineered 5d $\CalN=1$
gauge theory. We then introduce an Aganagic-Vafa brane supported on a special Lagrangian submanifold $\CalL\subset \CalX$, and review the refined open BPS invariants of the pair $(\CalX,\CalL)$. We analyze the associated half-BPS codimension-two defect of the 5d theory via the type IIB dual description in terms of a $(p,q)$-fivebrane web, in which the AV brane maps to
a D3-brane ending on a fivebrane segment. We present explicit expressions for the canonical codimension-two defect observables in both the Coulomb and Higgs phases, and propose that their
vacuum expectation values compute the generating function of refined open BPS invariants. We also investigate connections to open Gromov-Witten invariants and open Donaldson-Thomas invariants. In Section~\ref{sec:3dhalfindex}, we illustrate our method through the examples of $\BC^3$, resolved conifold, the resolved $A_1$-singularity, local $\BP^1\times \BP^1$ (i.e., local $F_0$), and local $F_1$. In each case, we identify the relevant 3d $\CalN=2$ $\U(1)$ gauge theory describing the defect, and the explicit defect observables it produces. We extract the refined open BPS invariants from the plethystic logarithm of the corresponding partition function and verify that they are non-negative integers, as predicted. We further perform non-trivial checks of the $S$-transformations relating different AV branes, providing additional evidence that the coupled 3d-5d partition function reproduces the open BPS invariants. We conclude in Section~\ref{sec:discussion} with discussions. Appendix~\ref{sec:table} collects tables of refined open BPS invariants computed by our method for all examples considered.

\paragraph{Acknowledgment} The authors thank Alba Grassi, Amihay Hanany, Alexander Hock,  Albrecht Klemm, Mauricio Romo, Raphael Senghaas, Valdo Tatitscheff, Rak-Kyeong Seong, and Masahito Yamazaki for discussions on relevant subjects. SB has been partially supported by ERC-SyG project “Recursive and Exact New Quantum Theory” (ReNewQuantum), which received funding from the European Research Council (ERC) under the European Union’s Horizon 2020 research and innovation program, grant agreement No. 810573. SB also acknowledges CERN and University of Geneva for providing good working conditions during various phases of the work. The work of NI was supported by the Research Start-up Fund of the Shanghai Institute for Mathematics and Interdisciplinary Sciences (SIMIS). NI also thanks IHES, Bures-sur-Yvette and IBS, Pohang for support and being host during some periods of this research.  The work of SJ is supported by the Institute for Basic Science under the project IBS-R003-Y3. SJ is also grateful to IHES for its hospitality while part of this work was carried out.

	\section{Refined BPS invariants from geometric engineering}
	\label{sec:bpsgeo}
We begin by reviewing refined closed and open BPS invariants arising from M-theory on a local Calabi-Yau threefold $\CalX$ in the presence of an Aganagic-Vafa brane supported on a special
Lagrangian $\CalL\subset\CalX$. We then give a field-theoretic description of the associated canonical half-BPS codimension-two defect and propose that its vacuum expectation value computes
the generating function of refined open BPS invariants. Next, we show that the resulting defect partition function is annihilated by a $q$-difference operator, yielding an unambiguous quantization of the mirror curve of $\CalX$. Finally, we discuss the connections to open Gromov-Witten and open Donaldson-Thomas invariants.
	
	\subsection{Refined closed BPS invariants from 5d \texorpdfstring{$\mathcal{N}=1$}{N=1 supersymmetric} gauge theory}

	\label{sec:closedengg}
	
	Consider the M-theory placed on $\BR^4 \times S^1 \times \mathcal{X}$, where $\mathcal{X}$ is a local Calabi-Yau threefold. The massive BPS particles arise from M2-branes wrapping holomorphic curves in $\CalX$. Quantizing the corresponding wrapped M2-brane moduli spaces yields a one-particle BPS Hilbert space $\CalH_{\rm BPS}$, graded by the curve class $\b\in H_2(\CalX;\BZ)$. The massive little group in five dimensions is $\text{Spin}(4)\cong \SU(2)_L \times \SU(2)_R$, hence each graded piece decomposes into irreducible representations as
\begin{align}
\CalH_{\rm BPS,\b} = \bigoplus_{j_L,j_R \in \frac{1}{2} \BZ_{\geq 0}}  V_{j_L}\otimes V_{j_R} \otimes \BC^{N_{\b}^{j_L,j_R}}, \qquad N_{\b}^{j_L,j_R}\in\BZ_{\ge0},
\end{align}
  where $V_j$ is the $(2j+1)$-dimensional representation of $\text{SU}(2)$. The degeneracies $N_\b ^{j_L,j_R} \in \BZ_{\geq 0}$ of BPS states with charge $\b \in H_2 (\mathcal{X};\BZ)$ and spin $(j_L,j_R)$ under $\mathrm{SU}(2)_L \times \mathrm{SU}(2)_R$ are called the refined closed BPS invariants of $\mathcal{X}$. The generating function of these refined closed BPS invariants can be written as \cite{Gopakumar:1998ii,Gopakumar:1998jq}
	\begin{align}
		\label{topstrZ}
		\CalZ^{\text{closed}}_{\text{BPS}} = \prod_{\b \in H_2 (X;\BZ)\setminus\{0\}} \prod_{j_L,j_R \in \frac{1}{2}\BZ_{\geq 0}} \text{PE} \left[(-1)^{2j_L + 2j_R} N_\b ^{j_L,j_R} \frac{\chi_{j_L} (q_1 q_2) \chi_{j_R} (q_1 q_2 ^{-1}) Q^\b}{\left(q_1^{ 1/ 2} - q_1 ^{- 1/ 2} \right) \left(q_2^{1/2} - q_2 ^{-1/2} \right)}  \right],
	\end{align}
	where we denoted the $\text{SU}(2)$ character by $\chi_j (q) = \text{Tr}_{V_j} q^J = q^{-j} + q^{-j+1} + \cdots + q^j$. The unrefined closed BPS invariants are recovered by taking $q_2 = q_1 ^{-1}$.
	
	Compactified on $\mathcal{X}$, the 11d supergravity multiplet yields the 5d supergravity multiplet, vector multiplets, and hypermultiplets. Since $\mathcal{X}$ is a local Calabi-Yau, the 5d $\CalN=1$ supergravity decouples from the gauge theory dynamics. The M2-branes supported on $S^1\times C$ where $C \subset \mathcal{X}$ is a compact holomorphic two-cycle lead to 5d BPS particles on $S^1$; W-bosons, charged hypermultiplets, and instanton particles. The first one is responsible for the nonabelian enhancement of the gauge group in the limit where the corresponding cycles shrink to zero. This construction of the 5d $\CalN=1$ gauge theory is called the geometric engineering \cite{Katz:1996fh}. Note that its contents, i.e., the gauge group, its representation attached to hypermultiplets, and the Chern-Simons level, are fully determined by the geometric data of $\mathcal{X}$.
	
	When $\mathcal{X}$ is a non-compact toric Calabi-Yau threefold, the M-theory is dual to the IIB string theory with a $(p,q)$-web of fivebranes, whose planar graph is given by the graph-dual of the toric diagram of $\mathcal{X}$ \cite{Leung:1997tw,Aharony:1997bh}.\footnote{We use the convention in which $\text{NS5}=(0,1)$ and $\text{D5}=(1,0)$.} The 5d $\CalN=1$ gauge theory is precisely the low-energy effective field theory realized on their five-dimensional intersection $\BR^4 \times S^1$.
	
	\subsubsection{5d \texorpdfstring{$\mathcal{N}=1$}{N=1} gauge theory and equivariant index}

	From the geometric engineering point of view, the partition function of the BPS degeneracies is calculated by the path integral of the 5d $\CalN=1$ gauge theory on the $\O$-background $\BR^2 _{\ve_1} \times \BR^2 _{\ve_2} \times S^1$ \cite{Nekrasov:2003rj,Nakajima:2005fg}. Throughout this work, we only consider the cases where the gauge group is $\U(n)$ unless it is trivial.
	
\begin{figure}[h!]
		\centering
		\begin{tikzpicture}[
			lab/.style={fill=white, inner sep=1pt}
			]
			
			\node (K) at (0,0) {$K$};
			\node (N) at (1.6,0) {$N$};
			
			\def\rad{0.30}  
			\def\dist{0.50} 
			
			\def\gap{45}    
			
			\coordinate (C1) at (125:\dist); 
			
			\draw[-stealth, thick] 
			(K) -- ($(C1) + ({305+\gap}:\rad)$)                 
			arc [start angle={305+\gap}, end angle={305-\gap+360}, radius=\rad] 
			node[lab, above left, pos=0.5] {$B_1$}              
			-- (K);                                             
			
			\coordinate (C2) at (235:\dist);
			
			\draw[-stealth, thick] 
			(K) -- ($(C2) + ({55+\gap}:\rad)$)
			arc [start angle={55+\gap}, end angle={55-\gap+360}, radius=\rad]
			node[lab, below left, pos=0.5] {$B_2$}
			-- (K);
			
			\draw[-stealth, thick]
			($(N)+(-0.22, 0.06)$) -- ($(K)+(0.22, 0.06)$)
			node[lab, above, pos=0.5] {$I$};
			
			\draw[-stealth, thick]
			($(K)+(0.22,-0.06)$) -- ($(N)+(-0.22,-0.06)$)
			node[lab, below, pos=0.5] {$J$};
			
		\end{tikzpicture}
		\caption{ADHM quiver with gauge node $K=\BC^k$ and framing node $N=\BC^n$.}
		\label{fig:nakajima-quiver}
	\end{figure}
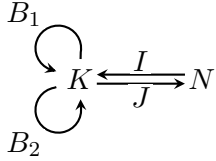

	The path integral computes the equivariant index of the space of BPS states. At each instanton sector $k = \frac{1}{8\pi^2} \int \text{Tr}\, F^2 \in \BZ_{\geq 0}$, the moduli space $\CalM_{n,k}$ of framed instantons is realized as the Nakajima quiver variety \cite{Nakajima:2005fg} for the framed Jordan quiver (i.e., the ADHM quiver \cite{Atiyah:1978ri}; see Figure \ref{fig:nakajima-quiver}), which comprises of a single gauge node $K=\BC^k$, a single framing $N=\BC^n$, two self-loops on the gauge node $B_{1,2}:K\to K$, and one arrow for each direction between the two nodes $I:N\to K$, $J:K \to N$. In the stability chamber where $K = \BC[B_1,B_2]\, \text{Im}\,I$, the moduli space of framed instantons is given by
	\begin{align}
		\CalM_{n,k} = \{(B_1,B_2,I,J) \,\vert \,  [B_1,B_2]+IJ = 0,\; K = \BC[B_1,B_2]\, \text{Im}\,I\}/\mathrm{GL}(k),
	\end{align}
	where the action of $\mathrm{GL}(k)$ is given by
	\begin{align}
		g: (B_1,B_2,I,J)\mapsto (g B_1 g^{-1},g B_2 g^{-1}, gI, Jg^{-1}),\qquad g\in \mathrm{GL}(k).
	\end{align}
	Note that the global symmetry group $\mathsf{T} = (\BC^\times)^n  \times (\BC^\times)^2 $ acts on $\mathcal{M}_{n,k}$ by
	\begin{align}
		(\ba,q_1,q_2): (B_1,B_2,I,J)\mapsto (q_1 B_1,q_2 B_2, I \ba^{-1},  q_1 q_2 \ba J),\qquad (\ba,q_1,q_2) \in \mathsf{T}.
	\end{align}
	
	The space of BPS states is given by the $\mathsf{T}$-equivariant cohomology of the moduli space $\mathcal{M}_{n,k}$ of framed instantons. Namely,
	\begin{align}
		\mathcal{H}_{\text{BPS}} = \bigoplus_{k=0}^\infty \CalH_k  := \bigoplus_{k=0} ^\infty H^\bullet _{\mathsf{T}} (\mathcal{M}_{n,k};\mathcal{E}),
	\end{align}
	where $\mathcal{E}$ is the bundle of zero-modes of the Dirac operator, twisted by the Chern-Simons line bundle. More specifically, the bundle of single-particle fermionic zero-modes is written as
	\begin{align}
		E = M^* \otimes K + \widetilde{M} \otimes K^*,
	\end{align}
	where $K$ is the tautological bundle and $M$ (resp. $\widetilde{M}$) is the flavor bundle for the fundamental (resp. anti-fundamental) hypermultiplets. $*$-symbol indicates the dual bundle. The bundle of multi-particle zero-modes is given by its total anti-symmetric exterior power, $\text{PE}[-E]:=\sum_{i\geq 0} (-1)^i\bigwedge^i E$. If we introduce a bare Chern-Simons level $k_{\text{CS}} \in \BZ$, this matter bundle is further twisted by the $k_{\text{CS}}$-th power of the determinant line bundle $L:=\det K$ of the tautological bundle; $L^{k_{CS}} = (\det K)^{k_{\text{CS}}}$. All in all, we have
	\begin{align}
		\mathcal{E} = \left( \sum_{i\geq 0} (-1)^i \bigwedge\nolimits^{\!i}  E \right) \otimes (\det K)^{k_{\text{CS}}}.
	\end{align}
	The equivariant index is the generating series of the equivariant Euler characteristics,
	\begin{align}
		\label{NekZ5d}
		\CalZ^{\text{inst}} = \sum_{k=0} ^\infty \qe^k \chi_\mathsf{T} \left(\CalM_{n,k} ; \mathcal{E} \right) = \sum_{k=0}^\infty \qe^k \int_{\CalM_k} \text{Ch}_\mathsf{T} (\text{PE}[-E]) \text{Ch}_\mathsf{T} (\det K)^{k_{\text{CS}}} \text{Td}_\mathsf{T}(T\mathcal{M}_{n,k}),
	\end{align}
	where the second equality follows from the Hirzebruch-Riemann-Roch theorem. The full partition function is obtained by incorporating the perturbative contribution, 
    \begin{align}
        \CalZ^{\text{5d}} (\ba,\mathbf{m},\qe;q_1,q_2) = \CalZ^{\text{1-loop}} (\ba,\mathbf{m};q_1,q_2) \times \CalZ ^{\text{inst}} (\ba,\mathbf{m},\qe;q_1,q_2),
    \end{align}
    where $\CalZ^{\text{1-loop}}$ is given by a product of refined MacMahon functions.
    \\
	
	The equivariant localization reduces each integral into a summation over the set $\CalM_{n,k} ^\mathsf{T}$ of discrete fixed points under the action of the global symmetry $\mathsf{T} = (\BC^\times)^n  \times (\BC^\times)^2 $. The fixed points in $\CalM_{n,k} ^\mathsf{T}$ are classified by $n$-tuples $\underline{\l} = (\l^{(1)} ,\l^{(2)} ,\cdots , \l^{(n)})$ of Young diagrams having $k$ boxes in total, $\vert \underline{\l} \vert = |\lambda^{(1)}| + \cdots + |\lambda^{(n)}|=k$. At a fixed point $\underline{\l}$, the equivariant Chern character of the tangent bundle can be obtained as
	\begin{align}
    \begin{split}
		\text{Ch}_{\mathsf{T}} (T_{\underline{\l}} \CalM_{n,k}) &= \text{Ch}_{\mathsf{T}}(N^*) \text{Ch}_{\mathsf{T}}(K _{\underline{\l}}) + q_1^{-1} q_2^{-1} \text{Ch}_{\mathsf{T}}(N) \text{Ch}_{\mathsf{T}}(K^* _{\underline{\l}}) \\
        &\quad- (1-q_1^{-1})(1-q_2^{-1}) \text{Ch}_{\mathsf{T}}(K_{\underline{\l}} )\text{Ch}_{\mathsf{T}}(K^*_{\underline{\l}}),
    \end{split}
	\end{align}
	where the equivariant Chern characters of the framing bundle $N$ and the tautological bundle $K$ are given by
	\begin{align}
		\text{Ch}_{\mathsf{T}}(N) = \sum_{\a=1} ^n a_\a,\qquad \text{Ch}_{\mathsf{T}}(K_{\underline{\l}} )= \sum_{\a=1} ^n \sum_{\Box \in \l^{(\a)}} \chi_\Box, 
	\end{align}
	where $\chi_{\Box} := a_\a q_1 ^{i-1} q_2 ^{j-1}$ for $\Box_{(i,j)} \in \l^{(\a)}$. The universal sheaf of instantons as an equivariant K-theory class is given by
	\begin{align}
		S = N - (1-q_1 )(1-q_2) K.
	\end{align}
	Moreover, the flavor bundles $M$ and $\widetilde{M}$ is acted on by the flavor symmetry $\mathsf{T}_F = (\BC^\times)^f \times (\BC^\times)^{\tilde{f}}$, whose weights are called the mass parameters. The equivariant Chern characters of these bundles are given by
	\begin{align}
		M = \sum_{i=1} ^f m_i,\qquad \widetilde{M} = \sum_{i=1} ^{\tilde{f}} \tilde{m}_i.
	\end{align}
	Thus, applying the Atiyah-Bott localization theorem to each integral in \eqref{NekZ5d}, we arrive at
	\begin{align}
		\CalZ^{\text{5d}} = \sum_{\uline\lambda} \qe^{|\uline\lambda|} \prod_{\Box \in \uline\lambda} \chi_{\Box} ^{k_{\text{CS}}} \,  \PE[-\text{Ch}_{\mathsf{T}} \left(E_{\uline\lambda} \right)] \PE[\text{Ch}_{\mathsf{T}} \left(T_{\uline\lambda}\cM_{n,k} \right)]. \label{Z5d}
	\end{align}
	Here, the plethystic exponential is defined as an operation on the equivariant Chern characters as
	\begin{equation}
		\PE\left[\sum_i {x_i} - \sum_j {y_j}\right] = \frac{\prod_j (1 - {y_j})}{\prod_i (1 - {x_i})}.
	\end{equation}
	
	The geometric engineering of the 5d $\CalN=1$ gauge theory identifies the generating function of the refined closed BPS invariants with the partition function of the 5d $\CalN=1$ gauge theory, up to an overall multiplicative factor contributed by the decoupled M2-branes and supergravity sector. Namely, we have the identity,
	\begin{align}    
		\label{MNOPeq}
		\CalZ^{\text{closed}}_{\text{BPS}} = \CalZ^{\text{decoupled}} \times \CalZ^{\text{5d}},
	\end{align}
	where the gauge coupling $\qe$, the Coulomb parameters $\ba$, and the mass parameters $\mathbf{m}$ are identified with the exponentiated K\"{a}hler moduli corresponding to the generators of $H_2 (\CalX;\BZ)$. In this work, we will neglect the decoupled part $\CalZ^{\text{decoupled}}$ and restrict our attention to the $\CalZ^{\text{5d}}$ part captured by the geometric engineering. 	
	
Note that the 5d gauge theory partition function $\CalZ^{\text{5d}}$ is defined as a (formal)
power series in the instanton counting parameter $\qe$. This is naturally associated with the large-base chamber in which $|\qe|<1$. Each coefficient of the series is an exact
rational function of the equivariant parameters $(\ba,\mathbf{m},q_1,q_2)$. To compare with the refined closed BPS generating function, which is written as a series in the K\"{a}hler parameters, we must choose a K\"{a}hler moduli chamber and expand it to yield a formal series. With this understood, the identity \eqref{MNOPeq} should be interpreted as an equality of the
corresponding chamber expansions (up to the standard Coulomb-independent decoupled
factors, if present). In this sense, $\CalZ^{\text{5d}}$ provides a partial resummation of the refined closed BPS generating function; it sums all contributions at fixed powers of the base parameter $\qe$ while keeping the remaining dependence exact, and the resulting meromorphic expression can be re-expanded in different regions of K\"ahler moduli to access other chamber expansions.

	\subsection{Refined open BPS invariants from half-BPS codimension-two defects}
	\label{sec:opengeomengg}
	
	Now, let us introduce an M5-brane supported on $\BR^2 _{\ve_1}\times S^1 \times \mathcal{L}$, where $\mathcal{L} \subset \mathcal{X}$ is a special Lagrangian submanifold. Due to the presence of the M5-brane, the M2-branes may now end on it, being supported on relative homology cycles in $H_2 (\mathcal{X},\mathcal{L};\BZ)$. Also, the rotation group is broken to the maximal torus $\U(1)_{\ve_1} \times \U(1)_{\ve_2}$. Note that the BPS particles are bound to $\BR^2_{\ve_1}$, which reduces the spin group to $\U(1)_{\ve_1}$ and promotes $\U(1)_{\ve_2}$ to a flavor symmetry.
	
	Let $N_\nu ^{r,s} \in \BZ_{\geq 0}$ be the degeneracy of BPS states with charge $\nu \in H_2 (\mathcal{X},\mathcal{L};\BZ)$ and weight $(r,s) \in \frac{\BZ}{2} \times \frac{\BZ}{2}$ under $\U(1)_{\ve_1} \times \U(1)_{\ve_2}$. We call them the refined open BPS invariants of the pair $(\mathcal{X},\CalL)$. The generating function of these refined open BPS invariants can be written as \cite{Dimofte:2010tz}
	\begin{align} \label{eq:openbps}
		\begin{split}
			\CalZ^{\text{open}} _{\text{BPS}} &= \prod_{\nu \in H_2 (\mathcal{X},\mathcal{L};\BZ)\setminus\{0\}} \prod_{r,s\in \frac{\BZ}{2}} \text{PE} \left[ \frac{(-1)^{2r} N^{r,s}_\nu q_1 ^r q_2 ^{s+\frac{1}{2}} z^\nu}{q_1 ^{1/2} - q_1 ^{-1/2} } \right] \\
			&= \prod_{\nu \in H_2(\mathcal{X},\mathcal{L};\BZ) \setminus \{0\}} \prod_{r,s\in \frac{\BZ}{2}} \prod_{n=0}^\infty \left(1-q_1 ^{r+n+\frac{1}{2}} q_2^{s+\frac{1}{2}} z^\nu \right)^{(-1)^{2r} N_{\nu}^{r,s}}.
		\end{split}
	\end{align}
	The generating function of the full BPS invariants is given by the product of the closed sector contribution \eqref{topstrZ} and the open sector contribution \eqref{eq:openbps}, 
	\begin{align}
		\CalZ_{\text{BPS}} = \CalZ_{\text{BPS}}^{\text{closed}} \times \CalZ_{\text{BPS}} ^{\text{open}}.
	\end{align}
	
	\subsubsection{Aganagic-Vafa brane and probe D3-brane}
	\label{Sec:D3brane}
	
	From the geometric engineering point of view, the M5-brane gives rise to a half-BPS codimension-two defect, supported on $\BR^2 _{\ve_1} \times S^1$, of the 5d $\mathcal{N}=1$ gauge theory. The contents of the codimension-two defect are in principle fully determined by the geometric data of the pair $(\mathcal{X},\mathcal{L})$. To focus on cases where the field-theoretic description of the codimension-two defect is tractable, we restrict $\mathcal{X}$ to be a non-compact toric Calabi-Yau threefold and $\mathcal{L}\subset \mathcal{X}$ to be a special Lagrangian of Aganagic-Vafa (AV) type \cite{Aganagic:2000gs,Aganagic:2001nx}. We call the corresponding M5-brane the Aganagic-Vafa brane. From the perspective of the topological A-model on $\mathcal{X}$, the AV brane is an A-brane, supported on an AV special Lagrangian submanifold $\mathcal{L}\subset \mathcal{X}$ with a flat $\U(1)$ bundle on it.
	
	Let us revisit the description of AV special Lagrangian brane briefly. A toric Calabi-Yau threefold is described by the symplectic quotient $\mathcal{X} = \mathbb{C}^{k+3}/\!/\,\U(1)^k$. It can be obtained as the vacuum moduli space of the 2d $\CalN=(2,2)$ gauged linear sigma model with gauge group $\U(1)^k$ and chiral multiplets $(\phi_i)_{i=1}^{k+3}$ charged under the $\a$-th $\U(1)$ factor by $Q^\alpha_i \in \BZ$, $\a=1,2,\cdots, k$. Then $\mathcal{X}$ is given by the locus of the moment map equations, 
	\begin{equation}
		\label{eq:DtermCY}
		\sum_{i=1}^{k+3} Q_i^\alpha \vert \phi_i\vert^2 = r^\alpha,\qquad \a=1,2,\cdots, k,
	\end{equation}
	where $r^\alpha \in \mathbb{R}_+$ are K\"ahler parameters  (we may take them to be positive by adjusting the overall sign of $Q^\a _i$ for fixed $\a$), modulo the gauge transformations. The Calabi-Yau condition further imposes $\sum_{i=1}^{k+3} Q_i^\alpha = 0$ for all $\a$. 
	
	Note that the charge matrix $Q$ is a $(k+3) \times k$ matrix. There are two independent primitive vectors $q^\b = (q^\b _i)_{i=1}^{k+3} \in \BZ^{k+3}$, $\b=1,2$, which satisfy $q^\b \in \text{ker}_\BZ\,Q^T$ and $\sum_{i=1}^{k+3} q^\b _i =0$. We obtain a special Lagrangian submanifold by imposing
	\begin{equation}
		\sum_{i=1}^{k+3} q_i^\beta \vert \phi_i\vert^2 = c^\beta,\qquad \beta=1,2, 
	\end{equation}
	where $c^\b\in \BR$, and
	\begin{align}
		\sum_{i=1} ^{k+3} \text{arg}\,\phi_i = \theta = \text{const}.
	\end{align}
	Recall that the non-compact toric CY threefold $\CalX$ can be viewed as a $T^3$-fibration over a non-compact convex polyhedron $B\subset \BR^3$.\footnote{Viewing each of $\mathbb{C}$, for $i=1,...,k$ as $S^1$-fibration over $\mathbb{R}_+$, $\mathcal{X}$ can be regarded as $T^3$-fibration over a non-compact (but linearly bounded) subspace $B \subset \mathbb{R}^3$. The moment map equation \eqref{eq:DtermCY} ensures that locally $\mathcal{X}$ is an $\mathbb{R}_+\times T^2$ fibration over $\mathbb{R}^3$.  Toric CY threefolds are obtained by gluing copies of $\mathbb{C}^3$ with coordinates $(z_1,z_2,z_3)$ and three real Hamiltonians $r_\alpha = \vert z_2\vert^2 - \vert z_1\vert^2, \, r_\beta = \vert z_3\vert^2-\vert z_1\vert^2, \, r_\gamma = {\mathrm{Im}}(z_1 z_2 z_3)$. Toric legs are defined as the loci when both the $S^1$'s of the $T^2$ pinch, corresponding to $r_\alpha$ and $r_\beta = 0$ (locally), and $r_\gamma$ being the proper time along the line defining the toric leg. In other words $L_{ij} = \{|\phi_i| = 0\} \cap \{|\phi_j|=0\}$ defines a toric leg at the intersection of two patches defined by $\{|\phi_i| = 0\}$ and $\{|\phi_j| = 0\}$ respectively.}  The edges of $B$ form the dual graph of the toric diagram of $\CalX$. In general, the special Lagrangian submanifold obtained above has topology of $\BR_+ \times T^2$, where the ray $\BR_+$ intersects a face of $B$. However, if the ray $\BR_+$ intersects an edge of $B$, i.e., an edge of the dual graph, one of the two circles in $T^2$ must pinch at the intersection and one obtains a special Lagrangian $\mathcal{L}$ having topology of $\mathbb{R}^2\times S^1$. The latter type of special Lagrangian submanifolds are called Aganagic-Vafa type.\\
	
	By construction, there is a real positive modulus $t\in \mathbb{R}_+$ of an AV special Lagrangian submanifold $\CalL$ given by its position on the edge. Moreover, the AV Lagrangian A-brane supported on $\CalL$ carries a $\U(1)$ local system; since $b_1 (\mathcal{L})=1$, there is an additional modulus $\th$ given by its holonomy. In turn, the moduli space of an AV brane becomes a complex algebraic surface $\S \subset (\mathbb{C}^\times)^2$ parametrized by the complexified coordinate $ te^{i\theta} \in \BC^\times$. This algebraic curve $\S$ is precisely the mirror curve of $\CalX$ \cite{Hori:2000ck}. In the mirror B-model with target $\CalX^\vee$, a conic fibration over $\S$, the B-brane dual of the AV Lagrangian A-brane is the one supported on the conic fiber over a point in the mirror curve $\S$.
	
	Moreover, the AV brane carries two discrete labels in addition to the continuous parameter $X$. First, there is a choice of the edge of the dual graph it ends on. There are two classes of edges: the external edges and the internal edges. The external edges are the ones extending to infinity. The internal edges are the ones introduced by a triangulation of the toric diagram, corresponding to a choice of crepant resolution $\r:\widetilde{\mathcal{X}}\to \mathcal{X}$ \cite{Batyrev:1993oya}. Second, there is a choice of the \textit{framing} $f\in \BZ$ \cite{Aganagic:2001nx,Aganagic:2003db}. Consider the boundary torus $\p \mathcal{L} \simeq T^2$ at infinity. Take a basis $H_1 (\p \mathcal{L};\BZ) \simeq \BZ^2 = \langle m,l_0 \rangle$, where $m$ is the \textit{meridian} which collapses in $\CalL$ while $l_0$ is the \textit{longitude} generating $H_1 (\mathcal{L};\BZ) \simeq \BZ$. The framing $f\in \BZ$ determines the image of the boundary of the open string worldsheet is lifted to a class in $H_1 (\p \CalL ;\BZ)$ as $l_f = l_0 + f m$. It does not affect the classical boundary holonomy, but it changes the polarization of the boundary phase space and thus the open string amplitudes. We will confirm the two discrete labels $-$ the edge and the framing $-$ indeed determine the polarization in which the mirror curve is quantized.
	
	\subsubsection{Canonical codimension-two defect from Aganagic-Vafa brane} \label{subsubsec:cc2def}
	
	To read off the 3d $\CalN=2$ theory engineered by the AV brane, it is convenient to pass to the dual IIB string theory with $(p,q)$-web of fivebranes. We fix the $SL(2;\BZ)$ frame of the IIB string theory in the way that, when the dual graph is rotated $90^\circ$-degree counterclockwise, the vertical lines are $(0,1)=\text{NS5}$ branes and horizontal lines are $(1,0)=\text{D5}$ branes (see Figure \ref{fig:toric-C3-shifted} for the case of $\CalX=\BC^3$). Under this duality, the AV brane passes to a single D3-brane ending on a fivebrane segment in the $(p,q)$-web \cite{Aganagic:2002qg}. The framing $f \in \BZ$ is implemented explicitly by introducing an additional fivebrane on which the D3-brane ends at the opposite boundary. By performing an overall $T$-transformation, we fix our reference frame so that the fivebrane at the opposite boundary is always an NS5-brane. Thus, the choice of the framing $f\in \BZ$ is reflected on the relative $T$-transformations of the $(p,q)$-web.

    Note that under the $T$-transformations, the D5-brane is invariant while $(p,1)$-branes maps to one another. Depending on the edge that the Aganagic-Vafa brane ends on, we distinguish the following two codimension-two defect observables:
    	\begin{itemize}
		\item $Q$-observable; if the D3-brane ends on a $(p,1)$-brane.
		
		\item $H$-observable; if the D3-brane ends on a D5-brane. 
	\end{itemize}
    Note both $Q$-observable and the $H$-observable may originate from a D3-brane ending on either an external leg or an internal leg. \\
    
	The effective field theory on the worldvolume of the D3-brane is always a 3d $\CalN=2$ $\U(1)$ gauge theory, with the Chern-Simons level assigned by the framing as $-f$ and the chiral multiplet content determined by the D3-D5 open string spectra. Recalling that the D5-D5 open strings give rise to the 5d $\CalN=1$ gauge theory, the codimension-two defect is defined by gauging a part of the 3d flavor symmetry by the 5d bulk gauge field restricted to the codimension-two locus. This codimension-two defect is the uplift of the canonical surface defect of 4d $\CalN=2$ field theories in \cite{Gaiotto:2011tf}. Adoptig the terminology, we refer to this defect as the \textit{canonical} codimension-two defect.
    
    For 3d $\mathcal N=2$ U(1) gauge theory on $\BR^2 _{\ve_1} \times S^1$, the real scalar in the vector multiplet gets augmented by the holonomy of the gauge field, yielding a complex scalar,
	\begin{align}
		\s^{\BC} = \s^\BR + \frac{ \ii}{R} \oint_{S^1} A,
	\end{align}
	which is periodic $\s^\BC \sim \s^\BC + \frac{2\pi \ii}{R}$ by the large gauge transformation. We define the $\BC^\times$-valued exponentiated Coulomb parameter by $X = \exp (R \s^{\BC})$. In the presence of the chiral multiplets, we may turn on the background gauge field for the flavor symmetry also, giving rise to $\BC^\times$-valued twisted masses in a similar way.
	
	Moreover, the real FI parameter is augmented by the holonomy of the background gauge field $A^{(T)}$ for the topological U(1) symmetry,\footnote{For any 3d abelian gauge field $A$, $\star \mathrm dA$ is a conserved current, generating the topological symmetry. Once the FI parameter is promoted to the bottom component of a background vector multiplet, the FI term coupling the the FI parameter and the dynamical vector multiplet scalar becomes part of a supermultiplet coupling that also contains the current term $A^{(T)} \wedge \star \mathrm dA$.} yielding a complexified FI parameter,
	\begin{align}
		\z^\BC = \z^\BR + \frac{\ii}{R} \oint_{S^1 } A^{(T)}.
	\end{align}
	It is periodic $\z^\BC \sim \z^{\BC} + \frac{2\pi \ii}{R}$ by the large gauge transformation for the background gauge field. We define the $\BC^\times$-valued exponentiated FI parameter by $Z = \exp R \z^\BC$. 
	
	A generic value of the Coulomb parameter $X$ can be turned on only when the complexified FI parameter $Z$ is turned off. Likewise, the complexified FI parameter can only be turned on only at special values of the Coulomb parameters. Accordingly, the $Q$-observable $Q_f(X)$ is obtained by the canonical codimension-two defect in the Coulomb phase, while the $H$-observable $H_f(Z)$ is obtained in the Higgs phase. The continuous parameter $t e^{i\th} \in \BC^\times $ of the Aganagic-Vafa brane is identified with the Coulomb parameter $X$ or the complexified FI parameter $Z$, respectively.

	\paragraph{$Q$-observable.} In the examples we consider, the 3d $\CalN=2$ $\U(1)$ gauge theory has $n_{\text{f}}$ chiral multiplets of charge $+1$ and $n$ chiral multiplets of charge $-1$. When the Coulomb parameter $X$ is turned on, its partition function on $\BR^2 \times S^1$ is simply given by the 1-loop contribution as
    \begin{align} \label{eq:1loopc}
        Q_f (X)_{\varnothing}=\th(X^{-1} ;q_1)^f \; \text{PE} \left[ \frac{X^{-1}(M-N)}{1-q_1} \right],
    \end{align}
    where $\th\left(x;q\right) = \left(x;q\right)_\infty \left(q x^{-1} ;q\right)_\infty$ is the $q$-theta function. The Chern-Simons level $-f$ induces an anomaly which is canceled by the anomaly inflow from the 2d fermi multiplet on the boundary torus $\p(\BR^2\times S^1) \simeq T^2$. The contribution of this 2d fermi multiplet is the above $q$-theta function.
    
   Now we couple the 3d $\CalN=2$ theory to the 5d $\CalN=1$ $\U(n)$ gauge theory by gauging the $\U(n)$ flavor symmetry for the chiral multiplets of charge $-1$. At each instanton sector, there is a 1-loop contribution of the 3d fields to the partition function, defining a codimension-two observable. The effect of gauging is reflected by replacing the framing bundle $N$ by the universal sheaf $S$ of instantons in the 1-loop contribution \eqref{eq:1loopc}. Thus, we arrive at the codimension-two defect observable given by
	\begin{align} \label{eq:qobs}
		\begin{split}
			Q_{f}(X) &= \th\left({X}^{-1};q_1\right)^f \, \text{PE}\left[  \frac{X^{-1} (M-S)}{1-q_1} \right],
		\end{split}
	\end{align}
	 At the fixed point $\underline{\l}$, its observable expression is given by
	\begin{align}
		Q_f (X) _{\underline{\l}} &= \frac{\th\left( X^{-1};q_1\right)^f}{ \prod_{i=1}^{n_{\text{f}}} \left( \frac{m_i}{X};q_1 \right)_\infty}  \prod_{\a=1}^n \left( \frac{a_\a q_1 ^{l(\l^{(\a)})}}{X};q_1 \right)_\infty \prod_{i=1} ^{l(\l^{(\a)})} \left( 1- \frac{a_\a q_1 ^{i-1} q_2 ^{\l^{(\a)}_i}}{X} \right).
	\end{align}
	The vacuum expectation value is therefore computed as
    \begin{align} \label{eq:qvev}
           \CalZ_\text{open} =\left\langle Q_f (X) \right\rangle =  \frac{\sum_{\uline\lambda} \qe^{|\uline\lambda|} \prod_{\Box \in \uline\lambda} \chi_{\Box} ^{k_{\text{CS}}} \,  \PE[-\text{Ch}_{\mathsf{T}} \left(E_{\uline\lambda} \right)] \PE[\text{Ch}_{\mathsf{T}} \left(T_{\uline\lambda}\cM_{n,k} \right)] Q_f (X)_{\underline{\l}}}{\sum_{\uline\lambda} \qe^{|\uline\lambda|} \prod_{\Box \in \uline\lambda} \chi_{\Box} ^{k_{\text{CS}}} \,  \PE[-\text{Ch}_{\mathsf{T}} \left(E_{\uline\lambda} \right)] \PE[\text{Ch}_{\mathsf{T}} \left(T_{\uline\lambda}\cM_{n,k} \right)]}.
    \end{align}

	\paragraph{$H$-observable.}
    The 3d $\CalN=2$ $\U(1)$ gauge theory engineered by an AV brane may transit into Higgs phase, where the Coulomb parameter $X$ is fixed at discrete loci (twisted masses of the chiral multiplets) and the complexified FI parameter $Z$ is turned on to open up a Higgs branch. Depending on whether $X$ is locked at the twisted mass of a chiral multiplet of charge $+1$ or charge $-1$, the complexified FI parameter can be turned on with either $\vert Z \vert>1$ or $\vert Z \vert<1$. In this work, we will only consider the latter case.
    
    In this Higgs phase, the path integral on $\BR^2 _{\ve_1} \times S^1$ localizes to vortex configurations. The resulting K-theoretic vortex partition function can be obtained as a $q_1$-lattice summation of the $Q$-observable.
    	\begin{equation}
		H^{(\alpha)}_f(Z)_\varnothing = \sum_{X \in a_\alpha q_1^{\mathbb Z}} \exp\left[-\frac{\log X }{\log q_1} \log\left(Z(-1)^f (X q_1^{-1})^{\frac{f}{2}}\right)\right] Q_0(Xq_1 ^{-1})_\varnothing,
	\end{equation}
    where $\a\in \{1,2,\cdots, n\}$ labels the isolated Higgs vacua. Note that the summation truncates from above due to the zeros of $Q_0 (Xq_1^{-1})_\varnothing = \frac{\prod_{\a=1}^n \left( \frac{a_\a}{X} q_1;q_1 \right)_\infty}{\prod_{i=1}^{n_{\mathrm{f}}} \left( \frac{m_i}{X} q_1 ;q_1 \right)_\infty }$, ensuring the series is convergent in the domain $\vert Z \vert <1$. The Chern-Simons level $-f$ only appears in a purely classical factor in the summand \cite{Dimofte:2017tpi,Bullimore:2020jdq}.

    Now, let us couple this 3d $\CalN=2$ gauge theory to the 5d $\CalN=1$ $\U(n)$ gauge theory. At each instanton sector, the 3d contribution to the partition function is still summed over the vortex configurations. Since the K-theoretic vortex partition function is written explicitly in terms of the $q_1$-lattice summation of the $Q$-observable at the zero-instanton sector, it is straightforward to generalize this expression to the non-zero instanton sector. Namely, the so-defined $H$-observable is obtained as
	\begin{equation} \label{eq:hobs}
		H^{(\alpha)}_f(Z) := \sum_{X \in a_\alpha q_1^{\mathbb Z}} \exp\left[-\frac{\log X }{\log q_1} \log\left(Z(-1)^f (X q_1^{-1})^{\frac{f}{2}}\right)\right] Q_0(Xq_1 ^{-1}),
	\end{equation}
    by simply adopting the $Q$-observables in the presence of the bulk instanton configuration. It should be noted that the summation remains to be truncated from above due to the zeros of $Q_0(X q_1^{-1})$, ensuring the convergence for $|Z|<1$. In contrast to the zero-instanton sector, however, the vortex number is bounded below by $-k$ instead of $0$, where $k\geq 0$ is the instanton number, indicating that both vortices and anti-vortices contribute to the partition function. This is the characteristic feature of the defect defined by GLSM in Higgs phase, which was already observed in the 2d-4d coupled system in \cite{Jeong:2018qpc}.

 Given the observable expression, the vacuum expectation value is computed to be
    \begin{align} \label{eq:hvev}
           \CalZ_\text{open} =\left\langle H_f ^{(\a)} (Z)  \right\rangle =  \frac{\sum_{\uline\lambda} \qe^{|\uline\lambda|} \prod_{\Box \in \uline\lambda} \chi_{\Box} ^{k_{\text{CS}}} \,  \PE[-\text{Ch}_{\mathsf{T}} \left(E_{\uline\lambda} \right)] \PE[\text{Ch}_{\mathsf{T}} \left(T_{\uline\lambda}\cM_{n,k} \right)] H_f ^{(\a)} (Z)_{\underline{\l}}}{\sum_{\uline\lambda} \qe^{|\uline\lambda|} \prod_{\Box \in \uline\lambda} \chi_{\Box} ^{k_{\text{CS}}} \,  \PE[-\text{Ch}_{\mathsf{T}} \left(E_{\uline\lambda} \right)] \PE[\text{Ch}_{\mathsf{T}} \left(T_{\uline\lambda}\cM_{n,k} \right)]}.
    \end{align}

	The main assertion of the present work is that the vacuum expectation values \eqref{eq:qvev} and \eqref{eq:hvev} are the generating functions \eqref{eq:openbps} of refined open BPS invariants of the pair $(\CalX,\CalL)$, where the continuous modulus of the AV brane is given by either the Coulomb parameter $X= t e^{i \th}$ or the complexified FI parameter $Z = t e^{i\th}$, depending on the phase of the canonical codimension-two defect, and the framing is given by $f$.

\subsection{Quantum mirror curves and their solutions}
Recall that the moduli space of the AV brane is given by the mirror curve $\S \subset (\BC^\times)^2$, parametrized by the continuous modulus $t e^{i \th} \in \BC^\times$. From the perspective of geometric engineering, we proposed that this modulus is identified with either the Coulomb parameter $X$ or the complexified FI parameter $Z$, depending on the toric leg on which the AV brane ends and hence on the phase of the canonical codimension-two defect theory. 

For the canonical surface defect in 4d $\CalN=2$ gauge theory in its Higgs phase, it is well known that the discrete Higgs vacua form a branched covering of the moduli space of the complexified FI parameter, which defines the Seiberg-Witten curve. Equivalently, the chiral ring equation governing local observables in the coupled 2d-4d system is given by the Seiberg-Witten curve.

In the present work, we consider the uplift to a coupled 3d-5d system. In this case, the discrete Higgs vacua over the moduli space of the complexified FI parameter again form the Seiberg-Witten curve of the 5d $\CalN=1$ gauge theory \cite{Nekrasov:1996cz}. The corresponding 5d Seiberg-Witten curve equation is therefore identified with the chiral ring relation among line defects wrapping the circle $S^1$ in the 3d-5d coupled system. In this uplift, the Coulomb and Higgs phases are treated democratically, differing only in that the defect parameter is identified with $X$ or with $Z$ giving conjugate parametrization of the same curve $\S$. 

This realization of the Seiberg-Witten curve is consistent with our proposal $X = t e^{i\th}$ or $Z = t e^{i\th}$, which identifies the continuous modulus of the mirror curve $\S$ with the defect parameter. This follows from the fact that the mirror curve of $\CalX$ coincides with the Seiberg-Witten curve of the geometrically engineered 5d $\CalN=1$ theory \cite{Hori:2000kt,Eguchi:2000fv}.\\

In this subsection, we introduce an additional layer involving the quantization of the mirror curve. We present a $q$-difference equation satisfied by the vacuum expectation value of the canonical codimension-two defect, which reduces to the classical mirror curve equation in the limit $q \to 1$. This \textit{quantum mirror curve} equation makes manifest that the Coulomb parameter $X$ or the complexified FI parameter $Z$ $-$ depending on the phase of the canonical codimension-two defect $-$ is identified with a parameter of the curve in a specific polarization, whose conjugate variable is promoted to a $q$-difference operator. Crucially, the polarization in which the mirror curve is quantized is determined by the discrete data of the AV brane, namely the toric edge on which it ends and the framing $f \in \BZ$.\footnote{An alternative prescription for quantization for a class of Calabi-Yau threefolds based on remodeling conjecture (which is already a theorem now) of \cite{Bouchard:2007ys} was provided in \cite{Banerjee:2025shz}. The open BPS partition function computed there depends on the choice of the basepoint ($X$ or $Z$), much in parallel to the discussion in this section. }

	\subsubsection{\texorpdfstring{$qq$}{qq}-Character}

    To derive the desired quantum mirror curve equation, we make use of a distinguished class of codimension-four observables: the $qq$-characters \cite{Nekrasov:2015wsu}. Let us first recall its definition and properties.
    
    In the 5d $\mathcal N=1$ $\text{U}(n)$ gauge theory compactified on a circle, there is a generalized Wilson line defined by
	\begin{align}
		W =\text{Tr} \,\text{Pexp} \,\oint_{S^1} \dd t (\phi +i A_0) =: \text{Tr}\, e^{\Phi},
	\end{align}
	where $\phi$ is the real scalar in the vector multiplet and $A_0$ is the component of the gauge field along the direction of $S^1$. A distinguished class of codimension-four defects generalizing this Wilson line is given by
	\begin{align} \label{eq:lineop}
		W_k := \text{Tr}\, e^{k \Phi},\qquad k \geq 0.
	\end{align}
	The generating function of these codimension-four defects is called the $\CalY$-observable. It is defined by
	\begin{align} 
		\CalY(X) := \exp \left[ -\sum_{k=1} ^\infty \frac{1}{k X^k} W_k \right].
	\end{align}
	By supersymmetric localization, the codimension-four observable $W_k$ turns into an equivariant K-theory class, given by the $k$-th Adams operation on the universal bundle at the origin, $\psi^k (S)$. Thus, the observable expression at the fixed point $\uline\lambda$ is given by its equivariant Chern character,
	\begin{align}
		\CalY (X)_{\uline\lambda} =&\; \PE\left[-X^{-1} S_{\uline\l} \right] .
		\label{Yatfixed}
	\end{align}
	
	The $qq$-characters are defined by certain Laurent polynomials in $\CalY$-observables with arguments shifted by $q_1$ and $q_2$ \cite{Nekrasov:2015wsu}. In our case of the $\text{U}(n)$ gauge theory with $f$ fundamental and $\tilde{f}$ anti-fundamental hypermultiplets, the fundamental $qq$-character is written as
	\begin{align} \label{eq:qqchar}
    \begin{split}
		\mathfrak{X}(X) : = \CalY(q_1 q_2 X) + {(-1)^n} a X^{-n+k_{\text{CS}}} \sfq \frac{ \prod_{i=1}^f \left(1 - \frac{{X}}{m_i}\right) \prod_{i=1}^{\wtd f} \left(1 - \frac{\wtd{m}_i}{X} \right)}{ \CalY(X)},
    \end{split}
	\end{align}
	where we denoted $a \equiv \prod_{\a=1}^n a_\a$. A crucial property of the $qq$-character is that its normalized vacuum expectation value,
	\begin{align}
		\begin{split}
			T(X) :=& \frac{1}{\CalZ^{\text{5d}}} \left\langle \mathfrak X(X) \right\rangle \\ 
			=&\frac{1}{\CalZ^{\text{5d}}} \left\langle \CalY(q_1 q_2 X) + {(-1)^n} a X^{-n+k_{\text{CS}}} \sfq \frac{ \prod_{i=1}^f \left(1 - \frac{{X}}{m_i}\right) \prod_{i=1}^{\wtd f} \left(1 - \frac{\wtd{m}_i}{X} \right)}{ \CalY(X)} \right\rangle,
			\label{T}
		\end{split}
	\end{align}
	is regular in $ X \in \BC^\times$, due to the compactness of the moduli space of spiked instantons \cite{Nekrasov:2015wsu,Nekrasov:2016qym}. Namely, $T(X)$ is a Laurent polynomial in $X$. By expanding \eqref{eq:qqchar} either in $X \to 0$ or $X\to \infty$, the coefficients of $T(X)$ are obtained as the correlation functions of the codimension-four line operators \eqref{eq:lineop}. These coefficients do not admit closed expressions in general, but can be expanded in series in the gauge coupling $\qe$, whose coefficients as rational functions of the equivariant parameters can be exactly determined up to arbitrary order.
	
	\subsubsection{TQ equations and \texorpdfstring{$q$}{q}-opers as quantum mirror curves}

    We give a derivation of the $q$-difference equation satisfied by the vev of the canonical codimension-two defect. For this, it is important to note that the $\CalY$-observable \eqref{Yatfixed} is expressed as a ratio of the $Q$-observables \eqref{eq:qobs},
	\begin{align}
		\CalY(X) = \left(-{X}\right)^f \prod_{i=1} ^{n_{\text{f}}} \left(1-\frac{m_i}{X} \right) \frac{Q_f(X)}{Q_f (q_1 ^{-1}X)}.
	\end{align}
	In the limit $q_2 \to 1$, the local and non-local observables supported on the $\BR^2_{\ve_1}$-plane can be arbitrarily separated on the topological $\BR^2_{\ve_2}$-plane, without affecting the correlation function. Thus, the correlation functions in \eqref{T} factorize into the product of individual vacuum expectation values, yielding a second-order $q_1$-difference equation,
	\begin{align} \label{eq:tqeq}
		\begin{split}
			0&=\left[ \left( - {q_1 X} \right)^f \prod_{i=1} ^{n_{\text{f}}} \left( 1- \frac{m_i}{q_1 X} \right) q_1 ^{D_X}  -T(X) \right. \\
			&\quad\quad + \left. (-1)^{n+n_{\text{f}}} \frac{a}{m} \left( - {X} \right)^{-f} X^{-n +n_{\text{f}} +k_{\text{CS}}} \mathfrak{q}  \prod_{i=1} ^{n_{\text{af}}} \left(1-\frac{\tilde{m}_i}{X}  \right) q_1 ^{-D_X}  \right] \left\langle Q_f (X) \right\rangle_{q_2=1} \\
			&= {\mathcal{D}} \left( X, q_1 ^{D_X} (-X)^f \right) \left\langle Q_f (X) \right\rangle_{q_2=1},
		\end{split}
	\end{align}
	where $\left\langle Q_f (X) \right\rangle_{q_2=1}$ denotes the $q_2 \to 1$ limit of the normalized vev at the vacuum specified by the Coulomb moduli $\ba$. We stress that this $q$-difference equation is derived here in the way that it is valid up to arbitrary order of the gauge coupling $\qe$ by construction.
    
    In the classical limit $q_1 \to 1$, the normalized vev behaves as $ \left\langle Q_f (X) \right\rangle_{q_2=1} = \exp \frac{\widetilde{\EuScript{W}}_f (\ba;X) }{\log q_1} + \cdots$ where $\widetilde{\EuScript{W}}_f (\ba;X)$ is the effective twisted superpotential of the KK-reduced 2d $\CalN=(2,2)$ defect theory. Here, $X$ is the complex scalar of the effective twisted chiral multiplet. Its dual variable $Z = \frac{\p \widetilde{\EuScript{W}}_f (\ba;X) }{\p \log X}$ is the effective complexified FI parameter. Parametrized by $(X,Z) \in( \BC^\times)^2$, the classical limit of the $q_1$-difference equation \eqref{eq:tqeq} yields a curve $\S \subset (\BC^\times)^2 $,
	\begin{align}
		\S: 0 = \mathcal{D}\left( X, Z (-X)^f \right). \label{eq:mirrorCurve}
	\end{align}
	This is exactly the Seiberg-Witten (SW) curve of the associated 5d $\CalN=1$ gauge theory. In the perspective of the geometric engineering for the toric CY threefold $\CalX$, it is also the mirror curve of $\CalX$ at the framing $f$. In this sense, the $q_1$-difference operator \eqref{eq:tqeq} we obtained is the $q_1 \neq 1$ uplift of the mirror curve. We call it the \textit{quantum mirror curve}. 
	
	Note that the framing dependence $f\in \BZ$ exclusively appears in the argument $q_1 ^{D_X} (-X)^f$ of the quantum mirror curve. In other words, turning on $f\in \BZ$ amounts to the transformation
	\begin{align} \label{eq:qfram}
		X \mapsto X,\qquad q_1 ^{D_X} \mapsto q_1 ^{D_X} (-X)^f.
	\end{align}
	This is precisely the quantum uplift of the reparametrization of the mirror curve under the framing transformation of the coordinates. The terminology for $f \in \BZ$ is justified in this sense.  We thus have the schematic correspondences:
    \begin{equation}
		\begin{tikzcd} 
			\substack{\text{Toric CY,  $\mathcal X$} \\ +\text{ AV A-brane on $\CalL$ with framing $f$}} \arrow[r, "\text{Mirror symmetry}"] \arrow[d, "\text{Compactify M-theory on $\CalX$}"] &[1.5cm]  \substack{\text{Mirror CY, $\CalX^\vee = \left\{uv= \mathcal{D} \left(X,Z(-X)^f\right)\right\}$} \\ + \text{B-brane at $\text{pt}\in \S = \{\mathcal{D}\left(X,Z(-X)^f\right) =0\} $}  } \arrow[d, "\text{Quantum mirror curve}"]
			\\
			\substack{\text{5d $\mathcal N=1$ $\mathrm{U}(n)$ theory}\\\text{with codim-2 defect}} \arrow[r, "\text{Quantum SW curve}"'] & \mathcal{D}\left(X, q_1^{D_X}(-X)^f\right) Q_f(X) = 0
		\end{tikzcd}
	\end{equation}
	
	Next, let us turn to the quantum mirror curve arising in the Higgs phase. Recall that the $H$-observable is defined as a Fourier transformation \eqref{eq:hobs} of the $Q$-observable. Using the $q_1$-shift invariance $q_1 \left( a_\a q_1 ^\BZ \right) = a_\a q_1 ^\BZ$ of the $q_1$-lattice, it is straightforward to show that the $q_1$-difference equation \eqref{eq:tqeq} for $  \left\langle Q_f (X) \right\rangle_{q_2=1} $ implies 
	\begin{align}
		{\mathcal{D}}\left(q_1 ^{-1-D_Z} , Z(-q_1^{-1-D_Z})^f \right) \left\langle H_f (Z) \right\rangle_{q_2=1} = 0, \label{eq:diffeqH}
	\end{align}
	where $\left\langle H_f (Z) \right\rangle_{q_2=1}$ is the $q_2\to 1$ limit of the normalized vev at the vacuum specified by $\ba$. By construction, the classical limit $q_1 \to 1$ of this $q_1$-difference equation yields the same mirror curve $\S$ \eqref{eq:mirrorCurve}. Compared to the previous \eqref{eq:tqeq}, this is quantum mirror curve associated with the dual polarization. The quantum uplift of the framing transformation changed accordingly as
    \begin{align}
        q_1 ^{-1-D_Z} \mapsto q_1 ^{-1-D_Z},\qquad Z \mapsto Z (-q_1 ^{-1-D_Z})^f.
    \end{align}
    
Spanning all framing transformations of both \eqref{eq:tqeq} and \eqref{eq:diffeqH}, we obtain quantum mirror curve associated with any polarization related to one another by the $SL(2;\BZ)$ transformations. We stress that the discrete labels of the AV brane $-$ the toric leg it ends on and the framing $f\in \BZ$ $-$ determines this polarization.	

\begin{remark}
The Coulomb branch of the 5d $\CalN=1$ gauge theory compactified on $T^2$ is given by the moduli space of multiplicative Higgs bundles on $\BC^\times$ ($\BP^1$ with framings at $\{0,\infty\}\subset \BP^1$) with ramifications at marked points \cite{Nekrasov:2012xe,Elliott:2018yqm}. From this perspective, the $q$-difference operators \eqref{eq:tqeq} and \eqref{eq:diffeqH} span an affine space of \textit{$q$-opers}, a holomorphic Lagrangian submanifold in the moduli space of $q$-connections \cite{Jeong:2025yys}. The canonical codimension-two defect realizes a Hecke operator associated with the fundamental coweight, explaining the notation for the $H$-observable. In terms of the associated quantum Hitchin integrable system, the $q$-difference equation is also identified with the Baxter TQ equation, where the canonical codimension-two defect gives a $Q$-operator, explaining the notation for the $Q$-observable. The two $q$-opers for $Q$- and $H$-observables are related by the Fourier-Mukai transformation between the moduli spaces of the multiplicative Higgs bundles, under which the rank of the base Lie algebra and the number of regular marked points get swapped; as Baxter TQ equations, this transformation realizes the bispectral duality \cite{Gorsky:1997jq} between the associated integrable models. See \cite{Jeong:2018qpc,Jeong:2023qdr,Jeong:2024hwf} for studies of opers and $\hbar$-opers in 4d $\CalN=2$ theory context. 
\end{remark}
	
\begin{remark}
We remark that the $q_2 \neq 1$ refinement of the $q_1$-difference equation \eqref{eq:tqeq}, which relates the vacuum expectation value of the $Q$-observable to its correlation function with codimension-four observables, is also derived in a recent work \cite{Jeong:2025yys}. In terms of the geometry of the mirror $\CalX^\vee$, this \textit{quantum $q$-oper} equation is therefore a \textit{double} quantization of the mirror curve $\S$. It would be interesting to investigate implications of the quantum $q$-opers on counting refined open BPS invariants. See also \cite{Jeong:2017pai,Jeong:2017mfh,Jeong:2019fgx,Jeong:2020uxz,Jeong:2021rll} for related works.
\end{remark}

	\subsubsection{Wave functions for general curves in the quantum torus}
	Generally speaking, given any algebraic curve in the torus
	\begin{equation}
		\Sigma := \{\mathcal D(X,Z) = 0\} \subset (\mathbb C^\times)^2 \label{curve}
	\end{equation}
	a difference equation $\mathcal D'(X, q_1^{D_X}) \psi(X)=0$ defines a quantization of the curve if the symbol of the difference operator coincides with the curve
	\begin{equation}
		\sigma(\mathcal D'(X, q_1^{D_X})) = \mathcal D(X,Z).
	\end{equation}
	Here $\si$ denotes the symbol we get by replacing $q_1^{D_X} \to Z$.
	
	Of course, there is nothing special about the coordinate $X$ and we could consider difference equations in the variable $Z$ as well. Alternatively, we can work in a more coordinate-independent fashion by using an operator formalism as follows. We consider the $q_1$-Weyl algebra generated by $\hat X, \hat Z$
	\begin{equation}
		q_1\text{-Weyl} := \mathbb C[\hat X, \hat Z]/\langle\hat Z \hat X - q_1 \hat X \hat Z\rangle \label{qWeyl}
	\end{equation}
	and their representations on a Hilbert space. A quantization of an algebraic curve \eqref{curve} is then an operator $\mathcal D(\hat X, \hat Z)$ whose symbol is $\mathcal D(X,Z)$. As usual there are ordering ambiguities when writing down the quantum operator and one has to make a choice. In case of the TQ equation \eqref{eq:tqeq}, the choice is made by the supersymmetric localization procedure as the output is an unambiguous difference operator with a well-defined ordering.
	
	Let us consider a generic situation: we start with an operator
	\begin{equation}
		\mathcal D_0(\hat X, \hat Y) = \sum_{\mu, \nu \in \mathbb Z} \gamma_{\mu,\nu} \hat X^\mu \hat Z^\nu \label{Op0}
	\end{equation}
	where only finitely many coefficients $\gamma_{\mu\nu}$ are non-zero, so that the symbol of the operator is a Laurent polynomial. The operator acts on a Hilbert space with a ``position basis'' $\{\ket{X}\}_{X \in \mathbb C^\times}$ and a ``momentum basis'' $\{|Z\rangle\!\rangle\}_{Z \in \mathbb C^\times}$. They are eigenvectors of $\hat X$ and $\hat Z$ respectively:
	\begin{equation}
		\hat X \ket{X} = X \ket{X}, \qquad \hat Z |Z\rangle\!\rangle = Z |Z\rangle\!\rangle.
	\end{equation}
	Since the labels are continuous, they are normalized with the Dirac delta function on $\mathbb C^\times$:
	\begin{equation}
		\langle X | X' \rangle = \delta(X-X'), \qquad \langle Z | Z' \rangle = \delta(Z-Z'), \qquad \int_{\mathbb C^\times} \mathrm dX \delta(X-X') =1.
	\end{equation}
	After a choice of normalization for the operators, the canonical relation \eqref{qWeyl} implies:\footnote{These normalizations are related to representations of $\hat X$ and $\hat Z$ in the $Z$ and the $X$-basis respectively. With these choices we get: $\langle X|\hat Z|\psi\rangle = \int \mathrm{d}X' \langle X|\hat Z|X'\rangle\langle X'|\psi\rangle = \int \mathrm dX' q_1^{-1} \delta(X-q_1^{-1}X') \psi(X') = \psi(q_1X) = q_1^{D_X} \psi(X)$ and similarly $\langle\!\langle Z|\hat X|\psi\rangle = q_1^{-1-D_Z} \widetilde\psi(Z)$. The choices are not unique and were made for convenience.}
	\begin{equation}
		\hat Z\ket{X} = q_1^{-1} \ket{q_1^{-1}X}, \qquad \hat X |Z\rangle\!\rangle = |q_1 Z\rangle\!\rangle.
	\end{equation}
	
	Let $\ket{\psi_0}$ be a state annihilated by the operator \eqref{Op0}:
	\begin{equation}
		\mathcal D_0(\hat X, \hat Z) \ket{\psi_0} = 0. \label{Op0Kernel}
	\end{equation}
	By taking inner product of the above with $\bra{X}$ and inserting the resolution of identity $\hat 1 = \int \mathrm{d}X \ket{X}\bra{X}$ between $\mathcal D_0$ and $\ket{\psi_0}$ we get the following difference equation for $\psi_0(X) := \langle X|\psi_0\rangle$
	\begin{equation}
		\mathcal D_0(X, q_1^{D_X}) \psi_0(X) = 0. \label{Op0DiffX}
	\end{equation}
	This difference equation represents the operator equation \eqref{Op0Kernel} in the $X$-polarization. To get the representation in the $Z$-polarization we take the inner product of \eqref{Op0Kernel} with $\langle\!\langle Z|$ and insert the resolution $\hat 1=\int \mathrm{d}Z |Z\rangle\!\rangle \langle\!\langle Z|$ between $\Sigma_0$ and $\ket{\psi_0}$ to get
	\begin{equation}
		\mathcal D_0(q_1^{-1-D_Z}, Z) \widetilde \psi_0(Z) = 0 \label{Op0DiffZ}
	\end{equation}
	where $\widetilde \psi_0(Z) := \langle\!\langle Z|\psi_0\rangle$. Both difference equations, \eqref{Op0DiffX} and \eqref{Op0DiffZ}, are equivalent to the operator equation \eqref{Op0Kernel},  expressed in terms of wave functions in different polarizations.

	\subsubsection{Defects as wave functions and automorphisms of the quantum torus} \label{sec:aut}
	Let us introduce two natural automorphisms $\mathrm{Ad}_{\hat S},\, \mathrm{Ad}_{\hat T}: q_1\text{-Weyl} \xrightarrow{\sim} q_1\text{-Weyl}$:
	\begin{subequations}
		\begin{gather}
			\hat S \hat X \hat S^{-1} = q_1^{-1} \hat Z^{-1}, \qquad \hat S \hat Z \hat S^{-1} = q_1 \hat X \label{S}
			\\
			\text{for } f,n \in \mathbb Z \qquad \hat T^f \hat X \hat T^{-f} = \hat X, \qquad \hat T^f \hat Z^n \hat T^{-f} = (-\hat X)^{fn} q_1^{\frac{f}{2}n(n+1)} \hat Z^n. \label{Tf}
		\end{gather}
	\end{subequations}
	The operator $\hat T$ can be constructed from the more elementary operator $\hat X$ as:
	\begin{equation}
		\hat T := \theta(\hat X^{-1};q_1) \label{Ttheta}
	\end{equation}
    where $\theta$ is the same $q$-theta function as in \eqref{eq:1loopc}.
	
	In the classical limit, $\hat S$ and $\hat T$ reduce to coordinate transformations of the torus. In terms of the log coordinates $\binom{z}{x} := \binom{\log Z}{\log X}$ they correspond to rotation by $-\pi/2$ and shearing up to affine shifts:
	\begin{equation}\begin{aligned}
			S: \begin{pmatrix} z \\ x \end{pmatrix} \mapsto&\; \begin{pmatrix} 0 & 1 \\ -1 & 0 \end{pmatrix} \begin{pmatrix} z \\ x \end{pmatrix} + \varepsilon_1 \begin{pmatrix} 1 \\ -1 \end{pmatrix}, 
			\\
			T^f : \begin{pmatrix} z \\ x \end{pmatrix} \mapsto&\; \begin{pmatrix} 1 & f \\ 0 & 1 \end{pmatrix} \begin{pmatrix} z \\ x \end{pmatrix} + f \begin{pmatrix} \ii \pi + \varepsilon_1 \\ 0 \end{pmatrix}.
			\label{STf}
	\end{aligned}\end{equation}
	When $\mathcal D(X,Z)=0$ corresponds to the mirror curve a toric Calabi-Yau $\mathcal X$, we shall depict the curve by the toric diagram of $\mathcal X$. This diagram coincides with the $(p,q)$-brane web whose low energy dynamics is described by the same 5d $\mathcal N=1$ theory engineered by compactifying M-theory on $\mathcal X$. The toric diagrams/$(p,q)$-webs are drawn in the real $xz$-plane and the classical limits of the $\hat S$ and $\hat T$ act as rotation and shearing on these diagrams.
	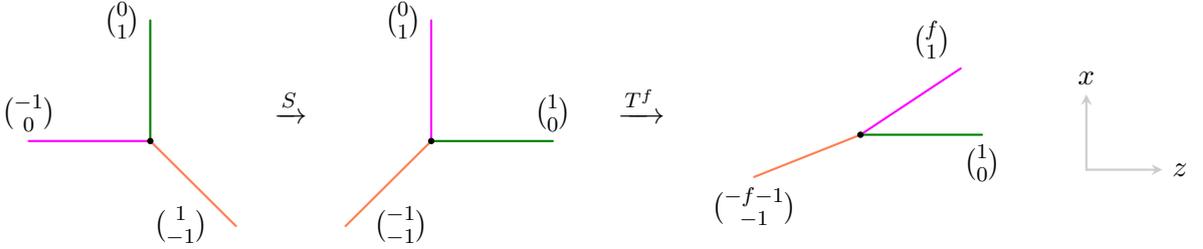
\begin{figure}[H]
		\begin{equation*}
			\begin{tikzpicture}[x=.8cm,y=.8cm,line cap=round,line join=round,baseline=(current bounding box.center)]
				\tikzset{
					edge/.style={thick},
				}
				\coordinate (o) at (0,0);
				\coordinate (v1) at (-2,0);
				\coordinate (v2) at (0,2);
				\coordinate (v3) at (1.41,-1.41);
				
				\draw[edge, Magenta] (o) -- (v1);
				\draw[edge, Green] (o) -- (v2);
				\draw[edge, Coral] (o) -- (v3);
				
				\fill[black] (o) circle (1.2pt);
				
				\node[left=0cm of v2] () {$\binom{0}{1}$};
				\node[above=0cm of v1] () {$\binom{-1}{0}$};
				\node[left=.3 of v3] () {$\binom{1}{-1}$};
			\end{tikzpicture}
			\quad
			\xrightarrow{S}
			\quad
			\begin{tikzpicture}[x=.8cm,y=.8cm,line cap=round,line join=round,baseline=(current bounding box.center)]
				\tikzset{
					edge/.style={thick},
				}
				\coordinate (o) at (0,0);
				\coordinate (v1) at (2,0);
				\coordinate (v2) at (0,2);
				\coordinate (v3) at (-1.41,-1.41);
				
				\draw[edge,Green] (o) -- (v1);
				\draw[edge,Magenta] (o) -- (v2);
				\draw[edge,Coral] (o) -- (v3);
				
				\fill[black] (o) circle (1.2pt);
				
				\node[left=0cm of v2] () {$\binom{0}{1}$};
				\node[above=0cm of v1] () {$\binom{1}{0}$};
				\node[right=.3 of v3] () {$\binom{-1}{-1}$};
			\end{tikzpicture}
			\quad
			\xrightarrow{T^f}
			\quad
			\begin{tikzpicture}[x=.8cm,y=.8cm,line cap=round,line join=round,baseline=(current bounding box.center)]
				\tikzset{
					edge/.style={thick},
				}
				\coordinate (o) at (0,0);
				\coordinate (v1) at (2,0);
				\coordinate (v2) at (1.65,1.1);
				\coordinate (v3) at (-1.75,-.7);
				
				\draw[edge,Green] (o) -- (v1);
				\draw[edge,Magenta] (o) -- (v2);
				\draw[edge,Coral] (o) -- (v3);
				
				\fill[black] (o) circle (1.2pt);
				
				\node[above left=0cm of v2] () {$\binom{f}{1}$};
				\node[below=0cm of v1] () {$\binom{1}{0}$};
				\node[below=0 of v3] () {$\binom{-f-1}{-1}$};
			\end{tikzpicture}
			\qquad
			\begin{tikzpicture}[x=.5cm,y=.5cm,line cap=round,line join=round,baseline=(current bounding box.center)]
				\tikzset{
					edge/.style={thick},
				}
				\coordinate (o) at (0,0);
				\coordinate (v1) at (2,0);
				\coordinate (v2) at (0,2);
				
				\draw[edge, -stealth, black!20] (o) -- (v1);
				\draw[edge, -stealth, black!20] (o) -- (v2);
				
				\node[right = 0cm of v1] () {$z$};
				\node[above = 0cm of v2] () {$x$};
			\end{tikzpicture}
		\end{equation*}
		\caption{Transformation of a classical curve (corresponding to the Calabi-Yau $\mathbb C^3$), represented by a toric diagram in the real $xz$-plane, under $S$ and $T$-transformations according to \eqref{STf}. The colors are only to help identify the same edge across transformations.}
	\end{figure}
	
	The operator $\mathcal D_0(\hat X, \hat Z)$ from \eqref{Op0Kernel} transforms under the automorphisms \eqref{S} and \eqref{Tf} simply by conjugation. We denote them by:
	\begin{equation}\begin{aligned}
			\mathcal D_f(\hat X, \hat Z) :=&\; \hat T^f \mathcal D_0(\hat X, \hat Z) \hat T^{-f} = \mathcal D_0\left(\hat X, \hat Z (-\hat X)^f\right), 
			\\
			\mathcal D_f^S(\hat X, \hat Z) :=&\; \hat S \mathcal D_f(\hat X, \hat Z) \hat S^{-1} = \mathcal D_f(q_1^{-1} \hat Z^{-1}, q_1 \hat X).
		\end{aligned}
		\label{transformedOp}
	\end{equation}
	Let us similarly define the transformed states and the corresponding wave functions:
	\begin{equation}\begin{aligned}
			\ket{\psi_f} :=&\; \hat T^f \ket{\psi_0}, \qquad&\; \ket{\psi_f^S} :=&\; \hat S \ket{\psi_f},
			\\
			\psi_f(X) :=&\; \langle X|\psi_f\rangle, \qquad&\; \psi_f^S(X) :=&\; \langle X|\psi_f^S\rangle,
			\\
			\widetilde\psi_f(Z) :=&\; \langle Z|\psi_f\rangle, \qquad&\; \widetilde\psi_f^S(Z) :=&\; \langle Z|\psi_f^S\rangle.
		\end{aligned}\label{transformedState}\end{equation}
	Then, starting from the annihilation of the original state \eqref{Op0Kernel} we can derive the following difference equations by taking inner products with appropriate states, inserting resolutions of identity, and using the canonical relations:
	\begin{equation}
		\begin{aligned}
			\mathcal D_f(X, q_1^{D_X}) \psi_f(X) =&\; 0,
			\quad&\;
			\mathcal D_f(q_1^{-1-D_X}, X) \psi_f^S(X) =&\; 0,
			\\
			\mathcal D_f(q_1^{-1-D_Z}, Z) \widetilde\psi_f(Z) =&\; 0,
			\quad&\;
			\mathcal D_f(q_1^{-1} Z^{-1}, q_1^{-D_Z}) \widetilde \psi^S_f(Z) =&\; 0.
		\end{aligned}
		\label{diffEqs}
	\end{equation}
	Also note that the operator relations \eqref{transformedOp} imply
	\begin{equation}
		\mathcal D_f(q_1^{-1-D_X}, X) = \mathcal D_f^S(X, q_1^{D_X}),
		\qquad
		\mathcal D_f(q_1^{-1} Z^{-1}, q_1^{-D_Z}) = \mathcal D_f^S(q_1^{-1-D_Z}, Z).
	\end{equation}
	Now, comparison with the difference equations \eqref{Op0DiffX} and \eqref{Op0DiffZ} for the reference curve shows that we can group curves, difference equations, and the solutions as follows:
	\begin{equation}\begin{aligned}
			\begin{array}{ccc}
				\text{Curve} & \text{Difference operators} & \text{Solution}
				\\\hline
				\multirow{ 2}{*}{$\sigma(\mathcal D_0(\hat X, \hat Z))=0$} & \mathcal D_0(X, q_1^{D_X}) & \psi_0(X)\\
				& \mathcal D_0(q_1^{-1-D_Z}, Z) & \widetilde\psi_0(Z)
				\\\hline
				\multirow{ 2}{*}{$\sigma(\mathcal D_f(\hat X, \hat Z))=0$} & \mathcal D_f(X, q_1^{D_X}) & \psi_f(X)\\
				& \mathcal D_f(q_1^{-1-D_Z}, Z) & \widetilde\psi_f(Z)
				\\\hline
				\multirow{ 2}{*}{$\sigma(\mathcal D_f^S(\hat X, \hat Z))=0$} & \mathcal D_f^S(X, q_1^{D_X}) & \psi_f^S(X)\\
				& \mathcal D_f(q_1^{-1-D_Z}, Z) & \widetilde\psi_f^S(Z)
				\\\hline
			\end{array}
	\end{aligned}\end{equation}

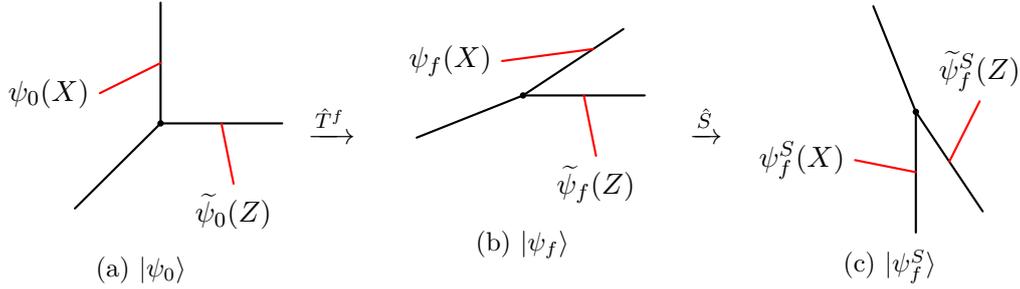
\begin{figure}[h] 
		\centering
		\raisebox{-0.5\height}{
			\begin{subfigure}{0.25\textwidth}
				\centering
				\begin{tikzpicture}[x=.8cm,y=.8cm,line cap=round,line join=round,baseline=0]
					\tikzset{
						edge/.style={thick},
					}
					\coordinate (o) at (0,0);
					\coordinate (v1) at (2,0);
					\coordinate (v2) at (0,2);
					\coordinate (v3) at (-1.41,-1.41);
					
					\fill[black] (o) circle (1.2pt);
					
					\draw[edge] (o) -- (v1);
					\draw[edge] (o) -- (v2);
					\draw[edge] (o) -- (v3);
					
					\coordinate (probe1i) at (0,1);
					\coordinate (probe1f) at ($(probe1i) + (-1,-.5)$);
					\draw[edge, Red] (probe1i) -- (probe1f);
					
					\coordinate (probe2i) at (1,0);
					\coordinate (probe2f) at ($(probe2i) + (.2,-1)$);
					\draw[edge, Red] (probe2i) -- (probe2f);
					
					\node[left=0cm of probe1f] () {$\psi_0(X)$};
					\node[below=0cm of probe2f] () {$\widetilde\psi_0(Z)$};
				\end{tikzpicture}
				\caption{$\ket{\psi_0}$}
			\end{subfigure}
		}
		$\xrightarrow{\hat T^f}$
		\raisebox{-0.5\height}{
			\begin{subfigure}{0.25\textwidth}
				\centering
				\begin{tikzpicture}[x=.8cm,y=.8cm,line cap=round,line join=round,baseline=0]
					\tikzset{
						edge/.style={thick},
					}
					\coordinate (o) at (0,0);
					\coordinate (v1) at (2,0);
					\coordinate (v2) at (1.65,1.1);
					\coordinate (v3) at (-1.75,-.7);
					
					\fill[black] (o) circle (1.2pt);
					
					\draw[edge] (o) -- (v1);
					\draw[edge] (o) -- (v2);
					\draw[edge] (o) -- (v3);
					
					\coordinate (probe1i) at ($0.3*(o) + 0.7*(v2)$);
					\coordinate (probe1f) at ($(probe1i) + (-1.5,-.2)$);
					\draw[edge, Red] (probe1i) -- (probe1f);
					
					\coordinate (probe2i) at ($0.5*(o) + 0.5*(v1)$);
					\coordinate (probe2f) at ($(probe2i) + (.2,-1)$);
					\draw[edge, Red] (probe2i) -- (probe2f);
					
					\node[left=0cm of probe1f] () {$\psi_f(X)$};
					\node[below=0cm of probe2f] () {$\widetilde\psi_f(Z)$};
				\end{tikzpicture}
				\caption{$\ket{\psi_f}$}
			\end{subfigure}
		}
		$\xrightarrow{\hat S}$
		\raisebox{-0.5\height}{
			\begin{subfigure}{0.25\textwidth}
				\centering
				\begin{tikzpicture}[x=.8cm,y=.8cm,line cap=round,line join=round,baseline=0]
					\tikzset{
						edge/.style={thick},
					}
					\coordinate (o) at (0,0);
					\coordinate (v1) at (0,-2);
					\coordinate (v2) at (1.1,-1.65);
					\coordinate (v3) at (-.7,1.75);
					
					\fill[black] (o) circle (1.2pt);
					
					\draw[edge] (o) -- (v1);
					\draw[edge] (o) -- (v2);
					\draw[edge] (o) -- (v3);
					
					\coordinate (probe1i) at ($0.5*(o) + 0.5*(v2)$);
					\coordinate (probe1f) at ($(probe1i) + (.5,1)$);
					\draw[edge, Red] (probe1i) -- (probe1f);
					
					\coordinate (probe2i) at ($0.5*(o) + 0.5*(v1)$);
					\coordinate (probe2f) at ($(probe2i) + (-1,.2)$);
					\draw[edge, Red] (probe2i) -- (probe2f);
					
					\node[above=0cm of probe1f] () {$\widetilde\psi_f^S(Z)$};
					\node[left=0cm of probe2f] () {$\psi_f^S(X)$};
				\end{tikzpicture}
				\caption{$\ket{\psi_f^S}$}
			\end{subfigure}
		}
		\caption{A $(p,q)$-web (black) represents a state for a quantized probe (red) constrained to the corresponding curve inside $(\mathbb C^\times)^2$. The three subfigures correspond to the states $\ket{\psi_0}$, $\ket{\psi_f}$, and $\ket{\psi_f^S}$ related by $\hat T$ and $\hat S$-transformations according to \eqref{transformedState}. A probe attached to a vertical (resp. horizontal) edge represents the associated wave function in the $X$ (resp. $Z$)-polarization.}
		\label{fig:wavefunctions}
	\end{figure}

	In this quantization picture, the algebraic torus $(\mathbb C^\times)^2$ serves as the phase space of a point probe and a curve $\Sigma = \{\mathcal D(X,Z)=0\}$ defines a subspace to which the probe is constrained. The state of the probe is captured by a state $\ket{\psi} \in \ker \mathcal D(\hat X, \hat Z)$. Projecting the the location of the probe either to the $X$-plane or to the $Z$-plane we can encode the state as a wave function in either polarization. Starting from a reference curve $\mathcal D_0(X,Z)=0$, we construct quantizations of transformed curves such as $\mathcal D_f(X,Z)=0$ and $\mathcal D^S_f(X,Z)=0$ by conjugating the corresponding operators with $\hat T$ and $\hat S$ as in \eqref{transformedOp}, and simultaneously transforming the states as in \eqref{transformedState}. In terms of $(p,q)$-web diagrams, we denote the wave functions by attaching the probe to either the vertical or the horizontal edges as in Fig. \ref{fig:wavefunctions}. Note in \eqref{diffEqs} that $\psi_f$ and $\widetilde\psi_f^S$ are solutions to the same difference equation, up to an inversion and rescaling of the coordinate,  and therefore must coincide up to endomorphisms of the kernel of the operator. The functions $\psi_f^S$ and $\widetilde\psi_f$ are solutions to exactly the same equation.  Concretely
	\begin{equation}
		\psi_f(X) \equiv \widetilde \psi^S_f\left((q_1X)^{-1}\right),
		\qquad
		\psi_f^S(X) \equiv \widetilde\psi_f(X).
	\end{equation}
	
\begin{figure}[h] 
		\centering
		\begin{equation*}
			\begin{tikzpicture}[x=.8cm,y=.8cm,line cap=round,line join=round,baseline=0]
				\tikzset{
					edge/.style={thick},
				}
				\coordinate (o) at (0,0);
				\coordinate (v1) at (2,0);
				\coordinate (v2) at (1.65,1.1);
				\coordinate (v3) at (-1.75,-.7);
				
				\fill[black] (o) circle (1.2pt);
				
				\draw[edge] (o) -- (v1);
				\draw[edge] (o) -- (v2);
				\draw[edge] (o) -- (v3);
				
				\coordinate (probe1i) at ($0.3*(o) + 0.7*(v2)$);
				\coordinate (probe1f) at ($(probe1i) + (-1.5,-.2)$);
				\draw[edge, Red] (probe1i) -- (probe1f);
				
				\coordinate (probe2i) at ($0.5*(o) + 0.5*(v1)$);
				\coordinate (probe2f) at ($(probe2i) + (.2,-1)$);
				\draw[edge, Red] (probe2i) -- (probe2f);
				
				\node[left=0cm of probe1f] () {$Q_f(X)$};
				\node[below=0cm of probe2f] () {$H_f(Z)$};
				
				\node[above=0cm of v2] () {{\scriptsize$(f,1)$}};
				\node[right=0cm of v1] () {{\scriptsize D5}};
				\node[below=0cm of v3] () {{\scriptsize $(-1-f,-1)$}};
			\end{tikzpicture}
			\qquad\xrightarrow{\hat S}\qquad
			\begin{tikzpicture}[x=.8cm,y=.8cm,line cap=round,line join=round,baseline=0]
				\tikzset{
					edge/.style={thick},
				}
				\coordinate (o) at (0,0);
				\coordinate (v1) at (0,-2);
				\coordinate (v2) at (1.1,-1.65);
				\coordinate (v3) at (-.7,1.75);
				
				\fill[black] (o) circle (1.2pt);
				
				\draw[edge] (o) -- (v1);
				\draw[edge] (o) -- (v2);
				\draw[edge] (o) -- (v3);
				
				\coordinate (probe1i) at ($0.5*(o) + 0.5*(v2)$);
				\coordinate (probe1f) at ($(probe1i) + (.5,1)$);
				\draw[edge, Red] (probe1i) -- (probe1f);
				
				\coordinate (probe2i) at ($0.5*(o) + 0.5*(v1)$);
				\coordinate (probe2f) at ($(probe2i) + (-1,.2)$);
				\draw[edge, Red] (probe2i) -- (probe2f);
				
				\node[right=0cm of probe1f] () {$\begin{aligned}&\;H_f^S(Z)\\=&\;Q_f\left((q_1Z)^{-1}\right)\end{aligned}$};
				\node[left=0cm and 0cm of probe2f] () {$\begin{aligned}&\;Q_f^S(X)\\=&\;H_f(X)\end{aligned}$};
				
				\node[below right=0cm of v2] () {{\scriptsize$(1,-f)$}};
				\node[below=0cm of v1] () {{\scriptsize NS5}};
				\node[above=0cm of v3] () {{\scriptsize $(-1,1+f)$}};
			\end{tikzpicture}
		\end{equation*}
		\caption{The abstract curves, probes, and wave functions of Fig. \eqref{fig:wavefunctions} realized as concrete $(p,q)$-web, D3 branes, and expectation values of codimension-2 defects in type IIB string theory. The $(p,q)$-branes engineer 5d $\mathcal N=1$ theories, the D3 branes (red) create codimesion-2 defects therein, and their expectation values realize the wave functions quantizing the Seiber-Witten curves. Expectation values of certain defects before and after S-duality coincide.}
		\label{fig:QHS}
	\end{figure}
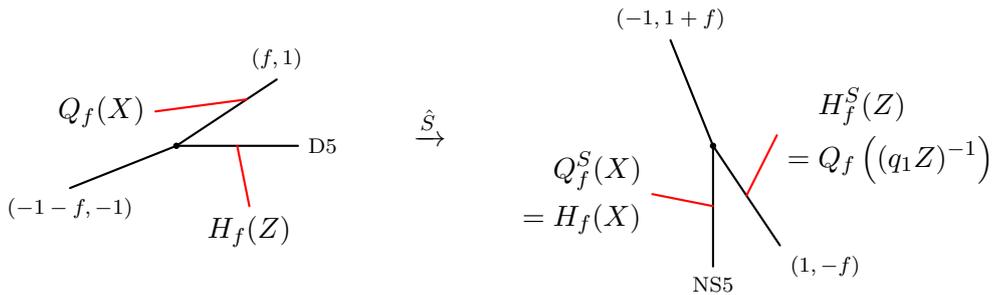

	One of the main proposals of this paper is that by interpreting the probe as a D3 brane ending on a $(p,q)$-web in type IIB string theory, the associated wave functions can be computed as expectation values of certain codimension-2 defects in the 5d $\mathcal N=1$ theories corresponding to the $(p,q)$-web. We compute the expectation values using supersymmetric localization and identify them as the appropriate wave functions by showing that they satisfy the difference equations \eqref{diffEqs} for the quantum Seiberg-Witten curves. From this perspective, certain wave functions related by $\hat S$-transformations being solutions to the same equation will be naturally tied to S-duality. More specifically, in a 5d $\mathcal N=1$ theory, we defined two codimension-2 defects whose expectation values are given by $Q_f(X)$ \eqref{eq:qobs} and $H_f(\alpha)(Z)$ \eqref{eq:hobs}. Their $q_2 \to 1$ limits are denoted by $Q_f(\mathbf a; X)$ and $H_f(\mathbf a;Z)$ respectively. We observe that the TQ equation \eqref{eq:tqeq} satisfied by $Q_f(\mathbf a;X)$ is the same as that of $\psi_f(X)$ from \eqref{diffEqs} and the difference equation \eqref{eq:diffeqH} satisfied by $H_f(\mathbf a;Z)$ coincides with $\widetilde\psi_f(X)$. $S$-transformation of the $(p,q)$-web corresponds to S-duality of type IIB string theory, which naturally swaps the role of the $Q$ and the $H$-observable, as evidenced by the matching difference equations according to \eqref{diffEqs}. We depict these information in diagrams as in Fig. \eqref{fig:QHS}.
	
	\subsection{Relations to enumerative invariants}
    In this section, we connect our results to the realms of the enumerative invariants. Particularly, in the limit $q_2=q_1^{-1}$, the closed BPS invariants are related to the so called Donaldson-Thomas (DT) invariants for the Calabi-Yau threefolds. The refined DT invariants have been understood mathematically rigorously and they are related to the `motivic DT invariants' of Kontsevich-Soibelman \cite{kontsevich2008stabilitystructuresmotivicdonaldsonthomas}. The refined Gromov-Witten (GW) invariants are also studied in the context of refined topological strings \cite{Aganagic:2011mi}. In \ref{sec:closedinv}, we comment on their origin in details and explain how they appear in our computations in this paper. 
    
    In contrast to refined open DT and refined open GW invariants, their unrefined open BPS counterparts are considerably better understood from a geometric perspective. In Section~\ref{sec:openGWDT}, we review the relationship between unrefined open DT invariants and open GW invariants. In particular, unrefined open DT invariants admit a geometric interpretation in terms of exponential networks \cite{Banerjee:2018syt}, which we briefly revisit; see also \cite{Aganagic:2009cg} for a complementary physical viewpoint. Within our framework, we additionally recall the standard setup of unrefined open GW invariants, following \cite{Fang_2013,MR3659939}.

Having established these connections, we emphasize a key novelty of our work. Guided by physical considerations, we propose definitions of both refined open DT invariants and refined open GW invariants. In particular, the refined open GW invariants are introduced indirectly, via the definition of refined LMOV invariants, providing a unified and physically motivated refinement of the open BPS counting problem.
	
	\subsubsection{Closed Gromov-Witten and closed Donaldson-Thomas invariants} \label{sec:closedinv}
	
	In this paper we are interested in local CY threefolds and the enumerative invariants associated with them. The Gromov-Witten (GW) invariants arise from counting stable maps from a Riemann surface $\Sigma_g$ of genus $g$ to a toric CY threefold $\mathcal{X}$ in the class $H_2(\mathcal{X};\mathbb{Z})$. By ``remodeling conjecture'' \cite{Bouchard:2007ys} (see \cite{Eynard:2012nj,Fang:2016svw,Banerjee:2025qgx} for the proof for toric CY threefolds/orbifolds), the all-genus open/closed GW potentials of $\mathcal X$
	are identified with the topological recursion invariants of the mirror curve of the mirror CY threefold $\mathcal X^\vee$. Physically, the generating function of the GW invariants is given by topological string partition function.
	
	On the other hand, the Donaldson-Thomas (DT) invariants are defined as counting of coherent sheaves in the CY threefold $\mathcal{X}$, where the latter are objects in the bounded derived category $D^b \text{Coh}(\mathcal{X})$ of coherent sheaves on $\CalX$, with a choice of stability data. By mirror symmetry, this problem can be reformulated in terms of the mirror $\mathcal{X}^\vee$ as well. In this context, the DT invariants correspond to the count of compactly supported stable objects in the Fukaya category $D^b \text{Fuk}(\mathcal{X}^\vee)$. A definition of the DT invariants that suffices for our purpose was given in \cite{Banerjee:2022oed,Banerjee:2023zqc}, purely based on the counting of the special Lagrangian A-branes which are the very objects of $D^b \text{Fuk}(\mathcal{X}^\vee)$. 
	
	In general, for an object $E\in D^b \text{Coh}(\mathcal{X})$ representing a D-brane system, there is a K-theoretic charge associated to it $\Gamma(E) = {\textrm{Ch}}(E)\sqrt{{\textrm{Td}}(\mathcal{X})} \in H^{\textrm{even}}(\mathcal{X};\mathbb{Q})$; physically, this vector identifies the conserved Ramond-Ramond (RR) charges of the state. Associated to the K-theory class is a complex number called the central charge $Z_\Gamma : \Gamma \rightarrow \mathbb{C}$, given as $Z_\Gamma = \int_{\mathcal{X}} e^{-B-iJ} {\textrm{Ch}}(E)\sqrt{{\textrm{Td}}(\mathcal{X})}$ where $B+iJ$ is the complexified K\"ahler form on $\mathcal{X}$. In the underlying $\mathcal{N}=2$ supersymmetry algebra, the magnitude of this charge determines the mass of the BPS state via $M_{\text{BPS}} = |Z_\Gamma|$. The stability condition can be phrased in terms of the phases of these central charges; this condition dictates whether a composite BPS state exists as a stable bound state or decays into constituents (wall-crossing). Consequently, the DT invariants are labeled by the charges $\Gamma$ and count the number of stable BPS states in a given chamber of the moduli space.

	 For this section, let us not get into the excruciating details of refined GW invariants and refined DT invariants, but review them briefly. For the given coherent sheaf $E$ from above and the Chern character $\Gamma = {\mathrm{Ch}}(E)$, one can define the moduli space stable coherent sheaf $\mathcal{M}_\Gamma$. There is a torus action $T=(\mathbb{C}^\times)^2$ on $\mathcal{X}$ which induces an action on $\mathcal{M}_\Gamma$. This provides a weight decomposition on the Ext-complexes that define the DT invariants and thus one has the refined version of the DT invariants.   On the other hand, the equation \eqref{topstrZ} re-expresses GW invariants in terms of Gopakumar-Vafa (GV) invariants \cite{Dijkgraaf:2006um, MR2222528}. The definition of refined GW invariants comes from the definition of the refined GV invariants. Consider a curve class $\beta \in H_2(\mathcal{X};\mathbb{Z})$, for which one can define the moduli space $\mathcal{M}_{\beta,n}$ as a proper Deligne-Mumford stack and we also choose a stability condition. Let $\pi : \mathcal{M}_{\beta,n} \rightarrow {\mathrm{Chow}}_\beta (\mathcal{X})$ be the Hilbert-Chow morphism.  The ${\mathrm{SU}}(2)_L\times {\mathrm{SU}}(2)_R$ action introduced before in the context of $\Omega$-deformation provides the cohomological gradings for $\mathcal{H}^*(\mathcal{M}_{\beta,n})$ to define the refined GV invariants \cite{Katz:1999xq}. Then expanding it in the form of refined topological string partition function provides the refined GW invariants. For a class of Calabi-Yau threefolds the closed topological string partition function was computed in \cite{Banerjee:2025qgx} using topological recursion using the remodeling conjecture of \cite{Bouchard:2007ys}, and subsequently related to GV invariants.
    
     Furthermore, the count of rank-zero DT invariants can be phrased as the count of D4-D2-D0 brane bound states for type IIA string theory on CY threefold $\mathcal{X}$. Lifting this to M-theory, in geometric engineering setup, as was mentioned before, one can consider M5-branes wrapping a holomorphic cycles $D \subset \mathcal{X}$, additionally. Compactifying M-theory on $\mathcal{X}$, in the 5d theory, as before there are BPS states corresponding to instanton particles, but also there are monopole strings now.  The BPS index of this theory $T^{[\mathrm{5d}]}(\mathcal{X})$ is then equal to the rank zero DT invariants of the CY threefold $\mathcal{X}$. However, the left hand side of the equation \eqref{MNOPeq} corresponds to the 5d BPS states in $T^{[\mathrm{5d}]}(\mathcal{X})$ which are instanton particles, whose count corresponds to GV invariants.  Based on the geometric engineering picture, then it is natural to expect that there is a correspondence between GW and DT invariants. Namely, if for some choice of gauge coupling parameters, there exists a chamber where the monopole strings are not BPS, one can expect the counts of GW and DT invariants are the equal \cite{Maulik:2003rzb,Maulik:2004txy}. 
	
	Indeed, there is such a remarkable (mathematically rigorous) correspondence which has been much under investigation for several years. There are two ways to formulate the GW/DT conjecture when $\mathcal{X}$ has compact four cycles. One is at the orbifold point where the CY threefold becomes $\mathbb{C}^3/\Gamma$, where $\Gamma \subset SL(3,\mathbb{Z})$ is a discrete subgroup. In this case one does not have any stable D4-brane. In the absence of a GW theory, one applies the ``crepant resolution conjecture'' \cite{article} to compute the orbifold GW invariants and match them with the DT invariants. This corresponds to strong coupling chamber in gauge theory. But our situation is different and falls into the second type. The equation \eqref{NekZ5d} is valid for small gauge coupling or equivalently the large volume point of the Calabi-Yau threefold $\mathcal{X}$, where one does have stable D4-branes. In particular, for toric Calabi-Yau threefolds it has been proven that the invariants associated to the ideal sheaves of $\mathcal{X}$, equivalently subschemes $Z\subset \mathcal{X}$, with one-dimensional support (the D0-D2 boundstates) the equality \eqref{MNOPeq} to hold.\footnote{Strictly speaking, the left hand side of \eqref{MNOPeq} computes GV invariants and the right hand side says how it is the same as the count of instanton particles in $T^{[\textrm{5d}]}(\mathcal{X})$. However, it is known that one can extend the left hand side to the partition function of closed BPS states associated to D4-branes in type IIA setting and the right hand side can be also extended to include monopole strings in $T^{[\textrm{5d}]}(\mathcal{X})$ such that the equality continues to hold \cite{Banerjee:2018syt,Banerjee:2019apt,Banerjee:2020moh,Closset:2019juk}. In this sense, in section \textsection \ref{sec:closedengg}, \eqref{MNOPeq} is subsumed in the extended setup involving M5-branes wrapping $D \subset \mathcal{X}$ and monopole strings. In small gauge coupling limit (large volume chamber of $\mathcal{X}$), one obtains \eqref{MNOPeq}, establishing the GW/DT correspondence of MNOP.}

	\subsubsection{Open Gromov-Witten and open Donaldson-Thomas invariants}
	\label{sec:openGWDT}
	
	To define the open GW invariants, now one considers the moduli space of stable maps from a Riemann surface $\Sigma_{g,n}$ of genus $g$ and $n$ marked points to $(\mathcal{X},\mathcal{L})$ where $\mathcal{L}$ corresponds to the toric Lagrangian that the Aganagic-Vafa M5-brane wraps. Let the relative homology class of the image be $\mu\in H_2 (\mathcal{X},\mathcal{L};\BZ)$. Each boundary map $\partial\Sigma_{g,n} \rightarrow \mathcal{L}$ carries a winding number $w_i \in H_1(\mathcal{L};\mathbb{Z})$. Let us denote the moduli stack of such stable maps as $\overline{\mathcal{M}}_{g,n}(\mathcal{X},\mathcal{L};\mu,w_i)$. The open GW invariants are then given as integrals over the virtual fundamental class of the moduli space ${\textrm{GW}}_{g,n,\mu,w_i} = \int_{[\overline{\mathcal{M}}_{g,n}(\mathcal{X},\mathcal{L};\mu,w_i)]^{\textrm{vir}}} 1$. To reduce clutter we will drop $(\CalX,\CalL)$ from the index. Once again by remodeling conjecture \cite{Bouchard:2007ys}, it can be shown that they correspond to periods corresponding to certain open paths on the mirror curve $\Sigma : F(X,Z)=0$, where the mirror CY threefold is given by $\mathcal{X}^\vee:=\{F(X,Z) = uv\} \in \mathbb{C}^\times_X\times\mathbb{C}^\times_Z\times \mathbb{C}^2_{u,v}$. These open paths lie in the relative homology $H_1(\Sigma,X;\BZ)$, where $X$ is a point on the mirror curve $\Sigma$ appearing as the mirror dual of the Aganagic-Vafa Lagrangian $\mathcal{L} \subset \CalX$.\footnote{More precisely, the mirror of the Aganagic-Vafa A-brane $\mathcal L$ is a holomorphic B-brane supported on a complex one-dimensional submanifold
		in the mirror $\CalX^\vee$, whose open modulus is encoded by a choice of point on the mirror curve $\Sigma$ \cite{Aganagic:2000gs}.}  More precisely, to define them we need to use a covering map over the $\BC^\times _X$-plane as 
	$\tilde\Sigma\xrightarrow{\tilde\pi} \Sigma \xrightarrow{\pi}\mathbb{C}^\times_X $. Denoting the sheets of $\tilde\pi$ by the logarithmic label $N\in \BZ$ and sheets of $\pi$ by roman letters $i,j,...$, one can define the lattice of relative homology \cite{Banerjee:2018syt} as 
	\begin{equation}
		\Gamma_{ij,n} (X) : = H_1(\Sigma,(i,N), (j,N+n);\mathbb{Z})
	\end{equation}
	and also a central charge associated to it obtained by integrating the one form $\lambda = \log Z \, \dd \log X$ over a given one chain. 
	
	Let us define $\mathcal{E}$-wall as, for an ordered pair $(ij,n)$, the interpolation between two preimages of $X\in \mathbb{C}^\times_X$, namely, $(X,\log Z_i(X) + 2\pi i N)$ and $(X,\log Z_j(x) + 2\pi i (N+n))$ for $n\in \mathbb{Z}$. Working on the $\BC^\times _X$-plane, this is achieved by paths crossing branch cuts and logarithmic cuts which were chosen while fixing trivialization. The shape of the wall is given the solution to the differential equation 
	\begin{equation}
		\label{nerweq}
		(\log Z_j(X) - \log Z_i(X) + 2\pi i n) \frac{\mathrm{d}\log X(t)}{\mathrm{d}t} \in e^{i\vartheta} \mathbb{R}_+
	\end{equation}
	where $t$ is a proper time along the path and for every $\vartheta$ there exists an exponential network $\mathcal{W}(\vartheta)$. 
	
	The $\mathcal{E}$-walls also carry certain combinatorial data. The paths in relative homology $a\in \tilde\Sigma$ connect one preimage of $X\in \BC^\times _X$ labeled by $(i,N)$ with another preimage $(j,N+n)$. We refer to them as soliton path. We also introduce the union of all such charge lattices for different solitons supported on the same wall as $\Gamma_{ij,n}(X) = \amalg_{N\in \mathbb{Z}} \Gamma_{ij,N,N+n} (X)$. The soliton data is an assignment $\nu(a;X)$ for each $a \in \Gamma_{ij,n}(X)$ (often we will drop the second argument $X$).  
	
	Denote by $\nabla^{ab}$ the abelian flat $GL(1)$ connection on $\tilde\Sigma$ and $\nabla^{na}$ the nonabelian $GL(N)$ flat connection on $\mathbb{C}^\times_X$. For an open path $\wp \in \mathbb{C}^\times_X$, associate the parallel transport map 
	\begin{equation}
		F(\wp) = P \exp \int_\wp \nabla^{na}
	\end{equation}
	and for open paths $a\in \tilde\Sigma$, the parallel transport 
	\begin{equation}
		X_a = P\exp \int_a \nabla^{ab}.
	\end{equation}
	
	There is a natural algebra of concatenation on the variables $X_a$ and one can express $F(\wp)$ in terms of $X_a$ using detour rules \cite{Banerjee:2018syt, Gaiotto:2012rg}. Flatness of $\nabla^{na}$ implies $F(\wp)$ depends only on the homotopy class of $\wp$. This can be leveraged to uniquely fix the data $\{\nu(a)\}$.\footnote{This is one of the steps to construct the nonabelianization map in \cite{Gaiotto:2012rg, Banerjee:2018syt, Banerjee:2024smk}. The second step to determine the 5d indices correspond to computing jumps in this map, corresponding to topology change of the network. An alternative viewpoint to that count was presented in \cite{Eager:2016yxd,Banerjee:2022oed,Banerjee:2023zqc,Banerjee:2024smk}.This computation has in fact origin in the CFIV index and $tt^*$ geometry \cite{Cecotti:1991qv,Cecotti:1992qh,Cecotti:1992rm}, as was explained for 3d-5d case in \cite{Banerjee:2018syt}.}

    Now, we come to an important issue. Given a $q$-difference equation, one can write a formal asymptotic in $\hbar$ solution to it. Then the $\nu(a)$ computed above corresponds to the $O(\hbar^{-1})$ term in that series. Given a difference equation, using the WKB ansatz, with the data of $\nu(a)$, one can write the formal solution to the associated Riccati equation \cite{DelMonte:2024dcr} to all order in $\hbar$. Then, the package of resurgence is expected to provide  in each angular sector defined by the complement of $\mathcal{W}(\vartheta)$ an analytic and convergent solution. Some aspects of this was studied using WKB method was studied in \cite{DelMonte:2024dcr}, and from alternative perspective of topological recursion in \cite{Banerjee:2025shz}. In this paper, we do not really study the details of the analytic continuation (in gauge coupling) of the solutions of the difference equations in details and we leave this important analysis for future work.

     As was described in \cite{Gupta:2024ics,Gupta:2024wos}, the  vortex BPS states however correspond to the $(ii,n)$ type soitons. These are some particular specialization of the solitons of type $(ij,n)$ where $i=j$ and were discussed in \cite{Banerjee:2018syt}. Their counting through nonabelianization techniques have been investigated and it was shown to match with the \eqref{eq:openbps} at the leading order in $\hbar$ expansion. In \cite{Gupta:2024ics}, this match with the K-theoretic vortex partition function was checked explicitly. Also in cases where the covering $\Sigma \xrightarrow{\pi} \mathbb{C}^\times_X$ is one-sheeted, there are only $(ii,n)$ type kinky vortices. These were considered in \cite{Grassi:2022zuk,Alim:2022oll} and the matches with exponential networks were established.  
	
	Crossing each of the $\mathcal{E}$-wall in $\mathcal{W}(\vartheta)$ one passes from one angular sector to another and the solutions are related by Stokes jump \cite{Grassi:2014zfa, Alim:2022oll, Kontsevich:2020piu}. The open DT invariants are associated to these jumps. In our framework, motivated by gauge theory, we are computing the open DT invariants in the large volume point for the associated mirror to the toric CY threefold. We can of course turn on the refinement and we report the invariants for some geometries later.

	\section{Refined open BPS invariants from 3d-5d partition functions} \label{sec:3dhalfindex}
	
	In this section, we apply our method of computing refined open BPS invariants from partiton functions of 3d-5d coupled systems at several examples: $\BC^3$, resolved conifold $\text{Tot}(\CalO(-1)^{\oplus 2} \to \BP^1)$, resolved $A_1$-singularity $\BC\times \widetilde{\BC^2/\BZ_2} = \text{Tot}(\mathcal{O}\oplus \mathcal{O}(-2) \to \BP^1 )$, local $\BP^1 \times \BP^1$, and finally local $F_1$.

\subsection{$\BC^3$}

	Let us begin with the simplest example, $\mathcal{X} =\BC^3$. Its toric diagram, in a chosen $SL(2;\BZ)$ frame, is drawn as in Figure \ref{fig:toric-C3-shifted}. Its dual graph, after $90^\circ$ counterclockwise rotation, yields the $(p,q)$-web comprising a trivalent junction of an NS5-brane, a D5-brane, and an $(1,1)$-brane as depicted in Figure \ref{fig:pqwebc3}. Since $H_2 (\BC^3 ;\BZ) = 0$, there is no closed BPS invariant associated with M2-branes wrapping closed 2-cycles. Correspondingly, there is no nontrivial 5d effective theory realized by the $(p,q)$-web regardless of the $SL(2;\BZ)$ frame.
	
	\subsubsection{3d \texorpdfstring{$\mathrm{U}(1)$}{U(1)} gauge theory with one chiral multiplet}

\begin{figure}[h!]
		\centering
		
		\begin{subfigure}{0.48\textwidth}
			\centering
			\begin{tikzpicture}[x=3cm,y=3cm,line cap=round,line join=round]
				\tikzset{
					toric vertex/.style={circle, fill=red, inner sep=1.7pt},
					toric edge/.style={blue, very thick},
					dual edge/.style={black, thick},
				}
				
				\coordinate (A) at (0,0);
				\coordinate (B) at (-1,0);
				\coordinate (C) at (0,1);
				
				\draw[toric edge] (A) -- (B) -- (C) -- cycle;
				
				\foreach \P in {A,B,C}{
					\node[toric vertex] at (\P) {};
				}
				
				\pgfmathsetmacro{\r}{1 - 1/sqrt(2)}
				\coordinate (J) at (-\r,\r);
				\fill[black] (J) circle (1.2pt);
				
				\draw[dual edge] (J) -- ++(0,-0.7);
				\draw[dual edge] (J) -- ++(0.7,0);
				\draw[dual edge] (J) -- ++(-0.5,0.5);
				
				\node at (0.12,-0.12) {$(0,0)$};
				\node at (-1.1,-0.12) {$(-1,0)$};
				\node at (0.12,1.12) {$(0,1)$};
			\end{tikzpicture}
			\caption{}
			\label{fig:toric-C3-shifted}
		\end{subfigure}
\begin{subfigure}{0.48\textwidth}
  \centering
  \begin{tikzpicture}[x=3cm,y=3cm,line cap=round,line join=round]
    \tikzset{
      dual edge/.style={black, thick},
      dthree/.style={red, very thick},
    }

    \pgfmathsetmacro{\r}{1 - 1/sqrt(2)}
    \coordinate (J) at (-\r,-\r);
    \fill[black] (J) circle (1.2pt);

    \draw[dual edge] (J) -- ++(0.7,0);      
    \draw[dual edge] (J) -- ++(0,0.7);      
    \draw[dual edge] (J) -- ++(-0.5,-0.5);  

    \node at (0.55,-0.3) {$\text{D5}$};
    \node at (-0.28,0.53) {$\text{NS5}$};
    \node at (-0.9,-0.9) {$(1,1)$};


    \coordinate (P1) at ($(J)+(0,0.35)$);
    \draw[dthree] (P1) -- ++(-0.35,-0.12) coordinate (TIP1);
    \node at ($(TIP1)+(-0.1,0)$) {$\text{(I)}$};

    \coordinate (P2) at ($(J)+(0.55,0)$);
    \draw[dthree] (P2) -- ++(-0.35,-0.12) coordinate (TIP2);
    \node at ($(TIP2)+(0,-0.1)$) {$\text{(III)}$};

    \coordinate (P3) at ($(J)+(-0.25,-0.25)$);
    \draw[dthree] (P3) -- ++(-0.35,-0.12) coordinate (TIP3);
    \node at ($(TIP3)+(-0.15,0)$) {$\text{(II)}$};

  \end{tikzpicture}
  \caption{} 
  \label{fig:pqwebc3}
\end{subfigure}
		\caption{(a) Toric diagram of $\BC^3$ at framing $f=0$ and its graph-dual; (b) $(p,q)$-web of fivebranes dual to $\BC^3$ and D3-branes dual to Aganagic-Vafa branes. The cases (I) and (II) give rise to a $Q$-observable, while the case (III) engineers an $H$-observable.}
	\end{figure}
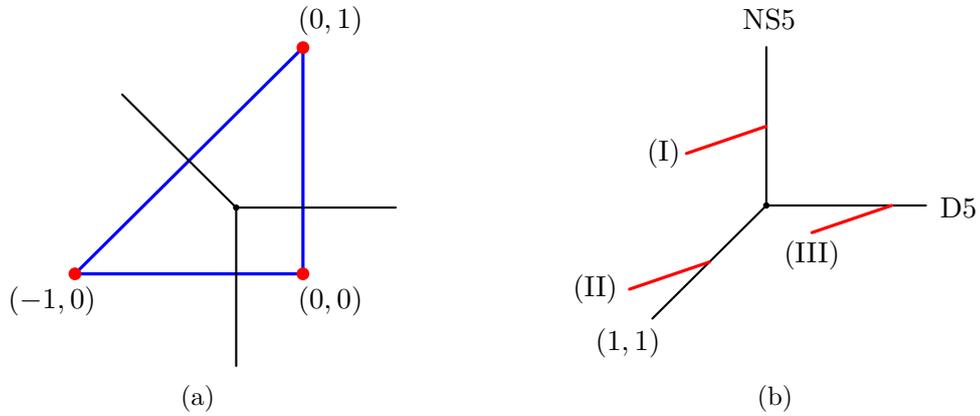

\begin{figure}[h!]
		\centering
		
		\begin{subfigure}{0.48\textwidth}
			\centering
			\begin{tikzpicture}[x=3cm,y=3cm,line cap=round,line join=round]
				\tikzset{
					toric vertex/.style={circle, fill=red, inner sep=1.7pt},
					toric edge/.style={blue, very thick},
					dual edge/.style={black, thick},
				}
				
				\coordinate (A) at (0,0);
				\coordinate (B) at (-1,0);
				\coordinate (C) at (-1,1);
				
				\draw[toric edge] (A) -- (B) -- (C) -- cycle;
				
				\foreach \P in {A,B,C}{
					\node[toric vertex] at (\P) {};
				}
				
				\pgfmathsetmacro{\r}{1 - 1/sqrt(2)}
				\coordinate (J) at (-1+\r,\r);
				\fill[black] (J) circle (1.2pt);
				
				\draw[dual edge] (J) -- ++(0,-0.7);
				\draw[dual edge] (J) -- ++(-0.7,0);
				\draw[dual edge] (J) -- ++(0.5,0.5);
				
				\node at (0.12,-0.12) {$(0,0)$};
				\node at (-1.1,-0.12) {$(-1,0)$};
				\node at (-1.1,1.12) {$(-1,1)$};
			\end{tikzpicture}
			\caption{}
			\label{fig:toric-tri-new}
		\end{subfigure}
\begin{subfigure}{0.48\textwidth}
  \centering
  \begin{tikzpicture}[x=3cm,y=3cm,line cap=round,line join=round]
    \tikzset{
      dual edge/.style={black, thick},
      dthree/.style={red, very thick},
    }

    \pgfmathsetmacro{\r}{1 - 1/sqrt(2)}
    \coordinate (J) at (-\r,-1+\r);
    \fill[black] (J) circle (1.2pt);

    \draw[dual edge] (J) -- ++(0.7,0);      
    \draw[dual edge] (J) -- ++(0,-0.7);     
    \draw[dual edge] (J) -- ++(-0.5,0.5);   

    \node at (0.54,-0.7) {$\text{D5}$};
    \node at (-0.28,-1.5) {$\text{NS5}$};
    \node at (-0.95,-0.11) {$(1,-1)$};


    \coordinate (P1) at ($(J)+(0,-0.35)$);
    \draw[dthree] (P1) -- ++(-0.3,-0.20) coordinate (TIP1);
    \node at (-0.73,-1.26) {$\text{(II)}$};

    \coordinate (P2) at ($(J)+(0.5,0)$);
    \draw[dthree] (P2) -- ++(-0.3,-0.20) coordinate (TIP2);
    \node at ($(TIP2)+(-0,-0.1)$) {$\text{(III)}$};

    \coordinate (P3) at ($(J)+(-0.25,0.25)$);
    \draw[dthree] (P3) -- ++(-0.3,-0.20) coordinate (TIP3);
    \node at ($(TIP3)+(-0.1,0.0)$) {$\text{(I)}$};

  \end{tikzpicture}
  \caption{} 
  \label{fig:pqweb-tri-new}
\end{subfigure}
		
		\caption{(a) Toric diagram of $\BC^3$ at framing $f=-1$ and its graph-dual; (b) $(p,q)$-web of fivebranes and D3-branes dual to Aganagic-Vafa branes. The cases (I) and (II) give a $Q$-observable, while the case (III) provides an $H$-observable.}
	\end{figure}
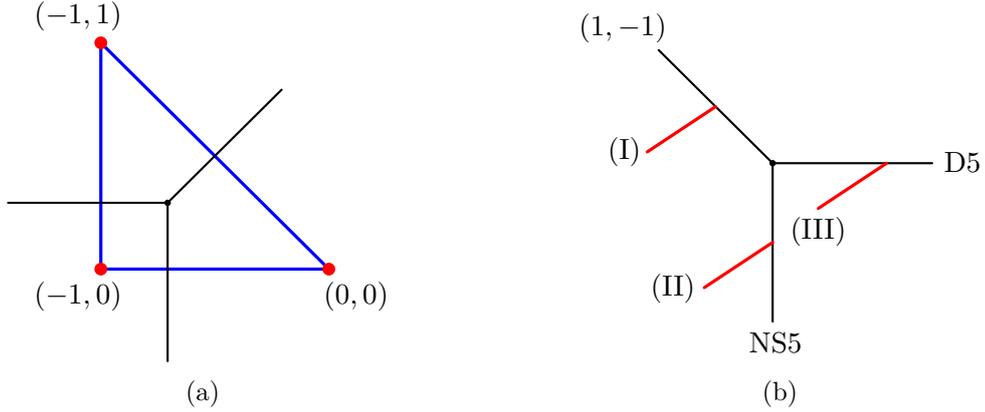

	Now we introduce an Aganagic-Vafa brane. Depending on which toric leg it ends on, there are three possible choices. In the type IIB description with the $(p,q)$-web, the D3-brane terminates on the corresponding fivebrane segment. There are three cases described in Figure \ref{fig:pqwebc3} $(f=0)$ and \ref{fig:pqweb-tri-new} $(f=-1)$. The cases (I) and (II) give $Q$-observables, while the case (II) gives rise to an $H$-observable.

	Let us first consider the $Q$-observable at framing $f\in \BZ$. Due to the open string stretched between the D3-brane and the D5-brane, the low-energy effective field theory is 3d $\CalN=2$ theory of a free chiral multiplet with the background gauge field for the $\U(1)$ flavor symmetry with the Chern-Simons level $-f$. Its partition function is simply given by
	\begin{align}
		Q_f (X) = \th (X^{-1};q_1) ^f \text{PE} \left[ -\frac{X^{-1}}{1-q_1} \right]= \th (X^{-1};q_1) ^f ( X^{-1} ;q_1 )_\infty.
	\end{align}
	It is immediate that it is a solution to the following $q_1$-difference equation,
	\begin{align} \label{eq:qmirrorc3}
		0 = \left[ q_1^{-1} X^{-1}+ (-q_1  X )^{f} q_1^{D_X}  -1 \right]Q_{f} (X).
	\end{align}
	
	In the limit $q_1 \to 1$, the partition function can be expanded as
	\begin{align}
		Q_f (X) = \exp \frac{\widetilde{\EuScript{W}}_f(X)}{\log q_1} + \cdots,
	\end{align}
	where $\widetilde{\EuScript{W}}_f (X) $ is the effective twisted superpotential of the KK-reduced 2d $\CalN=(2,2)$ effective theory. In this limit, the $q_1$-difference TQ equation reduces to
	\begin{align} \label{eq:mirrorc3}
		0 = X^{-1}+(-X)^f Z  -1,
	\end{align}
	where $Z = \frac{\p \widetilde{\EuScript{W}}_f (X)}{\p \log X}$ is the effective complexified FI parameter. This is precisely the mirror curve for $\BC^3$, i.e., the Newton polynomial equation determined by the toric diagram of $\BC^3$ (see Figure \ref{fig:toric-C3-shifted} for the case of $f=0$). Thus, the $q_1$-difference operator \eqref{eq:qmirrorc3} annihilating $Q_f (X)$ is indeed the quantum mirror curve for $\BC^3$. 
    
	Note that when $f \in \{-1,0\}$, $Q_f (X)$ is simply given by a single $q$-Pochhammer symbol,
	\begin{align}
		Q_0 (X) = ( X^{-1};q_1)_\infty,\quad Q_{-1}(X) = \left( q_1 X ;q_1 \right)_\infty ^{-1}.
	\end{align}
	
	Next, let us consider the $H$-observable at framing $f\in \BZ$ corresponding to the case (III). The 3d $\CalN=2$ theory is the $\U(1)$ gauge theory with a chiral multiplet of charge $+1$. The bare Chern-Simons level is given by the framing $f$. With the complexified FI parameter $Z \in \BC^\times$, the partition function on $\BR^2 \times S^1$ is given by the $q_1$-lattice summation \eqref{eq:hobs} of $Q_0 (X)$, which specializes in this case as
	\begin{align} \label{eq:hbosc3}
		\begin{split}
			H_f (Z)  &  = \sum_{X \in q_1 ^\BZ} \frac{\left(q_1 X^{-1};q_1 \right)_\infty}{(q_1;q_1)_\infty} \exp\left[-\frac{\log X  \log \left(Z (-1)^f (Xq_1^{-1})^{\frac{f}{2}} \right)} {\log q_1}\right] \\
			&=\sum_{n=0}^\infty \frac{1}{(q_1;q_1)_n} q_1 ^{-\frac{n(n+1)}{2}f} \left( (-1)^f Z \right)^n,   
		\end{split}
	\end{align}
	where the summation truncates from above due to the zeros at $X = q_1 ^{m+1}$, $m \geq 0$. This is indeed the K-theoretic vortex partition function of the 3d $\CalN=2$ gauge theory at hand.
	
	As derived in \eqref{eq:diffeqH}, the quantum mirror curve equation satisfied by $Q_0 (X)$ directly yields a Fourier-dual quantum mirror curve equation solved by $H_f (Z)$. In the current example, it reads
	\begin{align}
		0 = \left[ q_1 ^{D_Z} + Z \left( -q_1 ^{-1-D_Z} \right)^f  -1\right] H_f (Z).
	\end{align}
	It is in fact straightforward to see the K-theoretic vortex partition function \eqref{eq:hbosc3} satisfies this equation. In the limit $q_1 \to 1$, this quantum mirror curve equation reduces to the mirror curve \eqref{eq:mirrorc3} for $\CalX = \BC^3$ by construction, with $X = -\frac{\p \widetilde{\EuScript{W}}_f (Z)}{\p \log Z}$ identified with the effective complex scalar in the twisted chiral multiplet. 
	
	Note that, when $f\in \{-1,0\}$, the partition function $H_f(Z)$ simplifies into a single $q$-Pochhammer symbol. Namely,
	\begin{align}
		H_{-1} (Z) = ( q_1Z;q_1)_\infty,\quad H_0 (Z) = (Z;q_1)_\infty ^{-1} .
	\end{align}
	
	\subsubsection{Mapping Aganagic-Vafa branes by \texorpdfstring{$S$}{S}-transformation}

	For $\BC^3$, the two framings $f=0$ and $f=-1$ are special since they map to each other by an $S$-transformation of $SL(2;\BZ)$. When an Aganagic-Vafa brane is introduced, the $S$-transformation maps it to another AV-brane which ends on a different toric leg. Thus, the $S$-transformation should exchange the $Q$ and $H$ generating functions of the refined open BPS invariants. Indeed, there is obvious matching between them,
	\begin{align}
		Q_0 ^{\text{(I)}} (X = q_1^{-1}Z^{-1}) = H_{-1} ^{\text{(III)}} (Z),\qquad H_0 ^{\text{(III)}} (Z= X q_1 ) = Q_{-1} ^{\text{(II)}}(X),
	\end{align}
	which confirms the proposed exchange of the Aganagic-Vafa branes under the action of the $S$-transformation.

	\subsection{Resolved conifold}
	Let us consider the example of the resolved conifold, $\mathcal{X} = \text{Tot}\left(\mathcal{O}(-1)^{\oplus 2} \to \BP^1\right)$. When drawing the toric diagram, we begin from the two $SL(2;\BZ)$ frames drawn in Figure \ref{fig:resconifold1} and Figure \ref{fig:resconifold2}, related to each other by an $S$-transformation, and act $T$-transformations to each to span all possible frames. Recall that the $T$-transformation only shears the toric diagram horizontally, so that in the $(p,q)$-web of fivebranes the D5-branes are left invariant.

	\subsubsection{3d \texorpdfstring{$\mathrm{U}(1)$}{U(1)} gauge theory with two chiral multiplets}

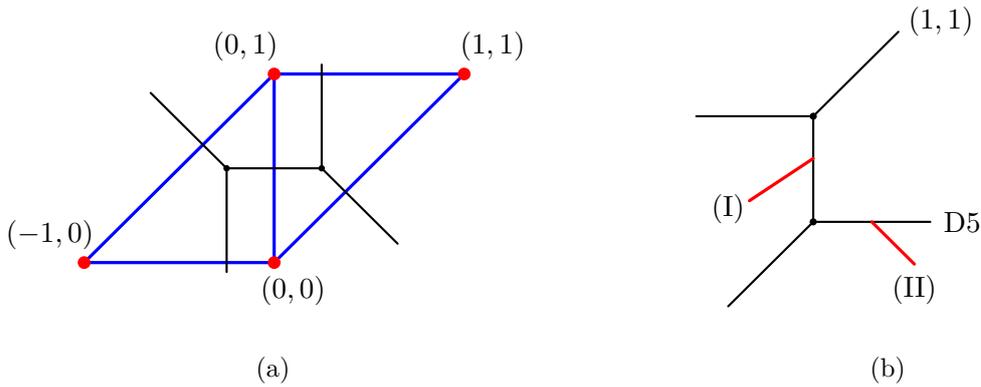
\begin{figure}[h!]
  \centering

  \begin{subfigure}[b]{0.48\textwidth}
    \centering
    \begin{tikzpicture}[
      x=2.5cm,y=2.5cm,
      line cap=round,line join=round,
      baseline={(0, 0.5)} 
    ]
      \useasboundingbox (-1.5,-0.4) rectangle (1.5,1.4);

      \tikzset{
        toric vertex/.style={circle, fill=red, inner sep=1.7pt},
        toric edge/.style={blue, very thick},
        dual edge/.style={black, thick},
      }

      \coordinate (A) at (-1,0);
      \coordinate (B) at ( 0,0);
      \coordinate (C) at ( 1,1);
      \coordinate (D) at ( 0,1);

      \draw[toric edge] (A) -- (B) -- (C) -- (D) -- cycle;
      \draw[toric edge] (B) -- (D); 

      \foreach \P in {A,B,C,D}{
        \node[toric vertex] at (\P) {};
      }

      \node at (-1.18, 0.15) {$(-1,0)$};
      \node at ( 0.10,-0.15) {$(0,0)$};
      \node at ( 1.15, 1.15) {$(1,1)$};
      \node at (-0.15, 1.15) {$(0,1)$};

      \coordinate (J2) at (-0.25, 0.50);
      \coordinate (J1) at ( 0.25, 0.50);
      \fill[black] (J1) circle (1.2pt);
      \fill[black] (J2) circle (1.2pt);

      \draw[dual edge] (J2) -- (J1);

      \draw[dual edge] (J2) -- ++( 0.00,-0.55);
      \draw[dual edge] (J2) -- ++(-0.40, 0.40);
      \draw[dual edge] (J1) -- ++( 0.00, 0.55);
      \draw[dual edge] (J1) -- ++( 0.40,-0.40);

    \end{tikzpicture}
    \caption{}
  \end{subfigure}
  \hfill
  \begin{subfigure}[b]{0.48\textwidth}
    \centering
    \begin{tikzpicture}[
      x=2.5cm,y=2.5cm,
      scale=1.12, 
      line cap=round,line join=round,
      baseline={(-0.5, 0)} 
    ]
      \useasboundingbox (-1.3,-0.8) rectangle (1.0,1.0);

      \tikzset{
        dual edge/.style={black, thick},
        dthree/.style={red, very thick},
      }

      \coordinate (K2) at (-0.50,-0.25); 
      \coordinate (K1) at (-0.50, 0.25); 
      \fill[black] (K1) circle (1.2pt);
      \fill[black] (K2) circle (1.2pt);

      \draw[dual edge] (K2) -- (K1);

      \draw[dual edge] (K1) -- ++(-0.55,0);        
      \draw[dual edge] (K1) -- ++(0.40,0.40);      
      \draw[dual edge] (K2) -- ++(0.55,0) coordinate (LegRight);         
      \draw[dual edge] (K2) -- ++(-0.40,-0.40) coordinate (LegBotLeft);  


      \coordinate (P_II) at ($(K2)!0.5!(LegRight)$);
      \draw[dthree] (P_II) -- ++(0.20,-0.20) coordinate (Tip_II);
      \node at ($(Tip_II)+(0,-0.12)$) {$\text{(II)}$};

      \coordinate (P_I) at ($(K2)!0.60!(K1)$);
      \draw[dthree] (P_I) -- ++(-0.30,-0.20);
      \node at (-0.9,-0.19) {$\text{(I)}$};

      \node at (0.2,-0.25) {$\text{D5}$};
      \node at (0.1,0.7) {$(1,1)$};

    \end{tikzpicture}
    \caption{}
  \end{subfigure}

  \caption{
    (a) Toric diagram of resolved conifold at framing $f=0$ with a chosen crepant resolution, and its graph-dual; (b) The corresponding $(p,q)$-web of fivebranes and D3-branes dual to Aganagic-Vafa branes. The case (I) gives a $Q$-observable, while the case (II) gives an $H$-observable.
  } 
  \label{fig:resconifold1}
\end{figure}

	First, we begin with the toric diagram in Figure \ref{fig:resconifold1} and its $T$-transformations. The 5d $\CalN=1$ theory engineered by the $(p,q)$-web is single free hypermultiplet theory, obtained by spectrum of the D5-D5 open string across the NS5-brane. Some partial results on the open BPS states partition function exists in \cite{Kashani-Poor:2006puz,Zhu:2019wew,Luo:2019rur}. The refined closed BPS invariants are encoded in its partition function, computed simply in 1-loop as
	\begin{align} \label{eq:closedres1}
		\CalZ_{\text{closed}} (m; q_1,q_2) =\text{PE}\left[ \frac{m}{(1-q_1)(1-q_2)} \right] = \prod_{i,j=0} ^\infty \left(1- m q_1 ^{i} q_2 ^j\right)^{-1},
	\end{align}
	where $m \in \BC^\times$ is the mass of the hypermultiplet.\footnote{The generating series of the refined Gopakumar-Vafa (GV) invariants \cite{Dimofte:2009bv,Mozgovoy:2020has} was computed and the factorized form has positive exponents. This is because their convention and ours differ by the domain of $q_2$, where for us, it is small in contrast to theirs being large. We also remark that we could have chosen any value of $q_2$. However, for computation, we choose small $q_2$ throughout this paper. To match with these invariants computed in the standard references \cite{Szendr_i_2008} we absorb additional sign in the exponent. So in this case, for the corresponding toric geometry, for $\beta= [\mathbb{C}\mathbb{P}^1] $ and $j_L=j_R=0$, $N^\beta_{0,0} = 1$, and zero otherwise.} The K\"{a}hler modulus associated with the generator of $H_2 (\CalX;\BZ)$ is identified with this mass. \\
	
	Now we insert an Aganagic-Vafa brane and study its refined open BPS invariants. As we gave a general description in section \ref{subsubsec:cc2def}, there are two kinds in terms of the fivebrane segment that the D3-brane ends on. Namely, the $Q$-observable is realized by a D3-brane ending on NS5-brane or $(1,1)$-brane (the case (I) in Figure \ref{fig:resconifold1}), while the $H$-observable is realized by a D3-brane ending on D5-brane (the case (II) in Figure \ref{fig:resconifold1}). 
	
	The 3d $\CalN=2$ $\U(1)$ gauge theory realized on the worldvolume of the D3-brane contain two chiral multiplets, one with charge $+1$ and the other with charge $-1$, from the spectra of D3-D5 open strings. Since the 5d theory is a free theory in the present case, this 3d theory does not couple to the 5d theory and the refined open BPS invariants are given purely by the 3d $\CalN=2$ partition functions. The $Q$-observable is obtained by turning on a generic value of the complex scalar $X$ in the twisted chiral multiplet, with the complexified FI parameter turned off. The $H$-observable is obtained by turning on the complexified FI parameter $Z$, at two special values of the complex scalar. In either case, $X$ or $Z$ is identified with the K\"{a}hler modulus $z$ associated with the relative cycle in $H_2 (\CalX,\CalL;\BZ)$. \\

	For the $Q$-observable, the partition function does not receive any contribution from non-perturbative field configuration, just as in the case of $\BC^3$. From the two chiral multiplets with opposite $\U(1)$ charges, we get
	\begin{align} \label{eq:Qobsconi}
		Q_f (X) = \th(X^{-1};q_1)^f (X^{-1};q_1)_\infty (m q_1 X;q_1)_\infty.
	\end{align}
    We identify $X$ with the K\"{a}hler modulus associated with the relative cycle in $H_2 (\CalX,\CalL;\BZ)$. The partition function gives the generating function of refined open BPS invariants. Note that this partition function is a product of finite number of $q$-Pochhammer symbols in any framing $f\in \BZ$. Thus, there are only finite non-zero refined open BPS invariants.
    
	Using the property of the $q$-Pochhammer symbol, it is straightforward to write that the quantum mirror curve equation satisfied by $Q_f (X)$,
	\begin{align} \label{eq:qcurveconi}
		0 = \left[ (-q_1 X)^f (1-mq_1 X)  q_1 ^{D_X} +q_1^{-1}X^{-1}-1 \right]Q_f (X).
	\end{align}
	In the limit $q_1 \to 1$, we recover the mirror curve of the resolved conifold
	\begin{align} \label{eq:mcurveconi}
		0 =  (1-m X) (-X)^f Z + X^{-1} -1,
	\end{align}
	which, at $f=0$, is indeed the Newton polynomial equation for the toric diagram in Figure \ref{fig:resconifold1}. The other cases with $f\neq 0$ are obtained by the action of $T^f$. \\
	
	Next, we turn to the other kind of Aganagic-Vafa brane which produces a $H$-observable. The 3d $\CalN=2$ gauge theory transits into the Higgs phase by turning on the complexified FI parameter $Z$, at a special value of the complex scalar in the twisted chiral multiplet. The partition function localizes onto the vortex configurations, yielding the K-theoretic vortex partition,
	\begin{align} \label{eq:hconi1}
		\begin{split}
			H_f (Z) &=  \sum_{X \in q_1 ^\BZ} \left(q_1 X^{-1};q_1 \right)_\infty \left(m X;q_1 \right)_\infty \, \exp\left[-\frac{\log X  \log \left(Z (-1)^f (Xq_1^{-1})^{\frac{f}{2}}\right) } {\log q_1}\right] \\
			& =(m;q_1)_\infty \sum_{n=0} ^\infty \frac{(m^{-1}q_1;q_1)_n}{(q_1;q_1)_n}q_1 ^{-\frac{f+1}{2} n(n-1)-fn} \left( Z (-1)^{f+1} q_1^{-1} \right)^n.
		\end{split}
	\end{align}
    The partition function is a series expanded in $\vert Z \vert <1$, with coefficients given by exact rational functions in $m$. Further expanding it in a given chamber, we identify it with the generating function of refined open BPS invariants there. We present the refined open BPS invariants at $f=0$ in the chamber $\vert Z \vert <\vert m \vert <1$ obtained in this way in Appendix \ref{app:conifold}.
	
	The quantum mirror curve equation \eqref{eq:qcurveconi} satisfied by $Q_0 (X)$ directly implies the quantum mirror curve equation satisfied by $H_f (Z)$, given by
	\begin{align}
		0 = \left[ Z \left(1-m q_1 ^{-1-D_Z} \right) \left(-q_1 ^{-1-D_Z} \right)^f +q_1 ^{D_Z} -1 \right] H_f (Z).
	\end{align}
	The limit $q_1 \to 1$ yields the same mirror curve \eqref{eq:mcurveconi} of the resolved conifold by construction.

\begin{remark} 
At a special choice of the framing $f=-1$, the partition function \eqref{eq:hconi1} simplifies to a product of $q$-Pochhammer symbols, $H_{f=-1} (Z) = (m;q_1)_\infty (mZ ;q_1)_\infty^{-1} (q_1 Z ;q_1)_\infty$. The refined open BPS invariants are easily read off in this case, with only finite number of them being non-zero. This is not the case for general framing $f \in \BZ$.
\end{remark}
	
	\subsubsection{3d \texorpdfstring{$\mathrm{U}(1)$}{U(1)} gauge theory with one chiral multiplet coupled to 5d pure \texorpdfstring{$\mathrm{U}(1)$}{U(1)} gauge theory}

\begin{figure}[h!]
		\centering
		
		\begin{subfigure}[b]{0.45\textwidth}
			\centering
			\begin{tikzpicture}[
				x=2.5cm,y=2.5cm,
				line cap=round,line join=round,
				baseline={(0,0)} 
				]
				\useasboundingbox (-1.5,-1.6) rectangle (0.5,1.6);
				
				\tikzset{
					toric vertex/.style={circle, fill=red, inner sep=1.7pt},
					toric edge/.style={blue, very thick},
					dual edge/.style={black, thick},
				}
				
				\coordinate (A) at (-1, 1);
				\coordinate (B) at (-1, 0);
				\coordinate (C) at ( 0, 0);
				\coordinate (D) at ( 0,-1);
				
				\draw[toric edge] (B) -- (A) -- (C) -- (D) -- cycle;
				
				\draw[toric edge] (B) -- (C);
				
				\foreach \P in {A,B,C,D}{
					\node[toric vertex] at (\P) {};
				}
				
				\node at (-1.18, 1.15) {$(-1,1)$};
				\node at (-1.32, 0.00) {$(-1,0)$};
				\node at ( 0.23, 0.00) {$(0,0)$};
				\node at ( 0.15,-1.15) {$(0,-1)$};
				
				\coordinate (J1) at (-0.50, 0.35);
				\coordinate (J2) at (-0.50,-0.35);
				\fill[black] (J1) circle (1.2pt);
				\fill[black] (J2) circle (1.2pt);
				
				\draw[dual edge] (J2) -- (J1);
				
				\draw[dual edge] (J1) -- ++(-0.60, 0);
				\draw[dual edge] (J1) -- ++( 0.45, 0.45);
				
				\draw[dual edge] (J2) -- ++( 0.60, 0);
				\draw[dual edge] (J2) -- ++(-0.45,-0.45);
				
			\end{tikzpicture}
			\caption{}
		\end{subfigure}%
		\hspace{1em}%
\begin{subfigure}[b]{0.45\textwidth}
  \centering
  \begin{tikzpicture}[
    x=2.5cm,y=2.5cm,
    scale=1.12,
    line cap=round,line join=round,
    baseline={(0,0)} 
  ]
    \useasboundingbox (-1.2,-1.43) rectangle (1.0,1.43);

    \tikzset{
      dual edge/.style={black, thick},
      dthree/.style={red, very thick},
    }

    \coordinate (K1) at (-0.35, 0); 
    \coordinate (K2) at ( 0.35, 0); 
    \fill[black] (K1) circle (1.2pt);
    \fill[black] (K2) circle (1.2pt);

    \draw[dual edge] (K1) -- (K2);

    
    \draw[dual edge] (K1) -- ++( 0.00,-0.60) coordinate (LegDownEnd);
    \draw[dual edge] (K1) -- ++(-0.45, 0.45) coordinate (LegTopLeftEnd);

    \draw[dual edge] (K2) -- ++( 0.00, 0.60);     
    \draw[dual edge] (K2) -- ++( 0.45,-0.45);     


    \coordinate (P_I) at ($(K1)!0.5!(LegDownEnd)$);
    \draw[dthree] (P_I) -- ++(-0.30,-0.15) coordinate (Tip_I);
    \node at ($(Tip_I)+(-0.15,0)$) {$\text{(II')}$};

    \coordinate (P3) at ($(K1)!0.60!(K2)$);
    \draw[dthree] (P3) -- ++(-0.30,-0.15);
    \node at ($(P3)+(-0.25,-0.25)$) {$\text{(I')}$};

  \end{tikzpicture}
  \caption{}
\end{subfigure}
		
		\caption{
			(a) Toric diagram of resolved conifold at framing $f=-1$ and a chosen crepant resolution, and its dual graph. (b) The corresponding $(p,q)$-web with D3-branes dual to AV branes. The case (I') gives an $H$-observable, while the case (II') gives a $Q$-observable.
		} \label{fig:resconifold2}
	\end{figure}
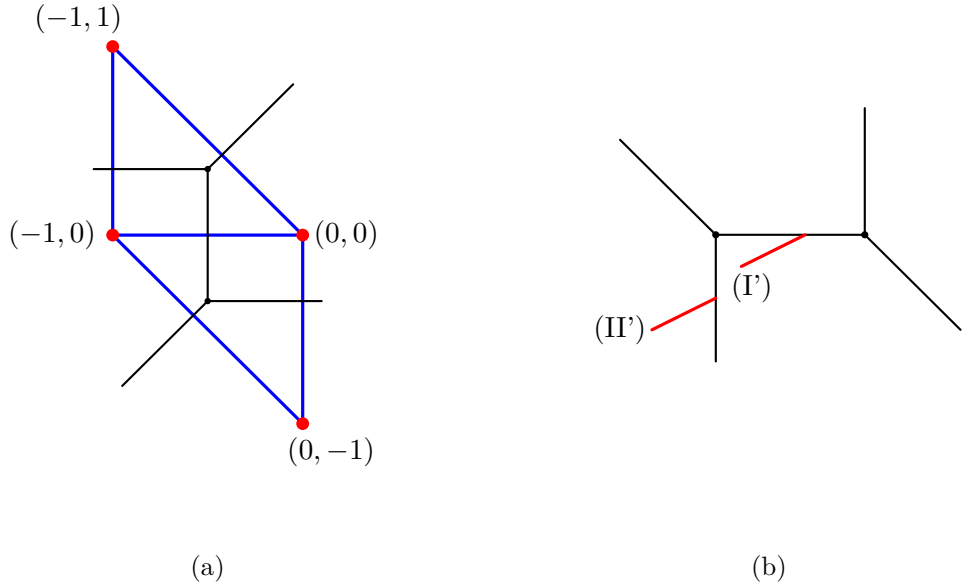

	In another $SL(2;\BZ)$ frame drawn in Figure \ref{fig:resconifold2}, the $(p,q)$-web of fivebranes contains a finite D5-brane segment, suspended between NS5-branes and $(1,1)$-branes. The 5d $\CalN=1$ theory engineered on this web is the pure $\U(1)$ gauge theory with zero Chern-Simons level. The refined closed BPS invariants are encoded in its partition function. As explained in \eqref{Z5d} for general case, the partition function is computed as the generating function of the equivariant Euler characteristics of the Hilbert scheme $\text{Hilb}^{[k]} (\BC^2)$ of $k$ points on $\BC^2$,
	\begin{align} \label{eq:closedres2}
		\begin{split}
			\mathcal{Z}_{\text{closed}} (\qe;q_1,q_2) &= \sum_{ k=0 } ^\infty \qe ^k \chi_{\mathsf{T}} \left(\text{Hilb}^{[k]}(\BC^2) \right)  \\
			&= \sum_{\{\l\}} \qe^{\vert \l \vert} \text{PE}\left[ \text{Ch}_{\mathsf{T}} \left( T_\l \text{Hilb}^{[k]} (\BC^2) \right) \right] = \prod_{i,j=0} ^\infty \left( 1- \qe q_1 ^{i+1} q_2 ^{j+1} \right)^{-1}.
		\end{split}
	\end{align}
	The last equality is combinatorially proven in \cite{Macdonald1995}. We identify the gauge coupling $\qe$ with the K\"{a}hler modulus assocaited with the generator of $H_2 (\CalX;\BZ)$. Note that there is precise match between \eqref{eq:closedres1} and \eqref{eq:closedres2}, under the identification of parameters $m= \qe q_1q_2$, as required by the independence of the refined closed BPS invariants on the $S$-transformation.\\
	
	Now we introduce an Aganagic-Vafa brane and study its refined open BPS invariants. As we discussed in general setup, the two kinds of codimension-two defects in the 5d theory arise from the AV brane. The $Q$-observable is engineered by a D3-brane ending on NS5-brane or $(1,1)$-brane (the case (II') in Figure \ref{fig:resconifold2}), while the $H$-observable is created by a D3-brane ending on the D5-brane segment (the case (I') in Figure \ref{fig:resconifold2}). 
	
	On the worldvolume of the D3-brane, the 3d $\CalN=2$ $\U(1)$ gauge theory with a chiral multiplet with charge $+1$ is realized. This theory couples to the 5d $\CalN=1$ pure $\U(1)$ gauge theory, yielding a codimension-two defect. The $Q$-observable is the defect observable for which the complex scalar $X$ is turn on with complexified FI parameter turned off. The $H$-observable is the defect observable for which the complexified FI parameter $Z$ is turned on at a special value of the complex scalar. In either case, the defect parameter $X$ or $Z$ is identified with the K\"{a}hler modulus for the relative cycle in $H_2 (\CalX,\CalL;\BZ)$. \\
	
	The expression of the $Q$-observable is written as
	\begin{align} \label{eq:qobsrk1}
		\begin{split}
			Q_f (X)_\l &= \th(X^{-1};q_1)^f \text{PE}\left[ - \frac{X^{-1} S_\l}{1-q_1} \right] \\
			&= \th(X^{-1};q_1) \left( X^{-1}q_1^{l(\l)} ;q_1 \right)_\infty\prod_{i=1}^{l(\l)} \left( 1- X^{-1}q_1^{i-1}q_2 ^{\l_i} \right)
		\end{split}
	\end{align}
	whose normalized vacuum expectation value is computed by
	\begin{align} \label{eq:qobsconi2}
		\CalZ_{\text{open}} (\qe,X;q_1,q_2) = \langle Q_f (X)\rangle =\frac{\sum_{\{\l\}} \qe^{\vert \l \vert} \text{PE} \left[ \text{Ch}_{\mathsf{T}} \left( T_\l \text{Hilb}^{[k]} (\BC^2) \right)\right] Q_f (X)_\l }{\sum_{\{\l\}} \qe^{\vert \l \vert} \text{PE} \left[ \text{Ch}_{\mathsf{T}} \left( T_\l \text{Hilb}^{[k]} (\BC^2) \right)\right] }.
	\end{align}
The vev is always expanded in $\vert \qe \vert <1$, where the series coefficients are exact rational functions in $X$. Further expanding it in a given chamber, we identify the partition function with the generating function of refined open BPS invariants there. At $f=-1$, we present the resulting refined open BPS invariants in the chamber $\vert X \vert <1 $ in Appendix \ref{app:conifold}.
    
	In the limit $q_2 \to 1$, this generating function solves the quantum mirror curve equation, as explained in \eqref{eq:tqeq} for general case. In the present case, where the gauge group is $\U(1)$, the only independent codimension-four observable is the basic one $W_1$, whose vacuum expectation value is simply the Coulomb parameter which we normalize to $1$. In turn, the coefficients of the $qq$-character \eqref{T} can be expressed in a closed form as
	\begin{equation}
		T (X) = 1   - q_1^{-1} q_2^{-1} X^{-1}. \label{eq:qqCon}
	\end{equation}
	With this, the quantum mirror curve equation \eqref{eq:tqeq} thus specializes in this case to
	\begin{equation}
		0=  \left[(-q_1 X)^f q_1^{D_X} - 1  + q_1 ^{-1} X^{-1} + (-X^{-1})^{f+1}  \mathfrak q q_1^{-D_X}\right] \left\langle Q_f(X) \right\rangle_{q_2=1}, \label{eq:qMirrorCon}
	\end{equation}
	In the limit $q_1 \to 1$, we recover the mirror curve of the resolved conifold,
	\begin{align} \label{eq:mirrorresconi}
		0 = (-X)^f Z -1 + X^{-1} + (-X^{-1})^{f+1} \qe Z^{-1}.
	\end{align}
	At $f=-1$ in particular, we obtain the Newton polynomial equation associated with the toric diagram in Figure \ref{fig:resconifold2}. All the other cases with $f\neq -1$ can be obtained by applying $T$-transformations.\\
	
	The $H$-observable, on the other hand, is obtained by turning on the complexified FI parameter. Its observable expression \eqref{eq:hobs} is obtained by
	\begin{align}
		\begin{split}
			H_f (Z)_\l &= \sum_{X \in q_1 ^\BZ} \exp\left[-\frac{\log X \log \left(Z (-1)^f (Xq_1^{-1})^{\frac{f}{2}}\right)}{\log q_1}\right] Q_0 (X q_1^{-1})_\l\\
			& =\sum_{n=0} ^\infty  \frac{\prod_{i=1} ^{l(\l)} \left(1-q_1 ^{n+i -l(\l)} q_2^{\l_i} \right)}{(q_1;q_1)_n} q_1 ^{-\frac{f}{2} (n-l(\l))(n-l(\l)+1)} \left( Z(-1)^f \right)^{n-l(\l)}.
		\end{split}
	\end{align}
	In the decoupling limit of the 5d gauge theory, $\qe \to 0$, the only nonvanishing contribution arises from $\l = \varnothing$, and the $H$-observable reduces to the K-theoretic vortex partition function \eqref{eq:hbosc3} of the 3d $\CalN=2$ gauge theory. From the geometry of the resolved conifold, this limit corresponds to degenerating the resolved conifold to $\BC^3$. When the 5d gauge coupling is turned on, $\qe \neq 0$, note that the vortex number is bounded below by $-k$, where $k\geq 0$ is the bulk instanton number, instead of $0$. Namely, the 3d-5d coupling allows not only vortex configurations but also anti-vortex configurations in sectors with nonzero bulk instanton number. Integrating out the 3d degrees of freedom at a fixed instanton configuration then yields the $H$-observable obtained above.
	
	The refined open BPS invariants are extracted from its vacuum expectation value,
	\begin{align} \label{eq:hvevconi2}
		\CalZ_{\text{open}} (\qe,Z;q_1,q_2) = \langle H_f (X)\rangle =\frac{\sum_{\{\l\}} \qe^{\vert \l \vert} \text{PE} \left[ \text{Ch}_{\mathsf{T}} \left(T_\l \text{Hilb}^{[k]} (\BC^2)\right) \right] H_f (X)_\l }{\sum_{\{\l\}} \qe^{\vert \l \vert} \text{PE} \left[ \text{Ch}_{\mathsf{T}} \left( T_\l \text{Hilb}^{[k]} (\BC^2) \right) \right] }.
	\end{align}
    Here, $\qe$ is still identified with the K\"{a}hler modulus for the generator of $H_2 (\CalX;\BZ)$, while $Z$ is the one for the relative cycle in $H_2 (\CalX,\CalL;\BZ)$. The vev is convergent in the domain $\vert \qe \vert < \vert Z \vert <1$, which gives the K\"{a}hler moduli chamber that the refined open BPS invariants are computed in.
    
	The quantum mirror curve equation satisfied by the $q_2 \to 1$ limit of this vev follows immediately as
	\begin{align}
		0 = \left[ (-q_1^{-D_Z})^f Z  -1 +  q_1 ^{D_Z}  + \qe   (-q_1^{1+D_Z})^{f+1} Z^{-1}  \right]\left\langle H_f (Z)\right\rangle_{q_2=1}.
	\end{align}
	The $q_1 \to 1$ limit yields the mirror curve \eqref{eq:mirrorresconi} of the resolved conifold by construction.
	
	\subsubsection{Mapping Aganagic-Vafa branes by \texorpdfstring{$S$}{S}-transformation}

	The two toric diagrams shown in Figures \ref{fig:resconifold1} and \ref{fig:resconifold2} are related by an $S$-transformation. Accordingly, the Aganagic–Vafa branes defined in the two setups are expected to map to each other under the same $S$-transformation, and the corresponding $Q$- and $H$-generating functions of refined open BPS invariants should match. 
	
	Indeed, we confirm the following matching at expansion in K\"{a}hler parameters:
	\begin{align}
		Q_{f=0} ^{\eqref{eq:Qobsconi}} (m ,X ;q_1,q_2)  = \left\langle H_{f=-1} ^\eqref{eq:hvevconi2} \right\rangle \left(\qe = m q_1 ^{-1} q_2 ^{-1},Z = (q_1X)^{-1}  ;q_1,q_2 \right),
	\end{align}
	in the K\"{a}hler moduli chamber $\vert m \vert <\vert X^{-1} \vert <1$, and
	\begin{align} \label{eq:rhsss}
		H_{f=0} ^{\eqref{eq:hconi1}} (m,Z;q_1,q_2) = \left\langle Q_{f=-1} ^{\eqref{eq:qobsconi2}} \right\rangle \left(\qe =m q_1^{-1 }q_2 ^{-1} ,X = Z m^{-1} q_1^{-1};q_1,q_2 \right),
	\end{align}
    in the K\"{a}hler moduli chamber $\vert Z \vert< \vert m \vert <1$. The matching of the two expressions involves nontrivial combinatorial identities, which can be verified order by order in series expansions. Note that the left hand sides in fact do not depend on $q_2$, and correspondingly the $q_2$-dependence of the right hand sides is entirely absorbed into the K\"{a}hler parameters. This nontrivial agreement provides a strong support for our main claim that the 3d-5d partition function furnishes the generating function of refined open BPS invariants.

    Note that the left hand side of \eqref{eq:rhsss} is a series in $Z$ whose coefficients are exact rational functions in $m$, while the right hand side is defined as a series in $\qe = m q_1 ^{-1} q_2 ^{-1}$ whose coefficients are exact rational functions in $X = Z m^{-1} q_1 ^{-1}$. Thus, the equality realized in the chamber $\vert Z \vert< \vert m \vert <1$ provides the analytic continuation of the generating function to other chambers.

	\subsection{Resolved \texorpdfstring{$A_1$}{A1}-singularity}

	The resolved $A_1$-singularity, $\mathcal{X} = \text{Tot}(\mathcal{O} \oplus \mathcal{O}(-2)\to \BP^1)$, admits toric diagram drawn in Figure \ref{fig:a1singtoric1} and \ref{fig:a1singtoric}. Similar to the resolved conifold case, we begin with the two $SL(2;\BZ)$ frames drawn in the two figures, related to each other by an $S$-transformation, and act $T$-transformations to each to span all possible frames. 
	
    \subsubsection{3d \texorpdfstring{$\mathrm{U}(1)$}{U(1)} gauge theory with two chiral multiplets}

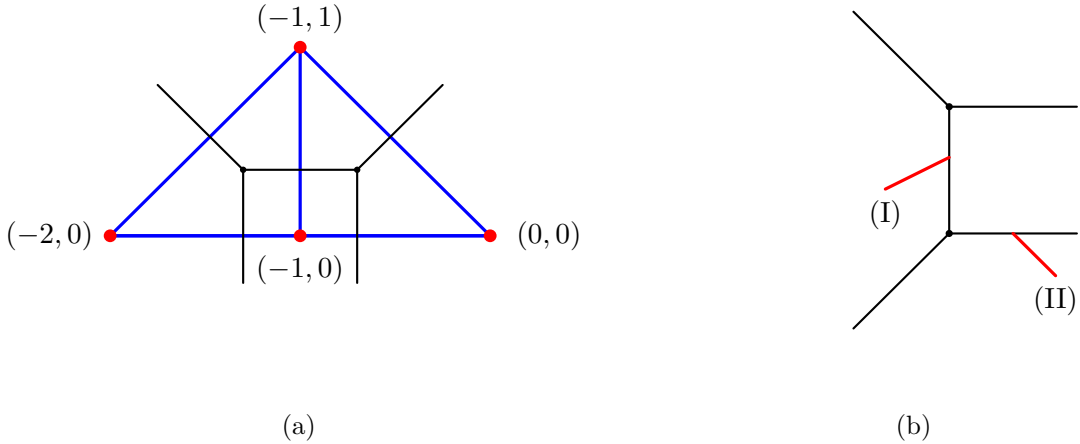
\begin{figure}[h!]
		\centering
		
		\begin{subfigure}[b]{0.48\textwidth}
			\centering
			\begin{tikzpicture}[
				x=2.5cm,y=2.5cm,
				line cap=round,line join=round,
				baseline={(0, 0.35)} 
				]
				\useasboundingbox (-2.5, -0.85) rectangle (0.5, 1.55);
				
				\tikzset{
					toric vertex/.style={circle, fill=red, inner sep=1.7pt},
					toric edge/.style={blue, very thick},
					dual edge/.style={black, thick},
				}
				
				\coordinate (A) at (0,0);
				\coordinate (B) at (-1,1);
				\coordinate (P) at (-1,0); 
				\coordinate (D) at (-2,0);
				
				\draw[toric edge] (D) -- (B) -- (A) -- (P) -- cycle;
				\draw[toric edge] (P) -- (B);
				\draw[toric edge] (P) -- (A);
				\draw[toric edge] (P) -- (D);
				
				\foreach \Q in {A,B,P,D}{
					\node[toric vertex] at (\Q) {};
				}
				
				\node at ( 0.31, 0.00) {$(0,0)$};
				\node at (-1.00, 1.15) {$(-1,1)$};
				\node at (-1.00,-0.18) {$(-1,0)$};
				\node at (-2.32, 0.00) {$(-2,0)$};
				
				\coordinate (J1) at (-0.70, 0.35);
				\coordinate (J2) at (-1.30, 0.35);
				\fill[black] (J1) circle (1.2pt);
				\fill[black] (J2) circle (1.2pt);
				
				\draw[dual edge] (J2) -- (J1);
				
				\draw[dual edge] (J1) -- ++(0, -0.60);
				\draw[dual edge] (J2) -- ++(0, -0.60);
				\draw[dual edge] (J1) -- ++(0.45, 0.45);
				\draw[dual edge] (J2) -- ++(-0.45, 0.45);
				
			\end{tikzpicture}
			\caption{}
		\end{subfigure}%
		\hfill
\begin{subfigure}[b]{0.48\textwidth}
  \centering
  \begin{tikzpicture}[
    x=2.5cm,y=2.5cm,
    scale=1.12,
    line cap=round,line join=round,
    baseline={(0,0)}
  ]
    \useasboundingbox (-1.5, -1.07) rectangle (1.5, 1.07);

    \tikzset{
      dual edge/.style={black, thick},
      dthree/.style={red, very thick},
    }

    \coordinate (K1) at (0,  0.30);
    \coordinate (K2) at (0, -0.30);
    
    \fill[black] (K1) circle (1.2pt);
    \fill[black] (K2) circle (1.2pt);

    \draw[dual edge] (K2) -- (K1);

    \draw[dual edge] (K1) -- ++(0.60, 0);        
    \draw[dual edge] (K1) -- ++(-0.45, 0.45);    
    
    \draw[dual edge] (K2) -- ++(0.60, 0) coordinate (LegBotRight);  
    \draw[dual edge] (K2) -- ++(-0.45, -0.45);   


    \coordinate (P) at ($(K2)!0.60!(K1)$);
    \draw[dthree] (P) -- ++(-0.30, -0.15);
    \node at ($(P)+(-0.3,-0.27)$) {$\text{(I)}$};

    \coordinate (P_II) at ($(K2)!0.5!(LegBotRight)$);
    \draw[dthree] (P_II) -- ++(0.20, -0.20) coordinate (Tip_II);
    \node at ($(Tip_II)+(0,-0.12)$) {$\text{(II)}$};

  \end{tikzpicture}
  \caption{}
\end{subfigure}
		
		\caption{
			(a) Toric diagram of resolved $A_1$-singularity at $f= -1$ with a chosen crepant resolution, and its dual graph. (b) The corresponding $(p,q)$-web D3-branes dual to AV branes. The case (I) provides a $Q$-observable, and the case (II) provides an $H$-observable.
		} \label{fig:a1singtoric1}
	\end{figure}

	In the $SL(2;\BZ)$ frame drawn in Figure \ref{fig:a1singtoric1}, the 5d $\CalN=1$ theory engineered on the $(p,q)$-web of fivebranes is just a W-boson of a decoupled gauge multiplet. The partition function, encoding the refined closed BPS invariants, is simply given by the 1-loop contribution
	\begin{align} \label{eq:closeda11}
		\CalZ_{\text{closed}}(m;q_1,q_2) = \text{PE}\left[ - \frac{m}{(1-q_1)(1-q_2)} \right] = \prod_{i,j=0} ^\infty \left( 1-m q_1 ^i q_2^j \right),
	\end{align}
	where $m\in \BC^\times$ is the mass of the W-boson. The K\"{a}hler modulus associated with the generator of $H_2 (\CalX;\BZ)$ is identified with this mass. \\

	Now we insert an Aganagic-Vafa brane and study the refined open BPS invariants associated to it. The worldvolume theory realized on the single D3-brane is the 3d $\CalN=2$ $\U(1)$ gauge theory with two chiral multiplets, both carrying charge $+1$. Since the 5d theory is free, the 3d $\CalN=2$ partition function itself provides the generating function of the refined open BPS invariants in this case. As before, we have the $Q$-observable from a D3-brane ending on NS5-brane, $(1,1)$-brane, or $(1,-1)$-brane (the case (I) in Figure \ref{fig:a1singtoric1}); while the $H$-observable from a D3-brane ending on D5-brane (the case (II) in Figure \ref{fig:a1singtoric1}). The K\"{a}hler modulus associated with the relative cycle in $H_2 (\CalX,\CalL;\BZ)$ is identified with $X$ or $Z$, respectively. \\
	
	The $Q$-observable is given by
	\begin{equation} \label{eq:qbosa11}
		Q_f(X) = \th(X^{-1};q_1)_\infty^f  (X^{-1};q_1)_\infty (m X^{-1};q_1)_\infty.
	\end{equation}
    In any framing $f\in \BZ$, this is a product of finite number of $q$-Pochhammer symbols. Thus, there are only finite number of non-zero refined open BPS invariants.
    
	The quantum mirror curve equation satisfied by $Q_f (X)$ is simply
	\begin{equation} \label{eq:qcurvea11}
		0= \left[(-q_1X)^f q_1^{D_X} - 1+q_1^{-1} (1+m)X^{-1} - q_1^{-2} m X^{-2} \right] Q_f(X) .
	\end{equation}
	In the limit $q_1 \to 1$, this reduces to the mirror curve of the resolved $A_1$-singularity,
	\begin{align} \label{eq:mcurvea12}
		0= ( -X)^f Z - 1 + (1+m) X^{-1} - m X^{-2}.
	\end{align}
	Indeed, at $f=-1$, it is the Newton polynomial equation for the toric diagram in Figure \ref{fig:a1singtoric1}. The other cases with $f\neq -1$ are simply obtained by the $T$-transformations.\\
	
	The $H$-observable can be obtained by a $q_1$-lattice summation \eqref{eq:hobs} of $Q_0 (X)$. It is computed as
	\begin{align} \label{eq:hfa1}
		\begin{split}
			H_f (Z) &= \sum_{X \in q_1 ^\BZ} \exp\left[-\frac{\log X \log \left(Z (-1)^f (Xq_1^{-1})^{\frac{f}{2}}\right)}{\log q_1}\right] Q_0 (Xq_1^{-1}) \\
			& = (mq_1;q_1)_\infty \sum_{n=0} ^\infty \frac{q_1 ^{-\frac{f}{2} n(n+1)} \left( Z (-1)^f \right)^n}{ (q_1;q_1)_n (mq_1 ;q_1)_n}.
		\end{split}
	\end{align}
Here, $m$ is still identified with the K\"{a}hler modulus for the generator of $H_2 (\CalX;\BZ)$, while $Z$ is the new K\"{a}hler modulus for the relative cycle in $H_2 (\CalX,\CalL;\BZ)$. The partition function is defined in $\vert Z \vert <1$ by construction, with the coefficients given as exact rational functions in $m$. By further expanding it in a given chamber, we obtain the generating function of refined open BPS invariants there. We present the refined open BPS invariants at $f=-1$ in the chamber $\vert m \vert<1$ in Appendix \ref{app:a1sing}.
    
	The quantum mirror curve equation satisfied by $H_f (Z)$ follows from \eqref{eq:qcurvea11} and \eqref{eq:hfa1} as
	\begin{align}
		0 = \left[ (-q_1 ^{-D_Z})^f Z - 1 + (1+m) q_1 ^{D_Z} - m q_1 ^{2D_Z} \right] H_f (Z).
	\end{align}
	In the limit $q_1\to 1$, this reduces to the mirror curve \eqref{eq:mcurvea12} by construction.
	
	\subsubsection{3d \texorpdfstring{$\mathrm{U}(1)$}{U(1)} gauge theory with one chiral multiplet coupled to 5d pure \texorpdfstring{$\mathrm{U}(1)$}{U(1)} gauge theory with \texorpdfstring{$k_{\text{CS}}=-1$}{k\_CS=-1}}

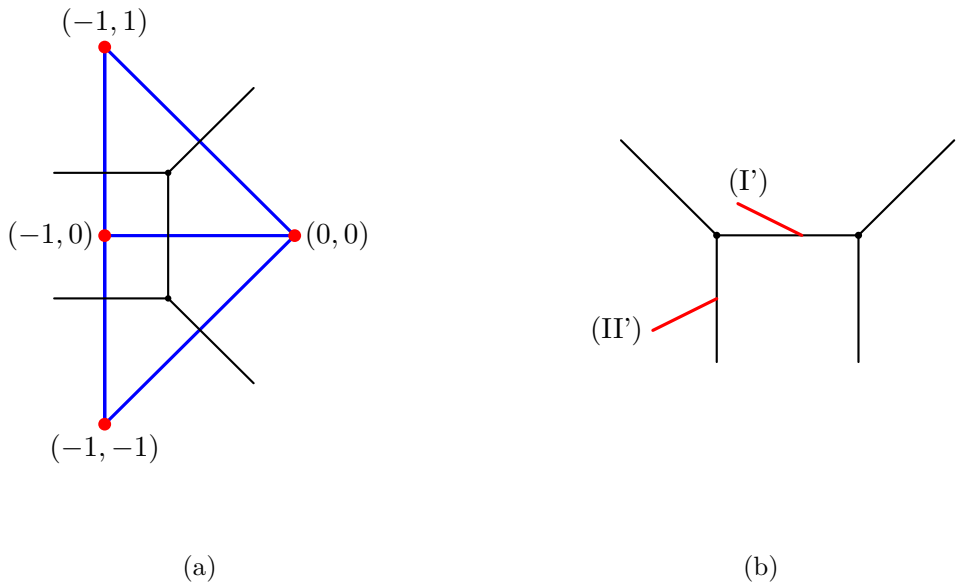
\begin{figure}[h!]
		\centering
		
		\begin{subfigure}[b]{0.45\textwidth}
			\centering
			\begin{tikzpicture}[
				x=2.5cm,y=2.5cm,
				line cap=round,line join=round,
				baseline={(0,0)} 
				]
				\useasboundingbox (-1.5,-1.6) rectangle (0.5,1.6);
				
				\tikzset{
					toric vertex/.style={circle, fill=red, inner sep=1.7pt},
					toric edge/.style={blue, very thick},
					dual edge/.style={black, thick},
				}
				
				\coordinate (A) at (0,0);
				\coordinate (N) at (-1, 1);
				\coordinate (S) at (-1,-1);
				\coordinate (P) at (-1, 0); 
				
				\draw[toric edge] (S) -- (A) -- (N) -- cycle; 
				\draw[toric edge] (N) -- (P) -- (S);          
				\draw[toric edge] (P) -- (A);                 
				
				\foreach \Q in {A,N,S,P}{
					\node[toric vertex] at (\Q) {};
				}
				
                \node[right=0cm of A] () {$(0,0)$};
				\node[above=0cm of N] () {$(-1,1)$};
				\node[below=0cm of S] () {$(-1,-1)$};
				\node[left=0cm of P] () {$(-1,0)$};
				
				\coordinate (J1) at (-0.666,  0.333); 
				\coordinate (J2) at (-0.666, -0.333); 
				\fill[black] (J1) circle (1.2pt);
				\fill[black] (J2) circle (1.2pt);
				
				\draw[dual edge] (J2) -- (J1);
				
				\draw[dual edge] (J1) -- ++( 0.45, 0.45); 
				\draw[dual edge] (J1) -- ++(-0.60, 0.00); 
				
				\draw[dual edge] (J2) -- ++( 0.45,-0.45); 
				\draw[dual edge] (J2) -- ++(-0.60, 0.00); 
				
			\end{tikzpicture}
			\caption{}
		\end{subfigure}%
		\hspace{1em}%
\begin{subfigure}[b]{0.45\textwidth}
  \centering
  \begin{tikzpicture}[
    x=2.5cm,y=2.5cm,
    scale=1.12,
    line cap=round,line join=round,
    baseline={(0,0)} 
  ]
    \useasboundingbox (-1.25,-1.43) rectangle (1.0,1.43);

    \tikzset{
      dual edge/.style={black, thick},
      dthree/.style={red, very thick},
    }

    \coordinate (K1) at (-0.333, 0);
    \coordinate (K2) at ( 0.333, 0);
    \fill[black] (K1) circle (1.2pt);
    \fill[black] (K2) circle (1.2pt);

    \draw[dual edge] (K1) -- (K2);

    
    \draw[dual edge] (K1) -- ++(-0.45, 0.45);        
    \draw[dual edge] (K1) -- ++( 0.00,-0.60) coordinate (LegLeftVert);

    \draw[dual edge] (K2) -- ++( 0.45, 0.45);        
    \draw[dual edge] (K2) -- ++( 0.00,-0.60);        


    \coordinate (P3) at ($(K1)!0.60!(K2)$);
    \draw[dthree] (P3) -- ++(-0.30,0.15);
    \node at ($(P3)+(-0.25,0.25)$) {$\text{(I')}$};

    \coordinate (P_IIp) at ($(K1)!0.5!(LegLeftVert)$);
    \draw[dthree] (P_IIp) -- ++(-0.30,-0.15) coordinate (Tip_IIp);
    \node at ($(Tip_IIp)+(-0.17,0)$) {$\text{(II')}$};

  \end{tikzpicture}
  \caption{}
\end{subfigure}
		
		\caption{
			(a) Toric diagram of resolved $A_1$-singularity at $f=-1$ with a chosen crepant resolution, and its graph-dual.	(b) The corresponding $(p,q)$-web with D3-branes dual to AV branes. The case (I') corresponds to an $H$-observable, while the case (II') gives a $Q$-observable.
		} \label{fig:a1singtoric}
	\end{figure}

	Next, we consider the $SL(2;\BZ)$ frame drawn in Figure \ref{fig:a1singtoric} and its $T$-transformations. The 5d $\CalN=1$ gauge theory engineered by the $(p,q)$-web is the pure $\U(1)$ gauge theory with bare Chern-Simons level $k_{\text{CS}} =-1$.\footnote{We may equivalently take $k_{\text{CS}}=+1$, which is related with our case $k_{\text{CS}} =-1$ by a $T$-transformation and a flip.} The partition function specializes in this case to the generating function of the equivariant Euler characteristics of $L^{-1}= \left(\det K\right)^{-1}$ over $\text{Hilb}^{[k]} (\BC^2)$,
	\begin{align} \label{eq:closeda12}
		\begin{split}
			\CalZ_{\text{closed}} (\qe;q_1,q_2) &= \sum_{k=0} ^\infty \qe^k \chi_{\mathsf{T}} (\text{Hilb}^{[k]} (\BC^2); L^{-1}) \\
			&=\sum_{\{\l\}} \qe^{\vert \l \vert} \left( \prod_{\Box \in \l} \chi_\Box ^{-1} \right) \text{PE}\left[ \text{Ch}_{\mathsf{T}} \left( T_\l \text{Hilb}^{[k]} (\BC^2) \right) \right] = \prod_{i,j=0} ^\infty \left( 1 + \qe q_1 ^{i+1} q_2 ^{j+1} \right),
		\end{split}
	\end{align}
	where the last equality was proven in \cite{Haiman2001VanishingTA}. The K\"{a}hler parameter associated with the generator of $H_2 (\CalX;\BZ)$ is identified with the gauge coupling $\qe$. Note that the partition function precisely agrees with \eqref{eq:closeda11}, given the identification of parameters $m = -\qe q_1q_2$. Such an agreement is expected from the independence of the refined closed BPS invariants on the $S$-transformation.\\

	Now, let us insert an Aganagic-Vafa brane. A single D3-brane ending on the $(p,q)$-web of fivebranes gives rise to 3d $\CalN=2$ $\U(1)$ gauge theory with one chiral multiplet with charge $+1$. This theory couples to the 5d $\CalN=1$ $\U(1)$ gauge theory with $k_{\text{CS}}=-1$ by gauging the 3d $\U(1)$ flavor symmetry by the 5d gauge field restricted to the locus of the 3d theory. We obtain a $Q$-observable by turning on the complex scalar $X$ in the twisted chiral multiplet (the case (II') in Figure \ref{fig:a1singtoric}), or a $H$-observable by turning on the complexified FI parameter $Z$ (the case (I') in Figure \ref{fig:a1singtoric}). These defect parameters are identified with the K\"{a}hler modulus for the relative cycle in $H_2 (\CalX,\CalL;\BZ)$. \\
	
	The $Q$-observable attains exactly the same expression with \eqref{eq:qobsrk1} for the resolved conifold. Namely, we have
	\begin{align} 
		\begin{split}
			Q_f (X) &= \th(X^{-1};q_1) ^f \text{PE}\left[ - \frac{X^{-1} S }{1-q_1} \right] \\
			& =\th(X^{-1};q_1)^f \left( X^{-1}q_1^{l(\l)} ;q_1 \right)_\infty\prod_{i=1}^{l(\l)} \left( 1- X^{-1}q_1^{i-1}q_2 ^{\l_i} \right).
		\end{split}
	\end{align}
	The only difference is that the 5d $\CalN=1$ $\U(1)$ gauge theory now gains the bare CS level $k_{\text{CS}}=-
	1$, modifying the vacuum expectation value by
	\begin{align} \label{eq:qobsa12}
		\CalZ_{\text{open}} (\qe,X;q_1,q_2) =  \langle Q_f (X)\rangle =\frac{\sum_{\{\l\}} \qe^{\vert \l \vert}  \left( \prod_{\Box \in \l} \chi_\Box ^{-1} \right)\text{PE} \left[ \text{Ch}_{\mathsf{T}} \left( T_\l \text{Hilb}^{[k]} (\BC^2) \right)\right] Q_f (X)_\l }{\sum_{\{\l\}} \qe^{\vert \l \vert}   \left( \prod_{\Box \in \l} \chi_\Box ^{-1} \right) \text{PE} \left[ \text{Ch}_{\mathsf{T}}\left( T_\l \text{Hilb}^{[k]} (\BC^2) \right) \right] }.
	\end{align}
    The vev is defined as a series in the gauge coupling in $\vert \qe \vert <1$ by construction. Each coefficient is an exact rational function in $X$. Further expanding it in a given chamber, we obtain the generating function of refined open BPS invariants. We present the so-obtained refined open BPS invariants at $f=-1$ in the chamber $\vert X \vert <1$ in Appendix \ref{app:a1sing}.
	
	The $qq$-character can be computed to be
	\begin{align}
		T(X) = 1+ \left(\qe- q_1 ^{-1} q_2 ^{-1}  \right) X^{-1}. \label{eq:qqResA1}
	\end{align}
	Thus, the quantum mirror curve equation \eqref{eq:tqeq} specializes in this case to
	\begin{align}
		0 = \left[ (-q_1 X)^f q_1 ^{D_X} -1 - \left(\qe -q_1^{-1} \right) X^{-1} -\qe  (-X)^{-f-2}  q_1 ^{-D_X} \right]\left\langle  Q_f (X) \right\rangle_{q_2=1}, \label{eq:qMirrorResA1}
	\end{align}
	which reduces to the mirror curve of the resolved $A_1$ singularity in the limit $q_1 \to 1$,
	\begin{align} \label{eq:mca1}
		0 = (-X)^f Z - 1 - (\qe -1) X^{-1}  - \qe  (-X)^{-f-2} Z^{-1}.
	\end{align}
	At $f=-1$, this indeed recovers the Newton polynomial equation associated to the toric diagram in Figure \ref{fig:a1singtoric}. All the other $f \neq -1$ can be obtained by taking the transformation by $T^f$.\\
	
	Next, let us turn to the $H$-observable. Its expression is again obtained by a $q_1$-lattice summation of $Q_0(X)$ as
	\begin{align} 
		\begin{split}
			H_f (Z)_\l & = \sum_{X \in q_1 ^\BZ}  \exp\left[-\frac{\log X \log \left(Z (-1)^f (Xq_1^{-1})^{\frac{f}{2}}\right)}{\log q_1}\right] Q_0 (Xq_1^{-1})_\l \\
			& = \sum_{n=0} ^\infty  \frac{\prod_{i=1} ^{l(\l)} \left(1-q_1 ^{n+i -l(\l)} q_2^{\l_i} \right)}{(q_1;q_1)_n} q_1 ^{-\frac{f}{2} (n-l(\l))(n-l(\l)+1)} \left( Z(-1)^f \right)^{n-l(\l)},
		\end{split}
	\end{align}
	which is the same with the one for the resolved conifold. The only difference is that the effect of the $k_{\text{CS}}=1$ when taking the vacuum expectation value,
	\begin{align}\label{eq:hobsa12}
		\CalZ_{\text{open}} (\qe,Z;q_1,q_2) = \langle H_f (Z)\rangle =\frac{\sum_{\{\l\}} \qe^{\vert \l \vert}  \left( \prod_{\Box \in \l} \chi_\Box \right) \text{PE} \left[ \text{Ch}_{\mathsf{T}} \left( T_\l \text{Hilb}^{[k]} (\BC^2) \right) \right] H_f (Z)_\l }{\sum_{\{\l\}} \qe^{\vert \l \vert}  \left( \prod_{\Box \in \l} \chi_\Box \right) \text{PE} \left[ \text{Ch}_{\mathsf{T}} \left( T_\l \text{Hilb}^{[k]} (\BC^2)\right) \right] }.
	\end{align}
This vev is a double series expanded in the domain $\vert \qe \vert < \vert Z \vert <1$. It is identified with the generating function of refined open BPS invariants in the corresponding chamber.
    
	The quantum mirror curve equation satisfied by $H_f(Z)$ is straightforward,
	\begin{align}
		0 = \left[ (-q_1 ^{-D_Z})^f Z -1 -(\qe q_1 -1) q_1 ^{D_Z} - \qe  (-q_1^{-1-D_Z})^{-f-2} Z^{-1}  \right]\langle H_f (Z) \rangle _{q_2 =1}.
	\end{align}
	In the limit $q_1 \to 1$, this reduces to the same mirror curve \eqref{eq:mca1} by construction.
	
	\subsubsection{Mapping Aganagic-Vafa branes by \texorpdfstring{$S$}{S}-transformation}

	The $S$-transformation relates the two $SL(2;\BZ)$ frames in Figure \ref{fig:a1singtoric1} and \ref{fig:a1singtoric}. The Aganagic-Vafa branes in two setups should be exchanged accordingly. In particular, the $Q$ and $H$ generating functions of refined open BPS invariants are expected to match with each other.
	
	Indeed, we confirm the following identities:
	\begin{align}
		Q_{f=-1} ^{\eqref{eq:qbosa11}} (m,X;q_1,q_2) = \left\langle H_{f=-1} ^{\eqref{eq:hobsa12}} \right\rangle \left(\qe= -m q_1^{-1} q_2 ^{-1},Z = (q_1X)^{-1} m;q_1,q_2)\right),
	\end{align}
    in the chamber $\vert m \vert< \vert X \vert <1$, and
	\begin{align} \label{eq:eqalal}
		H_{f=-1} ^{\eqref{eq:hfa1}} (m,Z;q_1,q_2) = \left\langle Q_{f=-1} ^{\eqref{eq:qobsa12}} \right\rangle \left(\qe= -m q_1  q_2 ^{-1},X = Z ;q_1 ^{-1},q_2\right),
	\end{align}
    in the chamber $\vert m \vert, \vert Z \vert <1$ at series expansions. Note that the left hand sides in fact is not dependent on $q_2$, and correspondingly the $q_2$-dependence of the right hand sides is entirely absorbed into the K\"{a}hler parameters.

Note that the right hand side of \eqref{eq:eqalal} is defined in $\vert \qe \vert <1$, while the left hand side is defined in $\vert Z \vert <1$. Thus, the matching in the domain $\vert m \vert, \vert Z \vert <1$ yields the analytic continuation of the generating function to different chambers. For instance, we may expand the left hand side in the chamber $\vert Z \vert <1 < \vert m \vert$ and extract the refined open BPS invariants there. This precisely agrees with the ones in \cite{Cheng:2021nex}, which were obtained by realizing the geometric transition in the topological vertex amplitudes.
    
	\subsection{Local $\BP^1 \times \BP^1$}
	Now, we consider the case where $\CalX$ contains a compact 4-cycle. The simplest example is the local $\BP^1 \times \BP^1$, i.e., $\CalX = \text{Tot}\left( K_{\BP^1 \times \BP^1}\right) $. The 5d $\CalN=1$ gauge theory realized by the $(p,q)$-web is the pure $\U(2)$ gauge theory. The partition function specializes to
	\begin{align} \label{eq:partp1p1}
		\begin{split}
			\mathcal{Z}_{\text{closed}}(a,\qe;q_1,q_2) &= \CalZ^{\text{1-loop}} (a;q_1,q_2) \sum_{k=0} ^\infty \qe^{k} \chi_{\mathsf{T}} \left( \CalM_{2,k} \right) \\
			&= \CalZ^{\text{1-loop}} (a;q_1,q_2) \sum_{\{\underline{\l}\}} \qe^{\vert \underline{\l} \vert} \text{PE}\left[ \text{Ch}_{\mathsf{T}} \left( T_{\underline{\l}} \CalM_{2,k}\right) \right],
		\end{split}
	\end{align}
	where the 1-loop part is independent of the gauge coupling $\qe$, determined solely by perturbative contributions of the W-bosons as
	\begin{align} \label{eq:1loop}
		\CalZ^{\text{1-loop}} (a;q_1,q_2) = \prod_{i,j=0}^\infty \left( 1- aq_1 ^i q_2 ^j \right)\left( 1- a q_1 ^{i+1} q_2 ^{j+1} \right).
	\end{align}
	To extract the refined closed BPS invariants of the local $\BP^1 \times \BP^1$, we identify the gauge coupling $\qe$ and the Coulomb parameter $a$ with the complexified K\"{a}hler parameters associated with the base $\BP^1$ and the fiber $\BP^1$ as $Q_b =  a$ and $Q_f= \qe a$. Note that the partition function \eqref{eq:partp1p1} is a series in $\qe$, with each coefficient given by a rational function of ($a$, $q_1$, $q_2$). To bring it into the Gopakumar-Vafa form, each coefficient should further be expanded in $a$. 
	
	The $\BP^1 \times \BP^1$ enjoys the symmetry of exchanging the base and the fiber. Thus, the refined closed BPS invariants should be symmetric under the exchange of $Q_b$ and $Q_f$. Such a symmetry is far from being obvious in the expression of the partition function \eqref{eq:partp1p1}, since the gauge coupling $\qe$ and the Coulomb parameter $a$ are not in the equal footing in the 5d $\CalN=1$ gauge theory. In particular, this symmetry swaps the perturbative contribution and the non-perturbative contribution to the partition function. Nevertheless, a straightforward check at series expansion shows that it is indeed symmetric.
	
	\begin{figure}[h!]
		\centering
		
		\begin{subfigure}{0.48\textwidth}
			\centering
			\begin{tikzpicture}[
				x=2.5cm,y=2.5cm,
				line cap=round,line join=round,
				baseline=(current bounding box.center)
				]
				\useasboundingbox (-1.6,-1.6) rectangle (1.6,1.6);
				
				\tikzset{
					toric vertex/.style={circle, fill=red, inner sep=1.7pt},
					toric edge/.style={blue, very thick},
					dual edge/.style={black, thick},
				}
				
				\coordinate (O) at (0,0);
				\coordinate (E) at (1,0);
				\coordinate (N) at (0,1);
				\coordinate (W) at (-1,0);
				\coordinate (S) at (0,-1);
				
				\draw[toric edge] (W) -- (N) -- (E) -- (S) -- cycle;
				
				\draw[toric edge] (O) -- (E);
				\draw[toric edge] (O) -- (N);
				\draw[toric edge] (O) -- (W);
				\draw[toric edge] (O) -- (S);
				
				\foreach \P in {O,E,N,W,S}{
					\node[toric vertex] at (\P) {};
				}
				
				\node (center) at ( -1.2, -.5) {$(-1,0)$};
                \draw[-stealth, black!50]
                (center)
                .. controls +(.5,0.5) and +(-0.20,-0.20) ..
                (-.02,-.02);
				\node at ( 1.12, 0.12) {$(0,0)$};
				\node at ( 0.12, 1.12) {$(-1,1)$};
				\node at (-1.18, 0.12) {$(-2,0)$};
				\node at ( 0.12,-1.18) {$(-1,-1)$};
				
				\coordinate (J1) at ( 1/3, 1/3);
				\coordinate (J2) at (-1/3, 1/3);
				\coordinate (J3) at (-1/3,-1/3);
				\coordinate (J4) at ( 1/3,-1/3);
				\foreach \Q in {J1,J2,J3,J4}{\fill[black] (\Q) circle (1.2pt);}
				
				\draw[dual edge] (J1) -- (J2);
				\draw[dual edge] (J2) -- (J3);
				\draw[dual edge] (J3) -- (J4);
				\draw[dual edge] (J4) -- (J1);
				
				\draw[dual edge] (J1) -- ++( 0.40, 0.40);
				\draw[dual edge] (J2) -- ++(-0.40, 0.40);
				\draw[dual edge] (J3) -- ++(-0.40,-0.40);
				\draw[dual edge] (J4) -- ++( 0.40,-0.40);
				
			\end{tikzpicture}
			\caption{}
		\end{subfigure}
\begin{subfigure}{0.48\textwidth}
  \centering
  \begin{tikzpicture}[
    x=2.5cm,y=2.5cm,
    line cap=round,line join=round,
    baseline=(current bounding box.center)
  ]
    \useasboundingbox (-1.6,-1.6) rectangle (1.6,1.6);

    \tikzset{
      dual edge/.style={black, thick},
      dthree/.style={red, very thick},
    }

    \coordinate (K1) at (-1/3, 1/3);   
    \coordinate (K2) at (-1/3,-1/3);   
    \coordinate (K3) at ( 1/3,-1/3);   
    \coordinate (K4) at ( 1/3, 1/3);   
    \foreach \Q in {K1,K2,K3,K4}{\fill[black] (\Q) circle (1.2pt);}

    \draw[dual edge] (K1) -- (K2);
    \draw[dual edge] (K2) -- (K3);
    \draw[dual edge] (K3) -- (K4);
    \draw[dual edge] (K4) -- (K1);

    \draw[dual edge] (K1) -- ++(-0.40, 0.40) coordinate (LegTopLeft);
    \draw[dual edge] (K2) -- ++(-0.40,-0.40) coordinate (LegBotLeft);
    \draw[dual edge] (K3) -- ++( 0.40,-0.40);
    \draw[dual edge] (K4) -- ++( 0.40, 0.40);


    \coordinate (P1) at ($(K2)!0.60!(K1)$);
    \draw[dthree] (P1) -- ++(-0.30,-0.15);
    \node at ($(P1)+(-0.45,-0.18)$) {$\text{(II)}$};

    \coordinate (P2) at ($(K1)!0.5!(LegTopLeft)$);
    \draw[dthree] (P2) -- ++(-0.30,-0.15);
    \node at ($(P2)+(-0.5,-0.15)$) {$\text{(III)}$};

    \coordinate (P3) at ($(K2)!0.5!(LegBotLeft)$);
    \draw[dthree] (P3) -- ++(-0.30, 0.10);
    \node at ($(P3)+(-0.44,0.13)$) {$\text{(I)}$};

    \coordinate (P4) at ($(K2)!0.5!(K3)$);
    \draw[dthree] (P4) -- ++(-0.20,-0.15);
    \node at ($(P4)+(-0.2,-0.28)$) {$\text{(IV)}$};

  \end{tikzpicture}
  \caption{}
\end{subfigure}
		
		\caption{
			(a) Toric diagram of local $\BP^1 \times \BP^1$ at $f=-1$, with a crepant resolution given by the triangulation, and its dual graph; (b) The corresponding $(p,q)$-web of fivebranes and the D3-branes dual to the AV branes. The cases (I), (II), (III) give a $Q$-observable, while the case (IV) provides an $H$-observable.
		} \label{fig:toricp1p1}
	\end{figure}
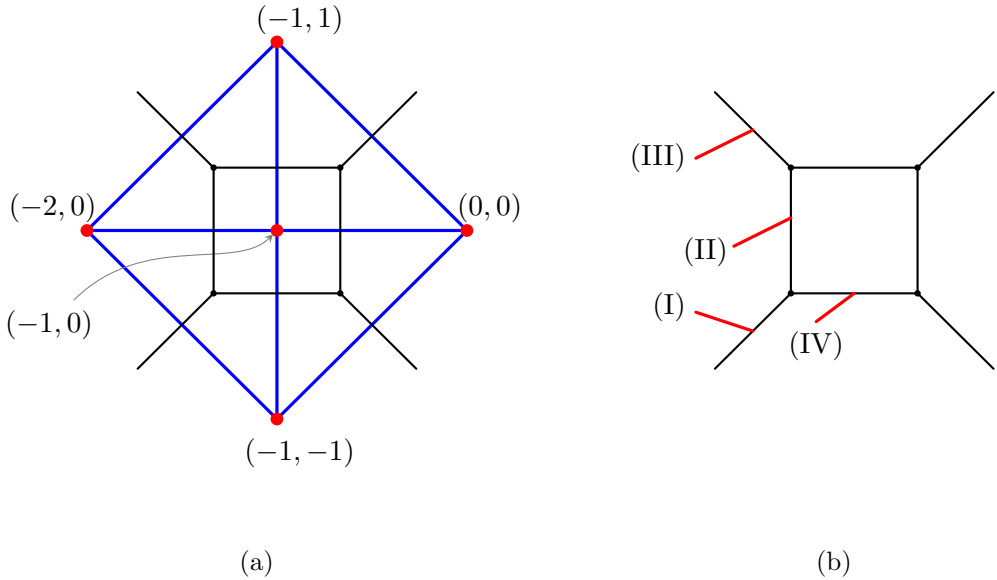

	\subsubsection{3d \texorpdfstring{$\mathrm{U}(1)$}{U(1)} gauge theory with two chiral multiplets coupled to 5d pure \texorpdfstring{$\mathrm{U}(2)$}{U(2)} gauge theory}

	Now, we insert an Aganagic-Vafa brane and study its refined open BPS invariants. On a single D3-brane, the 3d $\CalN=2$ $\U(1)$ gauge theory with two chiral multiplets with charge $+1$ is realized. This 3d $\CalN=2$ gauge theory couples to the 5d $\CalN=1$ pure $\U(2)$ gauge theory, through gauging the 3d  flavor symmetry by the 5d gauge field restricted to the locus of the 3d theory.\\
	
	The complex scalar $X$ is turned on for the $Q$-observable (the cases (I), (II), and (III) in Figure \ref{fig:toricp1p1}), which corresponds to the K\"{a}hler modulus for the relative cycle in $H_2 (\CalX,\CalL;\BZ)$. The $Q$-observable is written as
	\begin{align}
		Q_f (X)_{\underline{\l}} = \th(X^{-1} ;q_1) \prod_{\a=1,2} \left( \frac{a_\a q_1 ^{l(\l^{(\a)})}}{X};q_1 \right)_\infty \prod_{i=1} ^{l(\l^{(\a)})} \left( 1- \frac{a_\a q_1 ^{i-1} q_2 ^{\l_i ^{(\a)}}}{X} \right).
	\end{align}
	The observable only depends on the ratios of the Coulomb parameters $a_\a$ and the complex scalar $X$. We choose the normalization so that $a_1 =1$ and $a_2 =a$. The vacuum expectation value at the vacuum specified by the Coulomb parameter $a = a_2/a_1$ computed as
	\begin{align} \label{eq:coulombvvv}
		\CalZ_{\text{open}}(a,\qe,X;q_1,q_2) = \left\langle Q_f (X)\right\rangle = \frac{ \sum_{\{\underline{\l}\}} \qe^{\vert \underline{\l} \vert} \text{PE}\left[ \text{Ch}_{\mathsf{T}} \left( T_{\underline{\l}} \CalM_{2,k}\right)  \right] Q_f (X)_{\underline{\l}}}{ \sum_{\{\underline{\l}\}} \qe^{\vert \underline{\l} \vert} \text{PE}\left[ \text{Ch}_{\mathsf{T}} \left( T_{\underline{\l}} \CalM_{2,k}\right) \right]}.
	\end{align}
    By construction, this is a series in gauge coupling defined in the domain $\vert \qe \vert <1$. Each coefficient of the series is an exact rational function in $X$. Given a K\"{a}hler modulus chamber, we further expand the vev there to get the generating function of refined open BPS invariants. In Appendix \ref{sec:f0tables}, we present these invariants at $f=-1$ in the chambers (I) $\vert X \vert < \vert a \vert $; (II) $\vert a\vert < \vert X \vert <1$; and (III) $ 1 < \vert X \vert$.
	
	The quantum mirror curve equation is satisfied by the $q_2=1$ limit of this vev. To obtain this $q_1$-difference equation, first we compute coefficient of the $qq$-character \eqref{T} as 
	\begin{equation}
		\begin{split}
			&T(X)   = a q_1^{-2} q_2 ^{-2} X^{-2} +t(a,\qe;q_1,q_2) X^{-1} +1,\\ &t(a,\qe;q_1,q_2) :=  -\frac{1+a}{{q_1} {q_2}} + \frac{{q_1} {q_2}  a (1+a)}{({q_1} {q_2} -a) ({q_1} {q_2} a-1)} \mathfrak{q} + \mathcal O(\mathfrak q^2) ,
		\end{split}
        \label{eq:qqF0}
	\end{equation}
	where $t(a,\qe;q_1,q_2)$ cannot be written in a closed form, but only given as a series in $\qe$ whose coefficients can be determined up to arbitrary order.
	
	The quantum mirror curve equation is then given by
	\begin{equation} \label{eq:qcurvep1p1}
		0=\left[(-q_1X)^f q_1^{D_X} - a q_1 ^{-2} X^{-2} -t(a,\qe;q_1,1) X^{-1} -1 +  (-X)^{-f-2}  \qe a q_1^{-D_X} \right] \left\langle Q_f(X) \right\rangle_{q_2=1}.
	\end{equation}
	In the limit $q_1\to 1$, this reduces to the mirror curve of the local $\BP^1 \times \BP^1$,
	\begin{align}\label{eq:mcurvep1p1}
		0 = (-X)^f Z - a X^{-2} - t(a,\qe;1,1) X^{-1} -1 +(-X)^{-f-2}  \qe a Z^{-1} 
	\end{align}
	At $f=-1$, this indeed recovers the Newton polynomial equation associated to the toric diagram in Figure \ref{fig:toricp1p1}. All the other cases with $f\neq -1$ can be obtained by applying $T$-transformations. \\
	
	By turning on the complexified FI parameter $Z$ instead, we obtain the $H$-observable where $Z$ corresponds to the K\"{a}hler modulus associated with the relative cycle in $H_2 (\CalX,\CalL;\BZ)$ (the case (IV) in Figure \ref{fig:toricp1p1}). The $H$-observable is also obtained by applying \eqref{eq:hobs} as
	\begin{align} \label{eq:hobsp1p1}
		\begin{split}
			H_f (Z)_{\underline{\l}} &= \sum_{X\in q_1 ^\BZ} \exp\left[-\frac{\log X \log \left(Z (-1)^f (Xq_1^{-1})^{\frac{f}{2}}\right)}{\log q_1}\right] Q_0 (Xq_1^{-1})_{\underline{\l}} \\
			&= (aq_1;q_1)_\infty \sum_{n=0}^\infty \frac{\left( Z(-1)^f \right)^{n-l(\l^{(1)})} q_1^{-\frac{f}{2} \left( n-l(\l^{(1)}) \right) \left( n-l(\l^{(1)}) +1 \right)} }{(q_1;q_1)_n (a q_1 ;q_1 )_{n+l(\l^{(2)}) - l(\l^{(1)})  }}  \\
			&\qquad\qquad\qquad  \times \prod_{i=1} ^{l(\l^{(1)})} \left( 1- q_1 ^{n+i-l(\l^{(1)})} q_2 ^{\l_i ^{(1)}} \right) \prod_{i=1} ^{l(\l^{(2)})} \left( 1- a q_1 ^{n+i-l(\l^{(1)})} q_2 ^{\l_i ^{(2)}} \right).
		\end{split}
	\end{align}
	When the decoupling limit $\qe \to 0$ of the 5d $\CalN=1$ gauge theory is taken, the only contribution comes from the zero-instanton sector $\underline{\l}=\varnothing$, at which the $H$-observable reduces to the K-theoretic vortex partition function \eqref{eq:hfa1} of the 3d theory. From the geometric perspective, the 5d gauge coupling is identified with the K\"{a}hler parameter of the fiber $\BP^1$ as $Q_f = \qe a$ so that this limit corresponds to degenerating local $\BP^1 \times \BP^1$ to the resolved $A_1$-singularity (compare the toric diagrams and their dual graphs in Figure \ref{fig:toricp1p1} and \ref{fig:a1singtoric1}). With the 5d gauge coupling turned on, $\qe \neq 0$, the vortex number is bounded below by $-k$ where $k\geq 0$ is the bulk instanton number. Thus, both vortex and anti-vortex configurations on a fixed bulk instanton configuration contribute to the above $H$-observable.
	
	Its vacuum expectation value is identified with the generating function of refined open BPS invariants of the Aganagic-Vafa brane,
	\begin{align} \label{eq:higgsss}
		\CalZ_{\text{open}}(a,\qe,Z;q_1,q_2) = \left\langle H_f (Z)\right\rangle = \frac{ \sum_{\{\underline{\l}\}} \qe^{\vert \underline{\l} \vert} \text{PE}\left[ \text{Ch}_{\mathsf{T}} \left( T_{\underline{\l}} \CalM_{2,k}\right)  \right] H_f (Z)_{\underline{\l}}}{ \sum_{\{\underline{\l}\}} \qe^{\vert \underline{\l} \vert} \text{PE}\left[ \text{Ch}_{\mathsf{T}} \left( T_{\underline{\l}} \CalM_{2,k}\right) \right]}.
	\end{align}
    Note that this is a double series defined in the chamber $\vert \qe \vert<\vert Z \vert <1$. These refined open BPS invariants at $f=-1$ are presented in Appendix \ref{sec:f0tables}. 
    
	The quantum mirror curve equation satisfied by the $q_2 \to 1$ limit of the vev is also immediate from \eqref{eq:qcurvep1p1} and \eqref{eq:hobsp1p1}. Namely, we get
	\begin{align}
		0 = \left[ (-q_1 ^{-D_Z})^f Z - a q_1 ^{2D_Z} -t(a,\qe;q_1,1) q_1 ^{1+D_Z} -1 + (-q_1 ^{-1-D_Z})^{-f-2}  \qe a Z^{-1} \right] \left\langle H_f(Z) \right\rangle_{q_2=1}.
	\end{align}
	The $q_1 \to 1$ limit of this $q_1$-difference equation recovers the mirror curve \eqref{eq:mcurvep1p1} by construction.
	
	\subsubsection{Mapping Aganagic-Vafa branes by \texorpdfstring{$S$}{S}-transformation}

	The $S$-transformation maps the toric diagram in Figure \ref{fig:toricp1p1} to itself. It exchanges the toric legs corresponding to the base $\BP^1$ and the fiber $\BP^1$. Thus, the Aganagic-Vafa branes ending on those legs are swapped under the transformation. We are led to the expectation that the $Q$ generating function in the chamber (II) and $H$ generating function in the chamber (IV) must be exchanged with each other.
	
	Indeed, by explicit computations of the series in these chambers, we confirm the following identity between the two:
	\begin{align}
		\left.\left\langle Q_{f=-1} ^{\eqref{eq:coulombvvv}} \right\rangle \left(a,\qe, X ;q_1,q_2 \right)\right\vert_{ \substack{ X \to \sqrt{a}X }} =  \left.\left\langle H_{f=-1} ^{\eqref{eq:higgsss}} \right\rangle (a,\qe,Z;q_1,q_2) \right\vert_{\substack{ Z \to q_1(\sqrt{a}  X )^{-1}  \\  a \to \qe a q_1 q_2 \\ \qe \to \qe^{-1} \\ q_1 \to q_1 ^{-1}, \; q_2 \to q_2 ^{-1} }}.
	\end{align}
    We checked this matching up to the order of $\CalO (\qe^4 a^3)$ and all order in $X$. The combinatorial identities involved in the matching of coefficients are very nontrivial, as explicitly seen from the respective observable expressions. Note in particular that neither the left nor the right hand side has trivial $q_2$-dependence, unlike the previous examples. Such a surprising match strongly supports our claim that the 3d-5d partition functions provide the generating functions of refined open BPS invariants.
	
	\subsection{Local \texorpdfstring{$\mathbb{F}_1$}{F1}}

	Let us turn to the next-to-simplest Hirzebr\"uch surface, the local $F_1$, i.e., $\CalX = \text{Tot}(K_{F_1})$. The 5d $\CalN=1$ theory engineered by the associated $(p,q)$-web of fivebranes is the pure $\U(2)$ gauge theory with Chern-Simons level $k_{\text{CS}}=1$. The partition function specializes to
	\begin{align} \label{eq:partf1}
		\begin{split}
			\mathcal{Z}_{\text{closed}}(a,\qe;q_1,q_2) &= \CalZ^{\text{1-loop}} (a;q_1,q_2) \sum_{k=0} ^\infty \qe^{k} \chi_{\mathsf{T}} \left( \CalM_{2,k} ; L \right) \\
			&= \CalZ^{\text{1-loop}} (a;q_1,q_2) \sum_{\{\underline{\l}\}} \qe^{\vert \underline{\l} \vert} \left( \prod_{\Box \in \underline{\l}} \chi_\Box \right) \text{PE}\left[ \text{Ch}_{\mathsf{T}} \left( T_{\underline{\l}} \CalM_{2,k}\right) \right],
		\end{split}
	\end{align}
	where the 1-loop part is still coming from the W-bosons as in \eqref{eq:1loop}. The symmetry of exchanging base and fiber is now lost; accordingly, the partition function is no longer symmetric under the exchange of the associated K\"{a}hler parameters $Q_b = a$ and $Q_f = \qe a$.
	
	\begin{figure}[h!]
		\centering
		
		\begin{subfigure}[b]{0.45\textwidth}
			\centering
			\begin{tikzpicture}[
				x=1.8cm,y=1.8cm,   
				line cap=round,line join=round,
				baseline={(0,0)}
				]
				\useasboundingbox (-1.6,-1.6) rectangle (1.6,1.6);
				
				\begin{scope}[shift={(1, 0.4)}]
					
					\tikzset{
						toric vertex/.style={circle, fill=red, inner sep=1.7pt},
						toric edge/.style={blue, very thick},
						dual edge/.style={black, thick},
					}
					
					\coordinate (A) at (0,0);
					\coordinate (P) at (-1,0); 
					\coordinate (D) at (-2,0);
					\coordinate (B) at (-1,1);
					\coordinate (E) at (0,-1);
					
					\draw[toric edge] (D) -- (B) -- (A) -- (E) -- cycle;
					\draw[toric edge] (P) -- (A);
					\draw[toric edge] (P) -- (B);
					\draw[toric edge] (P) -- (D);
					\draw[toric edge] (P) -- (E);
					
					\foreach \Q in {A,P,D,B,E}{
						\node[toric vertex] at (\Q) {};
					}
					
					\node at ( 0.25, 0.15) {$(0,0)$};
                    \node (center) at ( -2.2, -1) {$(-1,0)$};
                    \draw[-stealth, black!50]
                    (center)
                    .. controls +(.3,0.75) and +(-0.20,-0.50) ..
                    (-1.02,-.04);
					\node at (-2.4, 0.12) {$(-2,0)$};
					\node at (-1.22, 1.2) {$(-1,1)$};
					\node at (0.1,-1.18) {$(0,-1)$};
					
					
					\coordinate (v1) at (-0.6,  0.4); 
					\coordinate (v2) at (-0.6, -0.4); 
					\coordinate (v4) at (-1.4,  0.4); 
					\coordinate (v3) at (-1.4, -1.2); 
					
					\foreach \V in {v1,v2,v3,v4}{\fill[black] (\V) circle (1.2pt);}
					
					\draw[dual edge] (v1) -- (v2) -- (v3) -- (v4) -- cycle;
					
					\draw[dual edge] (v1) -- ++( 0.45, 0.45);
					\draw[dual edge] (v2) -- ++( 0.60, 0.00);
					\draw[dual edge] (v3) -- ++(-0.25,-0.50);
					\draw[dual edge] (v4) -- ++(-0.45, 0.45);
					
				\end{scope}
			\end{tikzpicture}
			\caption{}
		\end{subfigure}%
		\hspace{1em}%
	\begin{subfigure}[b]{0.45\textwidth}
  \centering
  \begin{tikzpicture}[
    x=1.8cm,y=1.8cm,
    scale=1.12,
    line cap=round,line join=round,
    baseline={(0,0)}
  ]
    \useasboundingbox (-1.6,-1.43) rectangle (1.43,1.43);

    \begin{scope}[shift={(-0.4, 1.0)}]

      \tikzset{
        dual edge/.style={black, thick},
        dthree/.style={red, very thick},
      }

      \coordinate (w1) at (-0.4, -0.6); 
      \coordinate (w2) at ( 0.4, -0.6); 
      \coordinate (w4) at (-0.4, -1.4); 
      \coordinate (w3) at ( 1.2, -1.4); 
      
      \foreach \V in {w1,w2,w3,w4}{\fill[black] (\V) circle (1.2pt);}

      \draw[dual edge] (w1) -- (w2) -- (w3) -- (w4) -- cycle;

      
      \draw[dual edge] (w1) -- ++(-0.45, 0.45) coordinate (LegTopLeft);
      
      \draw[dual edge] (w4) -- ++(-0.45,-0.45) coordinate (LegBotLeft);
      
      \draw[dual edge] (w2) -- ++( 0.00, 0.60);
      \draw[dual edge] (w3) -- ++( 0.50,-0.25);


      \coordinate (P_Vert) at ($(w4)!0.60!(w1)$);
      \draw[dthree] (P_Vert) -- ++(-0.30,-0.15);
      \node at ($(P_Vert)+(-0.5,-0.2)$) {$\text{(II)}$};

      \coordinate (P_TopLeft) at ($(w1)!0.5!(LegTopLeft)$);
      \draw[dthree] (P_TopLeft) -- ++(-0.30,-0.15);
      \node at ($(P_TopLeft)+(-0.55,-0.25)$) {$\text{(III)}$};

      \coordinate (P_BotLeft) at ($(w4)!0.5!(LegBotLeft)$);
      \draw[dthree] (P_BotLeft) -- ++(-0.30, 0.10);
      \node at ($(P_BotLeft)+(-0.45,0.1)$) {$\text{(I)}$};

      \coordinate (P_BotHoriz) at ($(w4)!0.5!(w3)$);
      \draw[dthree] (P_BotHoriz) -- ++(-0.30,-0.15);
      \node at ($(P_BotHoriz)+(-0.4,-0.3)$) {$\text{(IV)}$};

    \end{scope}
  \end{tikzpicture}
  \caption{}
\end{subfigure}
		
		\caption{
			(a) Toric diagram of local $F_1$ at $f=-1$, with a crepant resolution chosen by the triangulation, and its dual graph. (b) The corresponding $(p,q)$-web with D3-branes dual to AV branes. The cases (I), (II), (III) correspond to a $Q$-observable, while the case (IV) gives an $H$-observable.
		} \label{fig:localf1}
	\end{figure}
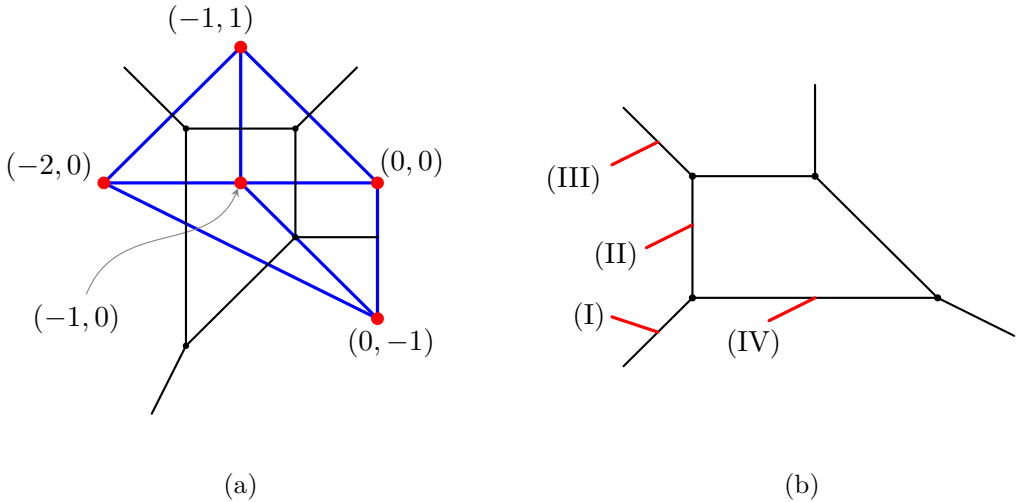
	
	\subsubsection{3d \texorpdfstring{$\mathrm{U}(1)$}{U(1)} gauge theory with two chiral multiplets coupled to 5d pure \texorpdfstring{$\mathrm{U}(2)$}{U(2)} gauge theory with \texorpdfstring{$k_{\text{CS}}=1$}{k\_CS=1}}

	Now, we insert an Aganagic-Vafa brane and study its refined open BPS invariants. The 3d $\CalN=2$ gauge theory realized on the D3-brane is still the $\U(1)$ gauge theory with two chiral multiplets with charge $+1$. The $Q$-observable and the $H$-observable assume exactly the same expression as in the case of the local $\BP^1 \times \BP^1$. Namely, we have
	\begin{align}
		Q_f (X)_{\underline{\l}} = \th(X^{-1} ;q_1) \prod_{\a=1,2} \left( \frac{a_\a q_1 ^{l(\l^{(\a)})}}{X};q_1 \right)_\infty \prod_{i=1} ^{l(\l^{(\a)})} \left( 1- \frac{a_\a q_1 ^{i-1} q_2 ^{\l_i ^{(\a)}}}{X} \right),
	\end{align}
	where we normalize $a_1 =1$ and $a_2=a$, and
	\begin{align}
		\begin{split}
			H_f (X)_{\underline{\l}} &= (aq_1;q_1)_\infty \sum_{n=0}^\infty \frac{\left( Z(-1)^f \right)^{n-l(\l^{(1)})} q_1^{-\frac{f}{2} \left( n-l(\l^{(1)}) \right) \left( n-l(\l^{(1)}) +1 \right)} }{(q_1;q_1)_n (a q_1 ;q_1 )_{n+l(\l^{(2)}) - l(\l^{(1)})  }}  \\
			&\qquad\qquad\qquad  \times \prod_{i=1} ^{l(\l^{(1)})} \left( 1- q_1 ^{n+i-l(\l^{(1)})} q_2 ^{\l_i ^{(1)}} \right) \prod_{i=1} ^{l(\l^{(2)})} \left( 1- a q_1 ^{n+i-l(\l^{(1)})} q_2 ^{\l_i ^{(2)}} \right).
		\end{split}
	\end{align}
	The only difference is that their vevs are computed with different measure due to the effect of the Chern-Simons level $k_{\text{CS}}=1$. Namely, the vevs are computed as
	\begin{subequations}
		\begin{align}
			&  \left\langle Q_f (X)\right\rangle =  \frac{ \sum_{\{\underline{\l}\}} \qe^{\vert \underline{\l} \vert} \left( \prod_{\Box \in \underline{\l}} \chi_\Box \right) \text{PE}\left[ \text{Ch}_{\mathsf{T}} \left( T_{\underline{\l}} \CalM_{2,k}\right)  \right ] Q_f (X)_{\underline{\l}}}{\sum_{\{\underline{\l}\}} \qe^{\vert \underline{\l} \vert} \left( \prod_{\Box \in \underline{\l}} \chi_\Box \right) \text{PE}\left[ \text{Ch}_{\mathsf{T}} \left( T_{\underline{\l}} \CalM_{2,k}\right) \right]}, \\
			& \left\langle H_f (X)\right\rangle= \frac{ \sum_{\{\underline{\l}\}} \qe^{\vert \underline{\l} \vert} \left( \prod_{\Box \in \underline{\l}} \chi_\Box \right) \text{PE}\left[ \text{Ch}_{\mathsf{T}} \left( T_{\underline{\l}} \CalM_{2,k}\right)  \right] H_f (X)_{\underline{\l}}}{\sum_{\{\underline{\l}\}} \qe^{\vert \underline{\l} \vert} \left( \prod_{\Box \in \underline{\l}} \chi_\Box \right) \text{PE}\left[ \text{Ch}_{\mathsf{T}} \left( T_{\underline{\l}} \CalM_{2,k}\right) \right]}.
		\end{align}        
	\end{subequations}
    The K\"{a}hler modulus associated with the relative cycle in $H_2 (\CalX,\CalL;\BZ)$ is identified with $X$ and $Z$, respectively. Note that the vev of the $Q$-observable is expanded in $\vert \qe \vert <1$ as a series in $\qe$, whose coefficients are exact rational functions in $X$. Expanding in a given chamber, we obtain the refined open BPS invariants in (I) $\vert X \vert < \vert a \vert$; (II) $\vert a \vert < \vert X \vert<1$; and (III) $\vert X\vert >1 $. The vev of the $H$-observable is a double series defined in the chamber $\vert \qe \vert < \vert Z \vert <1$, yielding the refined open BPS invariants there (the case (IV) in Figure \ref{fig:localf1}). Compared to the case of local $F_0$, the fiber-base duality is no longer present so that $Q$ and $H$ generating functions do not agree. We present some of these refined open BPS invariants at $f=-1$ in Appendix \ref{app:f1}.\\
	
	The $q_2 \to 1$ limit of these vevs satisfy the quantum mirror curve equations. For this, we first compute the coefficients of the $qq$-character as
	\begin{equation}
		\begin{split}
			&T(X) = a q_1^{-2} q_2 ^{-2} X^{-2} + t(a,\qe;q_1,q_2) X^{-1} +1,\\
			&t(a,\qe;q_1,q_2) = - \frac{1+a}{q_1 q_2} + \frac{(1+q_1 q_2) a^2}{(q_1q_2-a)(a q_1q_2-1)} \qe +\CalO(\qe^2),
		\end{split}
        \label{eq:qqf1}
	\end{equation}
	where $t(a,\qe;q_1,q_2)$ cannot be written in a closed form, but only given as a series in $\qe$ whose coefficients can be determined up to arbitrary order as exact rational functions in $(a,q_1,q_2)$.
	
	The quantum mirror curve equation is then obtained as
	\begin{subequations}
		\begin{align}
			& 0 = \left[(-q_1 X)^f q_1 ^{D_X} - a q_1 ^{-2} X^{-2} - t(a,\qe;q_1,1) X^{-1} -1 - (-X)^{-f-1} \qe a q_1 ^{-D_X} \right]\left\langle Q_f (X)\right\rangle_{q_2=1}, \label{eq:qcurvef1}\\
			& 0 = \left[ (-q_1^{-D_Z})^f Z - a q_1 ^{2D_Z} - t(a,\qe;q_1,1) q_1^{1+D_Z} -1 - (- q_1 ^{-1-D_Z})^{-f-1}  \qe a q_1 ^{-D_X} \right] \left\langle H_f (Z) \right\rangle _{q_2=1}.
		\end{align}
	\end{subequations}
	In the limit $q_1\to 1$, these equations reduce to the mirror curve of the local $F_1$,
	\begin{align}
		0 = (-X)^f Z - a X^{-2} - t(a,\qe;1,1) X^{-1} -1 -(-X)^{-f-1}  \qe a Z^{-1}.
	\end{align}
	Indeed, at $f=-1$ this recovers the Newton polynomial equation assocaited with the toric diagram in Figure \ref{fig:localf1}. All the other cases with $f\neq -1$ are obtained by applying $T$-transformations.
	
	\section{Discussion}\label{sec:discussion}

    One of the main achievements of this work is the computation of refined open BPS indices for the toric Calabi-Yau threefolds studied here. A rigorous mathematical definition of ``motivic open Donaldson-Thomas invariants'' is still lacking; nevertheless, our results provide nontrivial consistency checks of their expected structural properties, including integrality and positivity.

   The relations of the closed motivic DT invariants and 5d $\mathcal N=1$ BPS quivers have been already well understood, both from physical \cite{Manschot:2013sya, Manschot:2014fua, Beaujard:2020sgs, Mozgovoy:2020has, Closset:2018bjz} and mathematical points of view \cite{Kontsevich:2010px, Kontsevich:2008fj, reineke2005localizationquivermoduli}. These quivers were also studied from the perspective of mirror symmetry \cite{Banerjee:2024smk, Banerjee:2018syt, Hori:2000ck, Aganagic:2012au}. Given a toric Calabi-Yau threefold $\mathcal{X}$ and its mirror $\mathcal{X}^\vee$, these quivers have been constructed for both $\mathrm{Fuk}(\mathcal{X})$ and $D^b\mathrm{Coh} (\mathcal{X}^\vee)$ and the associated motivic DT invariants have been shown to reproduce refined DT invariants in several examples, e.g. \cite{Mozgovoy:2020has}. 
   
    It is therefore natural to expect the existence of an analogous quiver description encoding the 3d–5d BPS specta, from which motivic open DT invariants could be defined combinatorially. From a physical viewpoint, the appearance of such invariants is natural; however, a complete mathematical framework for their construction remains elusive. At present, quiver descriptions of 3d–5d BPS states are known only for limited classes of Calabi–Yau threefolds called strip geometries \cite{Panfil:2018faz, Cheng:2021nex,Banerjee:2025shz, Gupta:2024ics,hu2025skein,MR3366002}. Though even in these cases a satisfactory mathematical understanding of the refined invariants is lacking.  We hope that the computations presented here will provide further motivation for a systematic study of motivic open Donaldson–Thomas invariants.

    In the limit $q_2 \to 1 $, the $q_1$-difference equations studied in this paper have several mathematical significances which are yet to be investigated thoroughly. One of them is the connection to the cluster varieties and quantum Teichm\"uller theory. Some studies have been initiated in the works of \cite{MR4695875}. The procedure of nonabelianization in \cite{Banerjee:2018syt}, generalizing \cite{Gaiotto:2012rg}, provides a path lifting map, where the base is a $\mathbb{C}^\times$-plane. The moduli space of local systems on $\mathbb{C}^\times$ can be identified with the double Bruhat cells $\widehat{G}/\text{Ad}(\widehat{H})$, where $\widehat{G}$ is an affine Lie group determined by exponential networks on $\mathbb{C}^\times$ and $\widehat{H}$ is the Cartan. The cluster parametrization of the double Bruhat cells \cite{fomin1998doublebruhatcellstotal}, provides a direct connection between quantum Teichm\"uller theory and the difference equations.\footnote{We thank Valdo Tatitscheff to explain some aspects of his upcoming work with Raphael Senghaas and Johannes Walcher.} Indeed, one can view the exponential networks as defining the Stokes lines for the difference equations, some justifications for which were provided in \cite{Banerjee:2025shz} An alternative approach through WKB techniques was taken in \cite{DelMonte:2024dcr}, see also \cite{Kashani-Poor:2016edc, Grassi:2017qee,Grassi:2022zuk,Alim:2022oll}. Topological recursion is yet another method for solving difference equations \cite{Eynard:2012nj,Bouchard:2016obz}.\footnote{For the strip geometries, this was explicitly constructed in \cite{Banerjee:2025shz,hu2025skein}, through topological recursion. It was observed that the form of the quantum curve depends on the choice of the base point $X$ highlighting the aforementioned ambiguity in quantization.  } Given a classical curve, its quantization is not unique. There are in general multiple difference equations that reduce to the same curve in the classical limit. As far as we can tell, topological recursion generally produces a different quantization from the one given in terms of the TQ equation presented in this paper. It is also important to note that methods like WKB and topological recursion produce solutions that are perturbative in $\hbar$, whereas localization produces the solutions as exact functions in $\hbar=\varepsilon_1$. Our construction of the $q_1$-difference equation should be intimately connected to these approaches. Turning on  $q_2 \neq 1$, however remains highly mysterious. The results here in this regard should pave new directions for future research.

Finally, the study of BPS invariants admits a powerful algebraic recasting in terms of the quantum toroidal algebra and networks of its intertwiners \cite{Awata:2011ce}. A physical explanation of this picture was recently provided in \cite{Haouzi:2024qyo} within the framework of twisted M-theory \cite{Costello:2016nkh}, where the $(p,q)$-web of fivebranes is dualized to a network of holomorphic surface defects in the 5d holomorphic-topological non-commutative Chern-Simons theory. Moreover, motivated by
\cite{Gaiotto:2020dsq}, it was shown in \cite{Haouzi:2024qyo,Ishtiaque:2024orn} that the R-matrices intertwining certain representations can be identified with Miura operators for
$q$-deformed $W$- and $Y$-algebras. In a dual IIB frame, these R-matrices admit a brane realization in terms of D3-branes passing through the $(p,q)$-web. This perspective suggests that Aganagic-Vafa branes should admit a purely algebraic description, shedding light on the algebraic structures underlying open BPS invariants. See also \cite{Awata2016ExplicitEO,Bourgine:2017jsi,Bourgine:2019phm} for relevant studies.

	\appendix

	\section{Tables of refined open BPS invariants} \label{sec:table}
In this paper we proposed that we can compute the generating function \eqref{eq:openbps} of refined open BPS invariants associated with a pair $(\mathcal X, \mathcal L)$ consisting of a toric CY 3-fold and an Aganagic-Vafa (AV) Lagrangian brane as the expectation value of codim-2 defects $Q_f(X)$ \eqref{eq:qobs} and $H_f(Z)$ \eqref{eq:hobs}. The $H$-observables are related to the $Q$-observables of a (generically) different theory via combinations of $S$ and $T$-transforamtions (see Sec. \ref{sec:aut}, e.g. Fig. \ref{fig:QHS}). So, for computing invariants it suffices to focus on the $Q$-observables. Equating the normalized expectation value of $Q$ to the generating function we extract the invariants as
\begin{equation}\begin{gathered}
		\text{for charges} \quad r,s \in \frac{\mathbb Z}{2}, \quad \nu \in H_2(\mathcal X, \mathcal L; \mathbb Z) \backslash \{0\},
		\\
		N^{r,s}_\nu = (-1)^{2r} [q_1^r q_2^{s+\frac{1}{2}} z^\nu] \left((q_1^{1/2}-q_1^{-1/2}) \mathrm{PL}\left[\langle Q_f\rangle\right](z,q_1,q_2)\right).
\end{gathered}\end{equation}
Here $\mathrm{PL}(f)(z,\cdots)$ refers to the plethystic log of a function $f$ evaluated at $z,\cdots$ and for any monomial $z^\nu\cdots$ in any number of variables $[z^\nu\cdots](F)$ refers to the coefficient of $z^\nu\cdots$ in the expression $F$.

In the rest of this section we shall tabulate these integers for various local Calabi-Yaus. For brevity, we shall package the integers corresponding to different powers of $q_1$ and $q_2$ at a fixed power of the K\"ahler parameters into a single expression. Furthermore, since $q_1$ and $q_2$ can appear with half integer powers, we introduce the variables
\begin{equation}
	\lambda := q_1^{1/2}, \qquad \mu := q_2^{1/2}
\end{equation}
and write the expressions in terms of these variables. In particular, for a pair $(\mathcal X, \mathcal L)$ we define:
\begin{equation}
	\text{for } \nu \in H_2(\mathcal X, \mathcal L;\mathbb Z): \qquad \Lambda_\nu(\lambda, \mu) := \sum_{r,s \in \frac{\mathbb Z}{2}} N^{r,s}_\nu \lambda^{2r} \mu^{2s+1}.
\end{equation}
Note importantly that for any local Calabi-Yau and exponents $\nu$, the above expression is a Laurent polynomial with integer coefficients:
\begin{equation}
	\Lambda_\nu(\lambda, \mu) \in \mathbb Z[\lambda^\pm, \mu^\pm].
\end{equation}

There is in fact more constraint on the sign of the invariants. From the definition \eqref{eq:openbps} of their generating function as a supersymmetric index, we expect $N^{r,s}_\nu$ to be the dimensions of subspaces of the BPS sector of the theory with definite charges, in particular we expect them to be positive. We observe this in the computed invariants as follows. After choosing a $\mathbb Z$-basis of $H_2(\mathcal X, \mathcal L;\mathbb Z)$ corresponding to the appropriate K\"ahler parameters, let $X$ be the relative K\"ahler parameter corresponding to the location of the AV brane on a toric leg. For any $\nu \in H_2(\mathcal X, \mathcal L;\mathbb Z)$ let $\xi(\nu) \in \mathbb Z$ be the coefficient of the basis vector corresponding to $X$, so that we can write $z^\nu = X^{\xi(\nu)}\cdots$. Gauge theory provides the generating function \eqref{eq:openbps} as a power series in the 5d gauge coupling with coefficient being rational functions of $X$ (and other K\"ahler parameters) and we can expand it in different regions of the parameter space as long as the gauge coupling is small. A small $X$ expansion will contain only positive powers of $X$ and a large $X$ expansion only negative ones. More generally, there can be regions with mixed powers of $X$ (for geometries with internal vertical edges in their toric diagrams). We observe that
\begin{equation}
    \begin{array}{rl}
        \Lambda_\nu(\lambda, \mu) \in \mathbb N[\lambda^\pm, \mu^\pm] & \text{for } \xi(\nu) \ge 0,
        \\
        -\Lambda_\nu(\lambda, \mu) \in \mathbb N[\lambda^\pm, \mu^\pm] &  \text{for } \xi(\nu) < 0.
    \end{array}
\end{equation}
The exponent of $X$ is related to the oriented boundary 1-cycle $\partial \nu \in H_1(\mathcal L)$, i.e., to the winding of M2 branes/strings ending on $\mathcal L$. It is therefore expected that inverting $X$ corresponds to counting BPS states with the opposite sign, leading to the sign flip. This aspect of the refined invariants was also observed in \cite{Cheng:2021nex} in case of strip geometries.

In most of the examples we shall choose two out of the available K\"ahler parameters, one of them being $X$, labeling the position of the AV brane on a toric leg, and the other $\mathfrak q$, the 5d gauge coupling. Then for some set of exponents for the remaining K\"ahler parameters we shall make a table whose columns and rows will refer to powers of $\mathfrak q$ and $X$ respectively, and the corresponding entries will be the Laurent polynomials $\Lambda_\nu(\lambda, \mu)$. For example, suppose in an example we have the K\"ahler parameters $\mathfrak q$, $X$, and $a$. Then for various powers $a^k$, we shall make such as:\footnote{We are using the fact that $H_2(\mathcal X, \mathcal L;\mathbb Z)$ is torsion-free and we can choose a $\mathbb Z$-basis.}
{\setlength{\tabcolsep}{2pt} 
	\renewcommand{\arraystretch}{1.1}
	\begin{longtable}{>{\centering\arraybackslash}m{.5cm}||
			>{\centering\arraybackslash}m{1cm}|
			>{\centering\arraybackslash}m{3cm}|
			>{\centering\arraybackslash}m{1cm}|
		}
		\hline
        & \multicolumn{3}{c|}{\(a^k\)}
        \\\hline
		& \(\cdots\) & \(\mathfrak q^j\) & \(\cdots\) 
		\\\toprule
		\endfirsthead
		
		& \(\cdots\) & \(\mathfrak q^i\) & \(\cdots\) 
		\\\midrule
		\endhead
		
		\midrule
		\endfoot
		
		\toprule
		\endlastfoot
		
		\(\vdots\) & \(\ddots\) & \(\ddots\) & \(\ddots\) 
		\\\hline
		\(X^i\) & \(\cdots\) & \(\Lambda_{i,j,k}(\lambda, \mu)\) & \(\cdots\)
		\\\hline
		\(\vdots\) & \(\ddots\) & \(\ddots\) & \(\ddots\)

	\end{longtable}
}

There is a priori some ambiguity in these invariants stemming from the freedom to choose a basis of $H_2$ which corresponds to mixing the K\"ahler parameters, and even from mixing the K\"ahler parameters with $\lambda$ and $\mu$. The presentation of the mirror curve \eqref{eq:mirrorCurve} fixes the choice of the K\"ahler parameters. The quantum mirror curve, i.e., the TQ equation \eqref{eq:tqeq}, fixes any ambiguity about mixing the K\"ahler parameters with $\lambda$. The freedom to mix with $\mu$ can be fixed by the expression for the $qq$-character \eqref{T} in terms of the $Q$-observables. We shall therefore accompany our tables with the corresponding mirror curve and a perturbative expression for the $qq$-character.

\subsection{Resolved conifold} \label{app:conifold}
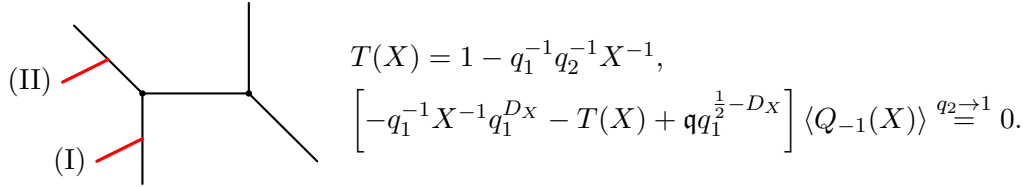
\begin{figure}[H]
	\centering
	\centering
	\renewcommand{\arraystretch}{1.5} 
	\begin{tabular}{c l}
		\begin{tikzpicture}[
			x=2cm,y=2cm,
			line cap=round,line join=round,
			baseline=(current bounding box.center)
			]
			\tikzset{
				dual edge/.style={black, thick},
				dthree/.style={red, very thick},
			}
			
			\coordinate (K1) at (-0.35, 0);
			\coordinate (K2) at ( 0.35, 0);
			\fill[black] (K1) circle (1.2pt);
			\fill[black] (K2) circle (1.2pt);
			
			\draw[dual edge] (K1) -- (K2);

			\coordinate (JSW) at ($(K1)+(0.00,-0.60)$);
			\draw[dual edge] (K1) -- (JSW);
			\coordinate (JNW) at ($(K1)+(-0.45, 0.45)$);
			\draw[dual edge] (K1) -- (JNW);
			
			\draw[dual edge] (K2) -- ++( 0.00, 0.60);
			\coordinate (JSE) at ($(K2)+(0.45,-0.45)$);
			\draw[dual edge] (K2) -- (JSE);
			
			\coordinate (P3) at ($(K1)!0.50!(JNW)$);
			\coordinate (P4) at ($(P3)+(-0.30,-0.15)$);
			\draw[dthree] (P3) -- (P4);
			\node[left=0cm of P4] () {(II)};
			
			\coordinate (Q3) at ($(K1)!0.50!(JSW)$);
			\coordinate (Q4) at ($(Q3)+(-0.30,-0.15)$);
			\draw[dthree] (Q3) -- (Q4);
			\node[left=0cm of Q4] () {(I)};
		\end{tikzpicture}
		&
		\begin{tabular}{@{}l@{}}
			\(T(X) = 1 - q_1^{-1} q_2^{-1}X^{-1} \),\\
			\(\left[-q_1^{-1} X^{-1} q_1^{D_X} - T(X) + \mathfrak q q_1^{\frac{1}{2}-D_X}\right]
			\langle Q_{-1}(X) \rangle \overset{q_2\to 1}{=} 0\).
		\end{tabular}
	\end{tabular}
	\caption{Toric diagram for the resolved conifold, the same diagram from Fig. \ref{fig:resconifold2}. The D3 brane inserts the $Q$-observable in the engineered 5d $\mathcal N=1$ $\mathrm{U}(1)$ gauge theory. The D3 brane in position I and II correspond to the small $X$ and large $X$ expansions of the resulting expectation value respectively. To the right are the corresponding $qq$-character \eqref{eq:qqCon} and the quantum mirror curve \eqref{eq:qMirrorCon} with $f=-1$ and $\mathfrak q \to \sqrt{q_1} \mathfrak q$.}
\end{figure}

Below we tabulate the BPS invariants extracted from $\langle Q_{-1}(X) \rangle$. In this theory there are only two K\"ahler parameters, the 5d gauge coupling $\mathfrak q$ and the defect moduli $X$. The defect partition function $\langle Q_{-1}(X) \rangle$ is analytic in $X$ and we extract two sets of invariants by expanding it in two domains, $|X| < 1$ and $|X| > 1$.

\paragraph{Case I: $|X| < 1$.}
{\setlength{\tabcolsep}{2pt} 
	\renewcommand{\arraystretch}{1.1}
	\begin{longtable}{>{\centering\arraybackslash}m{.5cm}||
			>{\centering\arraybackslash}m{.5cm}|
			>{\centering\arraybackslash}m{1cm}|
			>{\centering\arraybackslash}m{3cm}|
			>{\centering\arraybackslash}m{3.5cm}|
			>{\centering\arraybackslash}m{5cm}|
		}
		\hline
		& \(\mathfrak q^0\) & \(\mathfrak q^1\) & \(\mathfrak q^2\) & \(\mathfrak q^3\) & \(\mathfrak q^4\)
		\\\toprule
		\endfirsthead
		
		& \(\mathfrak q^0\) & \(\mathfrak q^1\) & \(\mathfrak q^2\) & \(\mathfrak q^3\) & \(\mathfrak q^4\)
		\\\midrule
		\endhead
		
		\midrule
		\endfoot
		
		\toprule
		\endlastfoot
		\(X^0\) & \( 0 \) & \( \lambda ^2 \mu ^2 \) & \( 0 \) & \( 0 \) & \( 0 \)
		\\\hline
		\(X^1\) & \( \lambda  \) & \( \lambda ^2 \mu ^2 \) & \( 0 \) & \( 0 \) & \( 0 \) 
		\\\hline
		\(X^2\) & \( 0 \) & \( \lambda ^2 \mu ^2 \) & \( \lambda ^3 \mu ^4 \) & \( 0 \) & \( 0 \) 
		\\\hline
		\(X^3\) & \( 0 \) & \( \lambda ^2 \mu ^2 \) & \( \lambda ^3 \mu ^4+\lambda  \mu ^4 \) & \( \lambda ^2 \mu ^6 \) & \( 0 \)
		\\\hline
		\(X^4\) & \( 0 \) & \( \lambda ^2 \mu ^2 \) & \( 2 \lambda ^3 \mu ^4+\lambda  \mu ^4+\frac{\mu ^4}{\lambda } \) & \( \lambda ^4 \mu ^6+2 \lambda ^2 \mu ^6+\frac{\mu ^6}{\lambda ^2}+\mu ^6 \) & \( \lambda ^3 \mu ^8+\frac{\mu ^8}{\lambda } \) 
		\\\hline
		\(X^5\) & \( 0 \) & \( \lambda ^2 \mu ^2 \) & \( 2 \lambda ^3 \mu ^4+\frac{\mu ^4}{\lambda ^3}+2 \lambda  \mu ^4+\frac{\mu ^4}{\lambda } \) & \( \frac{\mu ^6}{\lambda ^6}+2 \lambda ^4 \mu ^6+\frac{\mu ^6}{\lambda ^4}+4 \lambda ^2 \mu ^6+\frac{3 \mu ^6}{\lambda ^2}+3 \mu ^6 \) & \( \frac{\mu ^8}{\lambda ^7}+\lambda ^5 \mu ^8+\frac{\mu ^8}{\lambda ^5}+3 \lambda ^3 \mu ^8+\frac{2 \mu ^8}{\lambda ^3}+3 \lambda  \mu ^8+\frac{3 \mu ^8}{\lambda } \) 
		\\\hline
		\(X^6\) & \( 0 \) & \( \lambda ^2 \mu ^2 \) & \( \frac{\mu ^4}{\lambda ^5}+3 \lambda ^3 \mu ^4+\frac{\mu ^4}{\lambda ^3}+2 \lambda  \mu ^4+\frac{2 \mu ^4}{\lambda } \) & \( \frac{\mu ^6}{\lambda ^{10}}+\frac{\mu ^6}{\lambda ^8}+\frac{3 \mu ^6}{\lambda ^6}+3 \lambda ^4 \mu ^6+\frac{4 \mu ^6}{\lambda ^4}+7 \lambda ^2 \mu ^6+\frac{6 \mu ^6}{\lambda ^2}+6 \mu ^6 \) & \( \frac{\mu ^8}{\lambda ^{13}}+\frac{\mu ^8}{\lambda ^{11}}+\frac{3 \mu ^8}{\lambda ^9}+\frac{4 \mu ^8}{\lambda ^7}+2 \lambda ^5 \mu ^8+\frac{7 \mu ^8}{\lambda ^5}+8 \lambda ^3 \mu ^8+\frac{7 \mu ^8}{\lambda ^3}+8 \lambda  \mu ^8+\frac{11 \mu ^8}{\lambda } \) 
		\\\hline
		\(X^7\) & \( 0 \) & \( \lambda ^2 \mu ^2 \) & \( \frac{\mu ^4}{\lambda ^7}+\frac{\mu ^4}{\lambda ^5}+3 \lambda ^3 \mu ^4+\frac{2 \mu ^4}{\lambda ^3}+3 \lambda  \mu ^4+\frac{2 \mu ^4}{\lambda } \) & \( \frac{\mu ^6}{\lambda ^{14}}+\frac{\mu ^6}{\lambda ^{12}}+\frac{3 \mu ^6}{\lambda ^{10}}+\frac{4 \mu ^6}{\lambda ^8}+\frac{7 \mu ^6}{\lambda ^6}+5 \lambda ^4 \mu ^6+\frac{8 \mu ^6}{\lambda ^4}+10 \lambda ^2 \mu ^6+\frac{11 \mu ^6}{\lambda ^2}+10 \mu ^6 \) & \( \frac{\mu ^8}{\lambda ^{19}}+\frac{\mu ^8}{\lambda ^{17}}+\frac{3 \mu ^8}{\lambda ^{15}}+\frac{5 \mu ^8}{\lambda ^{13}}+\frac{8 \mu ^8}{\lambda ^{11}}+\frac{11 \mu ^8}{\lambda ^9}+\frac{16 \mu ^8}{\lambda ^7}+5 \lambda ^5 \mu ^8+\frac{19 \mu ^8}{\lambda ^5}+15 \lambda ^3 \mu ^8+\frac{22 \mu ^8}{\lambda ^3}+20 \lambda  \mu ^8+\frac{24 \mu ^8}{\lambda } \) 
	\end{longtable}
}

{\setlength{\tabcolsep}{2pt} 
	\renewcommand{\arraystretch}{1.1}
	\begin{longtable}{>{\centering\arraybackslash}m{.5cm}||
			>{\centering\arraybackslash}m{4.5cm}|
			>{\centering\arraybackslash}m{4.5cm}|
			>{\centering\arraybackslash}m{4.4cm}|
		}
		\hline
		& \(\mathfrak q^5\) & \(\mathfrak q^6\) & \(\mathfrak q^7\)
		\\\toprule
		\endfirsthead
		
		& \(\mathfrak q^5\) & \(\mathfrak q^6\) & \(\mathfrak q^7\)
		\\\midrule
		\endhead
		
		\midrule
		\endfoot
		
		\toprule
		\endlastfoot
		\(X^5\) & \( \frac{\mu ^{10}}{\lambda ^6}+\lambda ^4 \mu ^{10}+\lambda ^2 \mu ^{10}+\frac{\mu ^{10}}{\lambda ^2}+\mu ^{10} \) & \( 0 \) & \( 0 \) 
		\\\hline
		\(X^6\) & \( \frac{\mu ^{10}}{\lambda ^{14}}+\frac{\mu ^{10}}{\lambda ^{12}}+\frac{2 \mu ^{10}}{\lambda ^{10}}+\frac{3 \mu ^{10}}{\lambda ^8}+\lambda ^6 \mu ^{10}+\frac{5 \mu ^{10}}{\lambda ^6}+4 \lambda ^4 \mu ^{10}+\frac{5 \mu ^{10}}{\lambda ^4}+6 \lambda ^2 \mu ^{10}+\frac{7 \mu ^{10}}{\lambda ^2}+7 \mu ^{10} \) & \( \frac{\mu ^{12}}{\lambda ^{13}}+\frac{\mu ^{12}}{\lambda ^9}+\frac{\mu ^{12}}{\lambda ^7}+\lambda ^5 \mu ^{12}+\frac{2 \mu ^{12}}{\lambda ^5}+2 \lambda ^3 \mu ^{12}+\frac{\mu ^{12}}{\lambda ^3}+\lambda  \mu ^{12}+\frac{3 \mu ^{12}}{\lambda } \) & \( 0 \) 
		\\\hline
		\(X^7\) & \( \frac{\mu ^{10}}{\lambda ^{22}}+\frac{\mu ^{10}}{\lambda ^{20}}+\frac{3 \mu ^{10}}{\lambda ^{18}}+\frac{4 \mu ^{10}}{\lambda ^{16}}+\frac{8 \mu ^{10}}{\lambda ^{14}}+\frac{10 \mu ^{10}}{\lambda ^{12}}+\frac{15 \mu ^{10}}{\lambda ^{10}}+\frac{18 \mu ^{10}}{\lambda ^8}+3 \lambda ^6 \mu ^{10}+\frac{24 \mu ^{10}}{\lambda ^6}+12 \lambda ^4 \mu ^{10}+\frac{25 \mu ^{10}}{\lambda ^4}+20 \lambda ^2 \mu ^{10}+\frac{28 \mu ^{10}}{\lambda ^2}+26 \mu ^{10} \) & \( \frac{\mu ^{12}}{\lambda ^{23}}+\frac{\mu ^{12}}{\lambda ^{21}}+\frac{2 \mu ^{12}}{\lambda ^{19}}+\frac{3 \mu ^{12}}{\lambda ^{17}}+\frac{5 \mu ^{12}}{\lambda ^{15}}+\frac{7 \mu ^{12}}{\lambda ^{13}}+\frac{9 \mu ^{12}}{\lambda ^{11}}+\frac{11 \mu ^{12}}{\lambda ^9}+\lambda ^7 \mu ^{12}+\frac{14 \mu ^{12}}{\lambda ^7}+5 \lambda ^5 \mu ^{12}+\frac{16 \mu ^{12}}{\lambda ^5}+10 \lambda ^3 \mu ^{12}+\frac{16 \mu ^{12}}{\lambda ^3}+14 \lambda  \mu ^{12}+\frac{17 \mu ^{12}}{\lambda } \) & \( \frac{\mu ^{14}}{\lambda ^{22}}+\frac{\mu ^{14}}{\lambda ^{18}}+\frac{\mu ^{14}}{\lambda ^{16}}+\frac{2 \mu ^{14}}{\lambda ^{14}}+\frac{2 \mu ^{14}}{\lambda ^{12}}+\frac{3 \mu ^{14}}{\lambda ^{10}}+\frac{3 \mu ^{14}}{\lambda ^8}+\lambda ^6 \mu ^{14}+\frac{4 \mu ^{14}}{\lambda ^6}+2 \lambda ^4 \mu ^{14}+\frac{4 \mu ^{14}}{\lambda ^4}+3 \lambda ^2 \mu ^{14}+\frac{4 \mu ^{14}}{\lambda ^2}+4 \mu ^{14} \) 
	\end{longtable}
}

\paragraph{Case II: $|X| > 1$.}

{\setlength{\tabcolsep}{2pt} 
	\begin{longtable}{>{\centering\arraybackslash}m{.5cm}||
			>{\centering\arraybackslash}m{.6cm}|
			>{\centering\arraybackslash}m{1.1cm}|
			>{\centering\arraybackslash}m{2.9cm}|
			>{\centering\arraybackslash}m{3.5cm}|
			>{\centering\arraybackslash}m{4.9cm}|
		}
		\hline
		& \(\mathfrak q^0\) & \(\mathfrak q^1\) & \(\mathfrak q^2\) & \(\mathfrak q^3\) & \(\mathfrak q^4\)
		\\\toprule
		\endfirsthead
		
		& \(\mathfrak q^0\) & \(\mathfrak q^1\) & \(\mathfrak q^2\) & \(\mathfrak q^3\) & \(\mathfrak q^4\)
		\\\midrule
		\endhead
		
		\midrule
		\endfoot
		
		\toprule
		\endlastfoot
		\( X \) & \( \lambda \) & \( 0 \) & \( 0 \) & \( 0 \) & \( 0 \) 
		\\\hline
		\( X^0 \) & \( 0 \) & \( 0 \) & \( 0 \) & \( 0 \) & \( 0 \) 
		\\\hline
		\( \frac{1}{X} \) & \( -\frac{1}{\lambda } \) & \( -\lambda ^2 \mu ^2 \) & \( 0 \) & \( 0 \) & \( 0 \) 
		\\\hline
		\( \frac{1}{X^2} \) & \( 0 \) & \( -\lambda ^2 \mu ^2 \) & \( -\lambda ^5 \mu ^4 \) & \( 0 \) & \( 0 \) 
		\\\hline
		\( \frac{1}{X^3} \) & \( 0 \) & \( -\lambda ^2 \mu ^2 \) & \( -\lambda ^7 \mu ^4-\lambda ^5 \mu ^4 \) & \( -\lambda ^{10} \mu ^6 \) & \( 0 \) 
		\\\hline
		\( \frac{1}{X^4} \) & \( 0 \) & \( -\lambda ^2 \mu ^2 \) & \( - \lambda ^9 \mu ^4-\lambda ^7 \mu ^4-2 \lambda ^5 \mu ^4 \) & \( - \lambda ^{14} \mu ^6-\lambda ^{12} \mu ^6-2 \lambda ^{10} \mu ^6-\lambda ^8 \mu ^6 \) & \( - \lambda ^{17} \mu ^8-\lambda ^{13} \mu ^8 \) 
		\\\hline
		\( \frac{1}{X^5} \) & \( 0 \) & \( -\lambda ^2 \mu ^2 \) & \( - \lambda ^{11} \mu ^4-\lambda ^9 \mu ^4-2 \lambda ^7 \mu ^4-2 \lambda ^5 \mu ^4 \) & \( - \lambda ^{18} \mu ^6-\lambda ^{16} \mu ^6-3 \lambda ^{14} \mu ^6-3 \lambda ^{12} \mu ^6-4 \lambda ^{10} \mu ^6-2 \lambda ^8 \mu ^6 \) & \( - \lambda ^{23} \mu ^8-\lambda ^{21} \mu ^8-2 \lambda ^{19} \mu ^8-3 \lambda ^{17} \mu ^8-3 \lambda ^{15} \mu ^8-3 \lambda ^{13} \mu ^8-\lambda ^{11} \mu ^8 \) 
		\\\hline
		\( \frac{1}{X^6} \) & \( 0 \) & \( -\lambda ^2 \mu ^2 \) & \( - \lambda ^{13} \mu ^4-\lambda ^{11} \mu ^4-2 \lambda ^9 \mu ^4-2 \lambda ^7 \mu ^4-3 \lambda ^5 \mu ^4 \) & \( - \lambda ^{22} \mu ^6-\lambda ^{20} \mu ^6-3 \lambda ^{18} \mu ^6-4 \lambda ^{16} \mu ^6-6 \lambda ^{14} \mu ^6-6 \lambda ^{12} \mu ^6-7 \lambda ^{10} \mu ^6-3 \lambda ^8 \mu ^6 \) & \( - \lambda ^{29} \mu ^8-\lambda ^{27} \mu ^8-3 \lambda ^{25} \mu ^8-4 \lambda ^{23} \mu ^8-7 \lambda ^{21} \mu ^8-7 \lambda ^{19} \mu ^8-11 \lambda ^{17} \mu ^8-8 \lambda ^{15} \mu ^8-8 \lambda ^{13} \mu ^8-2 \lambda ^{11} \mu ^8 \) 
		\\\hline
		\( \frac{1}{X^7} \) & \( 0 \) & \( -\lambda ^2 \mu ^2 \) & \( - \lambda ^{15} \mu ^4-\lambda ^{13} \mu ^4-2 \lambda ^{11} \mu ^4-2 \lambda ^9 \mu ^4-3 \lambda ^7 \mu ^4-3 \lambda ^5 \mu ^4 \) & \( - \lambda ^{26} \mu ^6-\lambda ^{24} \mu ^6-3 \lambda ^{22} \mu ^6-4 \lambda ^{20} \mu ^6-7 \lambda ^{18} \mu ^6-8 \lambda ^{16} \mu ^6-11 \lambda ^{14} \mu ^6-10 \lambda ^{12} \mu ^6-10 \lambda ^{10} \mu ^6-5 \lambda ^8 \mu ^6 \) & \( - \lambda ^{35} \mu ^8-\lambda ^{33} \mu ^8-3 \lambda ^{31} \mu ^8-5 \lambda ^{29} \mu ^8-8 \lambda ^{27} \mu ^8-11 \lambda ^{25} \mu ^8-16 \lambda ^{23} \mu ^8-19 \lambda ^{21} \mu ^8-22 \lambda ^{19} \mu ^8-24 \lambda ^{17} \mu ^8-20 \lambda ^{15} \mu ^8-15 \lambda ^{13} \mu ^8-5 \lambda ^{11} \mu ^8 \) 
	\end{longtable}
}

{\setlength{\tabcolsep}{2pt} 
	\begin{longtable}{>{\centering\arraybackslash}m{.5cm}||
			>{\centering\arraybackslash}m{4.5cm}|
			>{\centering\arraybackslash}m{5cm}|
			>{\centering\arraybackslash}m{4cm}|
		}
		\hline
		& \(\mathfrak q^5\) & \(\mathfrak q^6\) & \(\mathfrak q^7\)
		\\\toprule
		\endfirsthead
		
		& \(\mathfrak q^5\) & \(\mathfrak q^6\) & \(\mathfrak q^7\)
		\\\midrule
		\endhead
		
		\midrule
		\endfoot
		
		\toprule
		\endlastfoot
		
		\( \frac{1}{X^5} \) & \( - \lambda ^{26} \mu ^{10}-\lambda ^{22} \mu ^{10}-\lambda ^{20} \mu ^{10}-\lambda ^{18} \mu ^{10}-\lambda ^{16} \mu ^{10} \) & \( 0 \) & \( 0 \) 
		\\\hline
		\( \frac{1}{X^6} \) & \( - \lambda ^{34} \mu ^{10}-\lambda ^{32} \mu ^{10}-2 \lambda ^{30} \mu ^{10}-3 \lambda ^{28} \mu ^{10}-5 \lambda ^{26} \mu ^{10}-5 \lambda ^{24} \mu ^{10}-7 \lambda ^{22} \mu ^{10}-7 \lambda ^{20} \mu ^{10}-6 \lambda ^{18} \mu ^{10}-4 \lambda ^{16} \mu ^{10}-\lambda ^{14} \mu ^{10} \) & \( - \lambda ^{37} \mu ^{12}-\lambda ^{33} \mu ^{12}-\lambda ^{31} \mu ^{12}-2 \lambda ^{29} \mu ^{12}-\lambda ^{27} \mu ^{12}-3 \lambda ^{25} \mu ^{12}-\lambda ^{23} \mu ^{12}-2 \lambda ^{21} \mu ^{12}-\lambda ^{19} \mu ^{12} \) & \( 0 \) 
		\\\hline
		\( \frac{1}{X^7} \) & \( - \lambda ^{42} \mu ^{10}-\lambda ^{40} \mu ^{10}-3 \lambda ^{38} \mu ^{10}-4 \lambda ^{36} \mu ^{10}-8 \lambda ^{34} \mu ^{10}-10 \lambda ^{32} \mu ^{10}-15 \lambda ^{30} \mu ^{10}-18 \lambda ^{28} \mu ^{10}-24 \lambda ^{26} \mu ^{10}-25 \lambda ^{24} \mu ^{10}-28 \lambda ^{22} \mu ^{10}-26 \lambda ^{20} \mu ^{10}-20 \lambda ^{18} \mu ^{10}-12 \lambda ^{16} \mu ^{10}-3 \lambda ^{14} \mu ^{10} \) & \( - \lambda ^{47} \mu ^{12}-\lambda ^{45} \mu ^{12}-2 \lambda ^{43} \mu ^{12}-3 \lambda ^{41} \mu ^{12}-5 \lambda ^{39} \mu ^{12}-7 \lambda ^{37} \mu ^{12}-9 \lambda ^{35} \mu ^{12}-11 \lambda ^{33} \mu ^{12}-14 \lambda ^{31} \mu ^{12}-16 \lambda ^{29} \mu ^{12}-16 \lambda ^{27} \mu ^{12}-17 \lambda ^{25} \mu ^{12}-14 \lambda ^{23} \mu ^{12}-10 \lambda ^{21} \mu ^{12}-5 \lambda ^{19} \mu ^{12}-\lambda ^{17} \mu ^{12} \) & \( - \lambda ^{50} \mu ^{14}-\lambda ^{46} \mu ^{14}-\lambda ^{44} \mu ^{14}-2 \lambda ^{42} \mu ^{14}-2 \lambda ^{40} \mu ^{14}-3 \lambda ^{38} \mu ^{14}-3 \lambda ^{36} \mu ^{14}-4 \lambda ^{34} \mu ^{14}-4 \lambda ^{32} \mu ^{14}-4 \lambda ^{30} \mu ^{14}-4 \lambda ^{28} \mu ^{14}-3 \lambda ^{26} \mu ^{14}-2 \lambda ^{24} \mu ^{14}-\lambda ^{22} \mu ^{14} \) 
	\end{longtable}
}

\subsection{Resolved $A_1$-singularity} \label{app:a1sing}
\begin{figure}[H]
	\centering
	\centering
	\renewcommand{\arraystretch}{1.5} 
	\begin{tabular}{c l}
		\begin{tikzpicture}[
			x=2cm,y=2cm,
			line cap=round,line join=round,
			baseline=(current bounding box.center)
			]
			\tikzset{
				dual edge/.style={black, thick},
				dthree/.style={red, very thick},
			}
			
			\coordinate (K1) at (-0.35, 0);
			\coordinate (K2) at ( 0.35, 0);
			\fill[black] (K1) circle (1.2pt);
			\fill[black] (K2) circle (1.2pt);
			
			\draw[dual edge] (K1) -- (K2);

			\coordinate (JNW) at ($(K1)+(-0.45, 0.45)$);
			\draw[dual edge] (K1) -- (JNW);
			\coordinate (JSW) at ($(K1)+(0, -0.6)$);
			\draw[dual edge] (K1) -- (JSW);
			
			\draw[dual edge] (K2) -- ++( 0.00, -0.60);
			\draw[dual edge] (K2) -- ++( 0.45,0.45);
			
			\coordinate (P3) at ($(K1)!0.50!(JSW)$);
			\coordinate (P4) at ($(P3)+(-0.30,-0.15)$);
			\draw[dthree] (P3) -- (P4);
			
			\coordinate (Q3) at ($(K1)!0.50!(JNW)$);
			\coordinate (Q4) at ($(Q3)+(-0.30,-0.15)$);
			\draw[dthree] (Q3) -- (Q4);
			
			\node[left=0cm of Q4] () {(II)};
			\node[left=0cm of P4] () {(I)};
		\end{tikzpicture}
		&
		\begin{tabular}{@{}l@{}}
			\(T(X) = 1 - \left(q_1^{-1} q_2^{-1} + \mathfrak q\right)X^{-1} \),\\
			\(\left[q_1^{-1} X^{-1} q_1^{D_X} + T(X) + \mathfrak q X^{-1} q_1^{-D_X}\right]
			\langle Q_{-1}(X) \rangle \overset{q_2\to 1}{=} 0\).
		\end{tabular}
	\end{tabular}
	\caption{Toric diagram for the resolved $A_1$-singularity, the same diagram from Fig. \ref{fig:a1singtoric}, with a D3 brane that inserts the $Q$-observable in the engineered 5d $\mathcal N=1$ $\mathrm{U}(1)$ gauge theory with Chern-Simons level $k_\mathrm{CS}=-1$. Position I and II of the D3 brane correspond to the small and large $X$ expansions of the expectation value. To the right are the corresponding $qq$-character \eqref{eq:qqResA1} and the quantum mirror curve \eqref{eq:qMirrorResA1} with $f=-1$ and $\mathfrak q \to -\mathfrak q$.}
\end{figure}

\paragraph{Case I: $|X| < 1$.}

{\setlength{\tabcolsep}{2pt} 
	\renewcommand{\arraystretch}{1.1}
	\begin{longtable}{>{\centering\arraybackslash}m{.5cm}||
			>{\centering\arraybackslash}m{.5cm}|
			>{\centering\arraybackslash}m{.75cm}|
			>{\centering\arraybackslash}m{2.5cm}|
			>{\centering\arraybackslash}m{4.25cm}|
			>{\centering\arraybackslash}m{5.25cm}|
		}
		\hline
		& \(\mathfrak q^0\) & \(\mathfrak q^1\) & \(\mathfrak q^2\) & \(\mathfrak q^3\) & \(\mathfrak q^4\)
		\\\toprule
		\endfirsthead
		
		& \(\mathfrak q^0\) & \(\mathfrak q^1\) & \(\mathfrak q^2\) & \(\mathfrak q^3\) & \(\mathfrak q^4\)
		\\\midrule
		\endhead
		
		\midrule
		\endfoot
		
		\toprule
		\endlastfoot
		\( X^0 \) & \( 0 \) & \( \lambda  \mu ^2 \) & \( 0 \) & \( 0 \) & \( 0 \) 
		\\\hline
		\( X \) & \( \lambda \) & \( \lambda  \mu ^2 \) & \( \lambda  \mu ^4 \) & \( \lambda  \mu ^6 \) & \( \lambda  \mu ^8 \) 
		\\\hline
		\( X^2 \) & \( 0 \) & \( \lambda  \mu ^2 \) & \( \lambda  \mu ^4+\frac{\mu ^4}{\lambda } \) & \( \frac{\mu ^6}{\lambda ^3}+2 \lambda  \mu ^6+\frac{\mu ^6}{\lambda } \) & \( \frac{\mu ^8}{\lambda ^5}+\frac{\mu ^8}{\lambda ^3}+2 \lambda  \mu ^8+\frac{2 \mu ^8}{\lambda } \) 
		\\\hline
		\( X^3 \) & \( 0 \) & \( \lambda  \mu ^2 \) & \( \frac{\mu ^4}{\lambda ^3}+2 \lambda  \mu ^4+\frac{\mu ^4}{\lambda } \) & \( \frac{\mu ^6}{\lambda ^7}+\frac{\mu ^6}{\lambda ^5}+\frac{3 \mu ^6}{\lambda ^3}+3 \lambda  \mu ^6+\frac{3 \mu ^6}{\lambda } \) & \( \frac{\mu ^8}{\lambda ^{11}}+\frac{\mu ^8}{\lambda ^9}+\frac{3 \mu ^8}{\lambda ^7}+\frac{4 \mu ^8}{\lambda ^5}+\frac{6 \mu ^8}{\lambda ^3}+5 \lambda  \mu ^8+\frac{5 \mu ^8}{\lambda } \) 
		\\\hline
		\( X^4 \) & \( 0 \) & \( \lambda  \mu ^2 \) & \( \frac{\mu ^4}{\lambda ^5}+\frac{\mu ^4}{\lambda ^3}+2 \lambda  \mu ^4+\frac{2 \mu ^4}{\lambda } \) & \( \frac{\mu ^6}{\lambda ^{11}}+\frac{\mu ^6}{\lambda ^9}+\frac{3 \mu ^6}{\lambda ^7}+\frac{4 \mu ^6}{\lambda ^5}+\frac{6 \mu ^6}{\lambda ^3}+5 \lambda  \mu ^6+\frac{5 \mu ^6}{\lambda } \) & \( \frac{\mu ^8}{\lambda ^{17}}+\frac{\mu ^8}{\lambda ^{15}}+\frac{3 \mu ^8}{\lambda ^{13}}+\frac{5 \mu ^8}{\lambda ^{11}}+\frac{8 \mu ^8}{\lambda ^9}+\frac{10 \mu ^8}{\lambda ^7}+\frac{14 \mu ^8}{\lambda ^5}+\frac{14 \mu ^8}{\lambda ^3}+8 \lambda  \mu ^8+\frac{12 \mu ^8}{\lambda } \) 
		\\\hline
		\( X^5 \) & \( 0 \) & \( \lambda  \mu ^2 \) & \( \frac{\mu ^4}{\lambda ^7}+\frac{\mu ^4}{\lambda ^5}+\frac{2 \mu ^4}{\lambda ^3}+3 \lambda  \mu ^4+\frac{2 \mu ^4}{\lambda } \) & \( \frac{\mu ^6}{\lambda ^{15}}+\frac{\mu ^6}{\lambda ^{13}}+\frac{3 \mu ^6}{\lambda ^{11}}+\frac{4 \mu ^6}{\lambda ^9}+\frac{7 \mu ^6}{\lambda ^7}+\frac{8 \mu ^6}{\lambda ^5}+\frac{10 \mu ^6}{\lambda ^3}+7 \lambda  \mu ^6+\frac{8 \mu ^6}{\lambda } \) & \( \frac{\mu ^8}{\lambda ^{23}}+\frac{\mu ^8}{\lambda ^{21}}+\frac{3 \mu ^8}{\lambda ^{19}}+\frac{5 \mu ^8}{\lambda ^{17}}+\frac{9 \mu ^8}{\lambda ^{15}}+\frac{12 \mu ^8}{\lambda ^{13}}+\frac{19 \mu ^8}{\lambda ^{11}}+\frac{22 \mu ^8}{\lambda ^9}+\frac{29 \mu ^8}{\lambda ^7}+\frac{30 \mu ^8}{\lambda ^5}+\frac{30 \mu ^8}{\lambda ^3}+14 \lambda  \mu ^8+\frac{21 \mu ^8}{\lambda } \) 
		\\\hline
		\( X^6 \) & \( 0 \) & \( \lambda  \mu ^2 \) & \( \frac{\mu ^4}{\lambda ^9}+\frac{\mu ^4}{\lambda ^7}+\frac{2 \mu ^4}{\lambda ^5}+\frac{2 \mu ^4}{\lambda ^3}+3 \lambda  \mu ^4+\frac{3 \mu ^4}{\lambda } \) & \( \frac{\mu ^6}{\lambda ^{19}}+\frac{\mu ^6}{\lambda ^{17}}+\frac{3 \mu ^6}{\lambda ^{15}}+\frac{4 \mu ^6}{\lambda ^{13}}+\frac{7 \mu ^6}{\lambda ^{11}}+\frac{9 \mu ^6}{\lambda ^9}+\frac{13 \mu ^6}{\lambda ^7}+\frac{13 \mu ^6}{\lambda ^5}+\frac{15 \mu ^6}{\lambda ^3}+9 \lambda  \mu ^6+\frac{12 \mu ^6}{\lambda } \) & \( \frac{\mu ^8}{\lambda ^{29}}+\frac{\mu ^8}{\lambda ^{27}}+\frac{3 \mu ^8}{\lambda ^{25}}+\frac{5 \mu ^8}{\lambda ^{23}}+\frac{9 \mu ^8}{\lambda ^{21}}+\frac{13 \mu ^8}{\lambda ^{19}}+\frac{21 \mu ^8}{\lambda ^{17}}+\frac{27 \mu ^8}{\lambda ^{15}}+\frac{37 \mu ^8}{\lambda ^{13}}+\frac{45 \mu ^8}{\lambda ^{11}}+\frac{54 \mu ^8}{\lambda ^9}+\frac{58 \mu ^8}{\lambda ^7}+\frac{59 \mu ^8}{\lambda ^5}+\frac{51 \mu ^8}{\lambda ^3}+20 \lambda  \mu ^8+\frac{36 \mu ^8}{\lambda } \) 
		\\\hline
		\( X^7 \) & \( 0 \) & \( \lambda  \mu ^2 \) & \( \frac{\mu ^4}{\lambda ^{11}}+\frac{\mu ^4}{\lambda ^9}+\frac{2 \mu ^4}{\lambda ^7}+\frac{2 \mu ^4}{\lambda ^5}+\frac{3 \mu ^4}{\lambda ^3}+4 \lambda  \mu ^4+\frac{3 \mu ^4}{\lambda } \) & \( \frac{\mu ^6}{\lambda ^{23}}+\frac{\mu ^6}{\lambda ^{21}}+\frac{3 \mu ^6}{\lambda ^{19}}+\frac{4 \mu ^6}{\lambda ^{17}}+\frac{7 \mu ^6}{\lambda ^{15}}+\frac{9 \mu ^6}{\lambda ^{13}}+\frac{14 \mu ^6}{\lambda ^{11}}+\frac{16 \mu ^6}{\lambda ^9}+\frac{20 \mu ^6}{\lambda ^7}+\frac{20 \mu ^6}{\lambda ^5}+\frac{21 \mu ^6}{\lambda ^3}+12 \lambda  \mu ^6+\frac{16 \mu ^6}{\lambda } \) & \( \frac{\mu ^8}{\lambda ^{35}}+\frac{\mu ^8}{\lambda ^{33}}+\frac{3 \mu ^8}{\lambda ^{31}}+\frac{5 \mu ^8}{\lambda ^{29}}+\frac{9 \mu ^8}{\lambda ^{27}}+\frac{13 \mu ^8}{\lambda ^{25}}+\frac{22 \mu ^8}{\lambda ^{23}}+\frac{29 \mu ^8}{\lambda ^{21}}+\frac{42 \mu ^8}{\lambda ^{19}}+\frac{53 \mu ^8}{\lambda ^{17}}+\frac{69 \mu ^8}{\lambda ^{15}}+\frac{80 \mu ^8}{\lambda ^{13}}+\frac{97 \mu ^8}{\lambda ^{11}}+\frac{102 \mu ^8}{\lambda ^9}+\frac{108 \mu ^8}{\lambda ^7}+\frac{98 \mu ^8}{\lambda ^5}+\frac{84 \mu ^8}{\lambda ^3}+30 \lambda  \mu ^8+\frac{54 \mu ^8}{\lambda } \) 
	\end{longtable}
}

{\setlength{\tabcolsep}{2pt} 
	\renewcommand{\arraystretch}{1.1}
	\begin{longtable}{>{\centering\arraybackslash}m{.5cm}||
			>{\centering\arraybackslash}m{4cm}|
			>{\centering\arraybackslash}m{4.5cm}|
			>{\centering\arraybackslash}m{5cm}|
		}
		\hline
		& \(\mathfrak q^5\) & \(\mathfrak q^6\) & \(\mathfrak q^7\) 
		\\\toprule
		\endfirsthead
		
		& \(\mathfrak q^5\) & \(\mathfrak q^6\) & \(\mathfrak q^7\) 
		\\\midrule
		\endhead
		
		\midrule
		\endfoot
		
		\toprule
		\endlastfoot
		\( X \) & \( \lambda  \mu ^{10} \) & \( \lambda  \mu ^{12} \) & \( \lambda  \mu ^{14} \) 
		\\\hline
		\( X^2 \) & \( \frac{\mu ^{10}}{\lambda ^7}+\frac{\mu ^{10}}{\lambda ^5}+\frac{2 \mu ^{10}}{\lambda ^3}+3 \lambda  \mu ^{10}+\frac{2 \mu ^{10}}{\lambda } \) & \( \frac{\mu ^{12}}{\lambda ^9}+\frac{\mu ^{12}}{\lambda ^7}+\frac{2 \mu ^{12}}{\lambda ^5}+\frac{2 \mu ^{12}}{\lambda ^3}+3 \lambda  \mu ^{12}+\frac{3 \mu ^{12}}{\lambda } \) & \( \frac{\mu ^{14}}{\lambda ^{11}}+\frac{\mu ^{14}}{\lambda ^9}+\frac{2 \mu ^{14}}{\lambda ^7}+\frac{2 \mu ^{14}}{\lambda ^5}+\frac{3 \mu ^{14}}{\lambda ^3}+4 \lambda  \mu ^{14}+\frac{3 \mu ^{14}}{\lambda } \) 
		\\\hline
		\( X^3 \) & \( \frac{\mu ^{10}}{\lambda ^{15}}+\frac{\mu ^{10}}{\lambda ^{13}}+\frac{3 \mu ^{10}}{\lambda ^{11}}+\frac{4 \mu ^{10}}{\lambda ^9}+\frac{7 \mu ^{10}}{\lambda ^7}+\frac{8 \mu ^{10}}{\lambda ^5}+\frac{10 \mu ^{10}}{\lambda ^3}+7 \lambda  \mu ^{10}+\frac{8 \mu ^{10}}{\lambda } \) & \( \frac{\mu ^{12}}{\lambda ^{19}}+\frac{\mu ^{12}}{\lambda ^{17}}+\frac{3 \mu ^{12}}{\lambda ^{15}}+\frac{4 \mu ^{12}}{\lambda ^{13}}+\frac{7 \mu ^{12}}{\lambda ^{11}}+\frac{9 \mu ^{12}}{\lambda ^9}+\frac{13 \mu ^{12}}{\lambda ^7}+\frac{13 \mu ^{12}}{\lambda ^5}+\frac{15 \mu ^{12}}{\lambda ^3}+9 \lambda  \mu ^{12}+\frac{12 \mu ^{12}}{\lambda } \) & \( \frac{\mu ^{14}}{\lambda ^{23}}+\frac{\mu ^{14}}{\lambda ^{21}}+\frac{3 \mu ^{14}}{\lambda ^{19}}+\frac{4 \mu ^{14}}{\lambda ^{17}}+\frac{7 \mu ^{14}}{\lambda ^{15}}+\frac{9 \mu ^{14}}{\lambda ^{13}}+\frac{14 \mu ^{14}}{\lambda ^{11}}+\frac{16 \mu ^{14}}{\lambda ^9}+\frac{20 \mu ^{14}}{\lambda ^7}+\frac{20 \mu ^{14}}{\lambda ^5}+\frac{21 \mu ^{14}}{\lambda ^3}+12 \lambda  \mu ^{14}+\frac{16 \mu ^{14}}{\lambda } \) 
		\\\hline
		\( X^4 \) & \( \frac{\mu ^{10}}{\lambda ^{23}}+\frac{\mu ^{10}}{\lambda ^{21}}+\frac{3 \mu ^{10}}{\lambda ^{19}}+\frac{5 \mu ^{10}}{\lambda ^{17}}+\frac{9 \mu ^{10}}{\lambda ^{15}}+\frac{12 \mu ^{10}}{\lambda ^{13}}+\frac{19 \mu ^{10}}{\lambda ^{11}}+\frac{22 \mu ^{10}}{\lambda ^9}+\frac{29 \mu ^{10}}{\lambda ^7}+\frac{30 \mu ^{10}}{\lambda ^5}+\frac{30 \mu ^{10}}{\lambda ^3}+14 \lambda  \mu ^{10}+\frac{21 \mu ^{10}}{\lambda } \) & \( \frac{\mu ^{12}}{\lambda ^{29}}+\frac{\mu ^{12}}{\lambda ^{27}}+\frac{3 \mu ^{12}}{\lambda ^{25}}+\frac{5 \mu ^{12}}{\lambda ^{23}}+\frac{9 \mu ^{12}}{\lambda ^{21}}+\frac{13 \mu ^{12}}{\lambda ^{19}}+\frac{21 \mu ^{12}}{\lambda ^{17}}+\frac{27 \mu ^{12}}{\lambda ^{15}}+\frac{37 \mu ^{12}}{\lambda ^{13}}+\frac{45 \mu ^{12}}{\lambda ^{11}}+\frac{54 \mu ^{12}}{\lambda ^9}+\frac{58 \mu ^{12}}{\lambda ^7}+\frac{59 \mu ^{12}}{\lambda ^5}+\frac{51 \mu ^{12}}{\lambda ^3}+20 \lambda  \mu ^{12}+\frac{36 \mu ^{12}}{\lambda } \) & \( \frac{\mu ^{14}}{\lambda ^{35}}+\frac{\mu ^{14}}{\lambda ^{33}}+\frac{3 \mu ^{14}}{\lambda ^{31}}+\frac{5 \mu ^{14}}{\lambda ^{29}}+\frac{9 \mu ^{14}}{\lambda ^{27}}+\frac{13 \mu ^{14}}{\lambda ^{25}}+\frac{22 \mu ^{14}}{\lambda ^{23}}+\frac{29 \mu ^{14}}{\lambda ^{21}}+\frac{42 \mu ^{14}}{\lambda ^{19}}+\frac{53 \mu ^{14}}{\lambda ^{17}}+\frac{69 \mu ^{14}}{\lambda ^{15}}+\frac{80 \mu ^{14}}{\lambda ^{13}}+\frac{97 \mu ^{14}}{\lambda ^{11}}+\frac{102 \mu ^{14}}{\lambda ^9}+\frac{108 \mu ^{14}}{\lambda ^7}+\frac{98 \mu ^{14}}{\lambda ^5}+\frac{84 \mu ^{14}}{\lambda ^3}+30 \lambda  \mu ^{14}+\frac{54 \mu ^{14}}{\lambda } \) 
		\\\hline
		\( X^5 \) & \( \frac{\mu ^{10}}{\lambda ^{31}}+\frac{\mu ^{10}}{\lambda ^{29}}+\frac{3 \mu ^{10}}{\lambda ^{27}}+\frac{5 \mu ^{10}}{\lambda ^{25}}+\frac{10 \mu ^{10}}{\lambda ^{23}}+\frac{14 \mu ^{10}}{\lambda ^{21}}+\frac{23 \mu ^{10}}{\lambda ^{19}}+\frac{31 \mu ^{10}}{\lambda ^{17}}+\frac{44 \mu ^{10}}{\lambda ^{15}}+\frac{54 \mu ^{10}}{\lambda ^{13}}+\frac{69 \mu ^{10}}{\lambda ^{11}}+\frac{75 \mu ^{10}}{\lambda ^9}+\frac{85 \mu ^{10}}{\lambda ^7}+\frac{80 \mu ^{10}}{\lambda ^5}+\frac{70 \mu ^{10}}{\lambda ^3}+25 \lambda  \mu ^{10}+\frac{45 \mu ^{10}}{\lambda } \) & \( \frac{\mu ^{12}}{\lambda ^{39}}+\frac{\mu ^{12}}{\lambda ^{37}}+\frac{3 \mu ^{12}}{\lambda ^{35}}+\frac{5 \mu ^{12}}{\lambda ^{33}}+\frac{10 \mu ^{12}}{\lambda ^{31}}+\frac{15 \mu ^{12}}{\lambda ^{29}}+\frac{25 \mu ^{12}}{\lambda ^{27}}+\frac{35 \mu ^{12}}{\lambda ^{25}}+\frac{53 \mu ^{12}}{\lambda ^{23}}+\frac{69 \mu ^{12}}{\lambda ^{21}}+\frac{94 \mu ^{12}}{\lambda ^{19}}+\frac{116 \mu ^{12}}{\lambda ^{17}}+\frac{145 \mu ^{12}}{\lambda ^{15}}+\frac{165 \mu ^{12}}{\lambda ^{13}}+\frac{190 \mu ^{12}}{\lambda ^{11}}+\frac{196 \mu ^{12}}{\lambda ^9}+\frac{200 \mu ^{12}}{\lambda ^7}+\frac{175 \mu ^{12}}{\lambda ^5}+\frac{140 \mu ^{12}}{\lambda ^3}+42 \lambda  \mu ^{12}+\frac{84 \mu ^{12}}{\lambda } \) & \( \frac{\mu ^{14}}{\lambda ^{47}}+\frac{\mu ^{14}}{\lambda ^{45}}+\frac{3 \mu ^{14}}{\lambda ^{43}}+\frac{5 \mu ^{14}}{\lambda ^{41}}+\frac{10 \mu ^{14}}{\lambda ^{39}}+\frac{15 \mu ^{14}}{\lambda ^{37}}+\frac{26 \mu ^{14}}{\lambda ^{35}}+\frac{37 \mu ^{14}}{\lambda ^{33}}+\frac{57 \mu ^{14}}{\lambda ^{31}}+\frac{78 \mu ^{14}}{\lambda ^{29}}+\frac{109 \mu ^{14}}{\lambda ^{27}}+\frac{141 \mu ^{14}}{\lambda ^{25}}+\frac{187 \mu ^{14}}{\lambda ^{23}}+\frac{227 \mu ^{14}}{\lambda ^{21}}+\frac{281 \mu ^{14}}{\lambda ^{19}}+\frac{326 \mu ^{14}}{\lambda ^{17}}+\frac{376 \mu ^{14}}{\lambda ^{15}}+\frac{407 \mu ^{14}}{\lambda ^{13}}+\frac{437 \mu ^{14}}{\lambda ^{11}}+\frac{428 \mu ^{14}}{\lambda ^9}+\frac{406 \mu ^{14}}{\lambda ^7}+\frac{336 \mu ^{14}}{\lambda ^5}+\frac{252 \mu ^{14}}{\lambda ^3}+66 \lambda  \mu ^{14}+\frac{144 \mu ^{14}}{\lambda } \) 
		\\\hline
		\( X^6 \) & \( \frac{\mu ^{10}}{\lambda ^{39}}+\frac{\mu ^{10}}{\lambda ^{37}}+\frac{3 \mu ^{10}}{\lambda ^{35}}+\frac{5 \mu ^{10}}{\lambda ^{33}}+\frac{10 \mu ^{10}}{\lambda ^{31}}+\frac{15 \mu ^{10}}{\lambda ^{29}}+\frac{25 \mu ^{10}}{\lambda ^{27}}+\frac{35 \mu ^{10}}{\lambda ^{25}}+\frac{53 \mu ^{10}}{\lambda ^{23}}+\frac{69 \mu ^{10}}{\lambda ^{21}}+\frac{94 \mu ^{10}}{\lambda ^{19}}+\frac{116 \mu ^{10}}{\lambda ^{17}}+\frac{145 \mu ^{10}}{\lambda ^{15}}+\frac{165 \mu ^{10}}{\lambda ^{13}}+\frac{190 \mu ^{10}}{\lambda ^{11}}+\frac{196 \mu ^{10}}{\lambda ^9}+\frac{200 \mu ^{10}}{\lambda ^7}+\frac{175 \mu ^{10}}{\lambda ^5}+\frac{140 \mu ^{10}}{\lambda ^3}+42 \lambda  \mu ^{10}+\frac{84 \mu ^{10}}{\lambda } \) & \( \frac{\mu ^{12}}{\lambda ^{49}}+\frac{\mu ^{12}}{\lambda ^{47}}+\frac{3 \mu ^{12}}{\lambda ^{45}}+\frac{5 \mu ^{12}}{\lambda ^{43}}+\frac{10 \mu ^{12}}{\lambda ^{41}}+\frac{16 \mu ^{12}}{\lambda ^{39}}+\frac{27 \mu ^{12}}{\lambda ^{37}}+\frac{39 \mu ^{12}}{\lambda ^{35}}+\frac{61 \mu ^{12}}{\lambda ^{33}}+\frac{85 \mu ^{12}}{\lambda ^{31}}+\frac{120 \mu ^{12}}{\lambda ^{29}}+\frac{158 \mu ^{12}}{\lambda ^{27}}+\frac{211 \mu ^{12}}{\lambda ^{25}}+\frac{263 \mu ^{12}}{\lambda ^{23}}+\frac{328 \mu ^{12}}{\lambda ^{21}}+\frac{390 \mu ^{12}}{\lambda ^{19}}+\frac{458 \mu ^{12}}{\lambda ^{17}}+\frac{512 \mu ^{12}}{\lambda ^{15}}+\frac{560 \mu ^{12}}{\lambda ^{13}}+\frac{580 \mu ^{12}}{\lambda ^{11}}+\frac{573 \mu ^{12}}{\lambda ^9}+\frac{527 \mu ^{12}}{\lambda ^7}+\frac{434 \mu ^{12}}{\lambda ^5}+\frac{312 \mu ^{12}}{\lambda ^3}+75 \lambda  \mu ^{12}+\frac{177 \mu ^{12}}{\lambda } \) & \( \frac{\mu ^{14}}{\lambda ^{59}}+\frac{\mu ^{14}}{\lambda ^{57}}+\frac{3 \mu ^{14}}{\lambda ^{55}}+\frac{5 \mu ^{14}}{\lambda ^{53}}+\frac{10 \mu ^{14}}{\lambda ^{51}}+\frac{16 \mu ^{14}}{\lambda ^{49}}+\frac{28 \mu ^{14}}{\lambda ^{47}}+\frac{41 \mu ^{14}}{\lambda ^{45}}+\frac{65 \mu ^{14}}{\lambda ^{43}}+\frac{93 \mu ^{14}}{\lambda ^{41}}+\frac{136 \mu ^{14}}{\lambda ^{39}}+\frac{184 \mu ^{14}}{\lambda ^{37}}+\frac{255 \mu ^{14}}{\lambda ^{35}}+\frac{331 \mu ^{14}}{\lambda ^{33}}+\frac{434 \mu ^{14}}{\lambda ^{31}}+\frac{542 \mu ^{14}}{\lambda ^{29}}+\frac{676 \mu ^{14}}{\lambda ^{27}}+\frac{810 \mu ^{14}}{\lambda ^{25}}+\frac{968 \mu ^{14}}{\lambda ^{23}}+\frac{1107 \mu ^{14}}{\lambda ^{21}}+\frac{1261 \mu ^{14}}{\lambda ^{19}}+\frac{1376 \mu ^{14}}{\lambda ^{17}}+\frac{1477 \mu ^{14}}{\lambda ^{15}}+\frac{1507 \mu ^{14}}{\lambda ^{13}}+\frac{1498 \mu ^{14}}{\lambda ^{11}}+\frac{1379 \mu ^{14}}{\lambda ^9}+\frac{1204 \mu ^{14}}{\lambda ^7}+\frac{924 \mu ^{14}}{\lambda ^5}+\frac{630 \mu ^{14}}{\lambda ^3}+132 \lambda  \mu ^{14}+\frac{330 \mu ^{14}}{\lambda } \) 
		\\\hline
		\( X^7 \) & \( \frac{\mu ^{10}}{\lambda ^{47}}+\frac{\mu ^{10}}{\lambda ^{45}}+\frac{3 \mu ^{10}}{\lambda ^{43}}+\frac{5 \mu ^{10}}{\lambda ^{41}}+\frac{10 \mu ^{10}}{\lambda ^{39}}+\frac{15 \mu ^{10}}{\lambda ^{37}}+\frac{26 \mu ^{10}}{\lambda ^{35}}+\frac{37 \mu ^{10}}{\lambda ^{33}}+\frac{57 \mu ^{10}}{\lambda ^{31}}+\frac{78 \mu ^{10}}{\lambda ^{29}}+\frac{109 \mu ^{10}}{\lambda ^{27}}+\frac{141 \mu ^{10}}{\lambda ^{25}}+\frac{187 \mu ^{10}}{\lambda ^{23}}+\frac{227 \mu ^{10}}{\lambda ^{21}}+\frac{281 \mu ^{10}}{\lambda ^{19}}+\frac{326 \mu ^{10}}{\lambda ^{17}}+\frac{376 \mu ^{10}}{\lambda ^{15}}+\frac{407 \mu ^{10}}{\lambda ^{13}}+\frac{437 \mu ^{10}}{\lambda ^{11}}+\frac{428 \mu ^{10}}{\lambda ^9}+\frac{406 \mu ^{10}}{\lambda ^7}+\frac{336 \mu ^{10}}{\lambda ^5}+\frac{252 \mu ^{10}}{\lambda ^3}+66 \lambda  \mu ^{10}+\frac{144 \mu ^{10}}{\lambda } \) & \( \frac{\mu ^{12}}{\lambda ^{59}}+\frac{\mu ^{12}}{\lambda ^{57}}+\frac{3 \mu ^{12}}{\lambda ^{55}}+\frac{5 \mu ^{12}}{\lambda ^{53}}+\frac{10 \mu ^{12}}{\lambda ^{51}}+\frac{16 \mu ^{12}}{\lambda ^{49}}+\frac{28 \mu ^{12}}{\lambda ^{47}}+\frac{41 \mu ^{12}}{\lambda ^{45}}+\frac{65 \mu ^{12}}{\lambda ^{43}}+\frac{93 \mu ^{12}}{\lambda ^{41}}+\frac{136 \mu ^{12}}{\lambda ^{39}}+\frac{184 \mu ^{12}}{\lambda ^{37}}+\frac{255 \mu ^{12}}{\lambda ^{35}}+\frac{331 \mu ^{12}}{\lambda ^{33}}+\frac{434 \mu ^{12}}{\lambda ^{31}}+\frac{542 \mu ^{12}}{\lambda ^{29}}+\frac{676 \mu ^{12}}{\lambda ^{27}}+\frac{810 \mu ^{12}}{\lambda ^{25}}+\frac{968 \mu ^{12}}{\lambda ^{23}}+\frac{1107 \mu ^{12}}{\lambda ^{21}}+\frac{1261 \mu ^{12}}{\lambda ^{19}}+\frac{1376 \mu ^{12}}{\lambda ^{17}}+\frac{1477 \mu ^{12}}{\lambda ^{15}}+\frac{1507 \mu ^{12}}{\lambda ^{13}}+\frac{1498 \mu ^{12}}{\lambda ^{11}}+\frac{1379 \mu ^{12}}{\lambda ^9}+\frac{1204 \mu ^{12}}{\lambda ^7}+\frac{924 \mu ^{12}}{\lambda ^5}+\frac{630 \mu ^{12}}{\lambda ^3}+132 \lambda  \mu ^{12}+\frac{330 \mu ^{12}}{\lambda } \) & \( \frac{\mu ^{14}}{\lambda ^{71}}+\frac{\mu ^{14}}{\lambda ^{69}}+\frac{3 \mu ^{14}}{\lambda ^{67}}+\frac{5 \mu ^{14}}{\lambda ^{65}}+\frac{10 \mu ^{14}}{\lambda ^{63}}+\frac{16 \mu ^{14}}{\lambda ^{61}}+\frac{29 \mu ^{14}}{\lambda ^{59}}+\frac{43 \mu ^{14}}{\lambda ^{57}}+\frac{69 \mu ^{14}}{\lambda ^{55}}+\frac{101 \mu ^{14}}{\lambda ^{53}}+\frac{151 \mu ^{14}}{\lambda ^{51}}+\frac{211 \mu ^{14}}{\lambda ^{49}}+\frac{300 \mu ^{14}}{\lambda ^{47}}+\frac{402 \mu ^{14}}{\lambda ^{45}}+\frac{545 \mu ^{14}}{\lambda ^{43}}+\frac{709 \mu ^{14}}{\lambda ^{41}}+\frac{921 \mu ^{14}}{\lambda ^{39}}+\frac{1155 \mu ^{14}}{\lambda ^{37}}+\frac{1449 \mu ^{14}}{\lambda ^{35}}+\frac{1759 \mu ^{14}}{\lambda ^{33}}+\frac{2127 \mu ^{14}}{\lambda ^{31}}+\frac{2501 \mu ^{14}}{\lambda ^{29}}+\frac{2917 \mu ^{14}}{\lambda ^{27}}+\frac{3313 \mu ^{14}}{\lambda ^{25}}+\frac{3728 \mu ^{14}}{\lambda ^{23}}+\frac{4069 \mu ^{14}}{\lambda ^{21}}+\frac{4385 \mu ^{14}}{\lambda ^{19}}+\frac{4566 \mu ^{14}}{\lambda ^{17}}+\frac{4645 \mu ^{14}}{\lambda ^{15}}+\frac{4515 \mu ^{14}}{\lambda ^{13}}+\frac{4235 \mu ^{14}}{\lambda ^{11}}+\frac{3696 \mu ^{14}}{\lambda ^9}+\frac{3024 \mu ^{14}}{\lambda ^7}+\frac{2184 \mu ^{14}}{\lambda ^5}+\frac{1386 \mu ^{14}}{\lambda ^3}+245 \lambda  \mu ^{14}+\frac{679 \mu ^{14}}{\lambda } \)
	\end{longtable}
}

\paragraph{Case II: $|X| > 1$.}

{\setlength{\tabcolsep}{2pt} 
	\begin{longtable}{>{\centering\arraybackslash}m{.5cm}||
			>{\centering\arraybackslash}m{.5cm}|
			>{\centering\arraybackslash}m{1cm}|
			>{\centering\arraybackslash}m{5cm}|
			>{\centering\arraybackslash}m{7cm}|
		}
		\hline
		& \(\mathfrak q^0\) & \(\mathfrak q^1\) & \(\mathfrak q^2\) & \(\mathfrak q^3\) 
		\\\toprule
		\endfirsthead
		
		& \(\mathfrak q^0\) & \(\mathfrak q^1\) & \(\mathfrak q^2\) & \(\mathfrak q^3\) 
		\\\midrule
		\endhead
		
		\midrule
		\endfoot
		
		\toprule
		\endlastfoot
		\( X \) & \( \lambda \) & \( 0 \) & \( 0 \) & \( 0 \) 
		\\\hline
		\( X^0 \) & \( 0 \) & \( 0 \) & \( 0 \) & \( 0 \) 
		\\\hline
		\( \frac{1}{X} \) & \( 0 \) & \( -\lambda  \mu ^2 \) & \( 0 \) & \( 0 \) 
		\\\hline
		\( \frac{1}{X^2} \) & \( 0 \) & \( -\lambda  \mu ^2 \) & \( 0 \) & \( 0 \) 
		\\\hline
		\( \frac{1}{X^3} \) & \( 0 \) & \( -\lambda  \mu ^2 \) & \( -\lambda ^3 \mu ^4 \) & \( 0 \) 
		\\\hline
		\( \frac{1}{X^4} \) & \( 0 \) & \( -\lambda  \mu ^2 \) & \( - \lambda ^5 \mu ^4-\lambda ^3 \mu ^4 \) & \( -\lambda ^5 \mu ^6 \) 
		\\\hline
		\( \frac{1}{X^5} \) & \( 0 \) & \( -\lambda  \mu ^2 \) & \( - \lambda ^7 \mu ^4-\lambda ^5 \mu ^4-2 \lambda ^3 \mu ^4 \) & \( - \lambda ^9 \mu ^6-\lambda ^7 \mu ^6-2 \lambda ^5 \mu ^6 \) 
		\\\hline
		\( \frac{1}{X^6} \) & \( 0 \) & \( -\lambda  \mu ^2 \) & \( - \lambda ^9 \mu ^4-\lambda ^7 \mu ^4-2 \lambda ^5 \mu ^4-2 \lambda ^3 \mu ^4 \) & \( - \lambda ^{13} \mu ^6-\lambda ^{11} \mu ^6-3 \lambda ^9 \mu ^6-3 \lambda ^7 \mu ^6-3 \lambda ^5 \mu ^6 \) 
		\\\hline
		\( \frac{1}{X^7} \) & \( 0 \) & \( -\lambda  \mu ^2 \) & \( - \lambda ^{11} \mu ^4-\lambda ^9 \mu ^4-2 \lambda ^7 \mu ^4-2 \lambda ^5 \mu ^4-3 \lambda ^3 \mu ^4 \) & \( - \lambda ^{17} \mu ^6-\lambda ^{15} \mu ^6-3 \lambda ^{13} \mu ^6-4 \lambda ^{11} \mu ^6-6 \lambda ^9 \mu ^6-5 \lambda ^7 \mu ^6-5 \lambda ^5 \mu ^6 \) 
	\end{longtable}
}

{\setlength{\tabcolsep}{2pt} 
	\begin{longtable}{>{\centering\arraybackslash}m{.5cm}||
			>{\centering\arraybackslash}m{6.5cm}|
			>{\centering\arraybackslash}m{5cm}|
			>{\centering\arraybackslash}m{1.5cm}|
			>{\centering\arraybackslash}m{.5cm}|
		}
		\hline
		& \(\mathfrak q^4\) & \(\mathfrak q^5\) & \(\mathfrak q^6\) & \(\mathfrak q^7\)
		\\\toprule
		\endfirsthead
		
		& \(\mathfrak q^4\) & \(\mathfrak q^5\) & \(\mathfrak q^6\) & \(\mathfrak q^7\)
		\\\midrule
		\endhead
		
		\midrule
		\endfoot
		
		\toprule
		\endlastfoot
		
		\( \frac{1}{X^5} \) & \( -\lambda ^7 \mu ^8 \) & \( 0 \) & \( 0 \) & \( 0 \) 
		\\\hline
		\( \frac{1}{X^6} \) & \( - \lambda ^{13} \mu ^8-\lambda ^{11} \mu ^8-2 \lambda ^9 \mu ^8-2 \lambda ^7 \mu ^8 \) & \( -\lambda ^9 \mu ^{10} \) & \( 0 \) & \( 0 \) 
		\\\hline
		\( \frac{1}{X^7} \) & \( - \lambda ^{19} \mu ^8-\lambda ^{17} \mu ^8-3 \lambda ^{15} \mu ^8-4 \lambda ^{13} \mu ^8-6 \lambda ^{11} \mu ^8-5 \lambda ^9 \mu ^8-5 \lambda ^7 \mu ^8 \) & \( - \lambda ^{17} -\mu ^{10}-\lambda ^{15} \mu ^{10}-2 \lambda ^{13} \mu ^{10}-2 \lambda ^{11} \mu ^{10}-3 \lambda ^9 \mu ^{10} \) & \( -\lambda ^{11} \mu ^{12} \) & \( 0 \) 
	\end{longtable}
}

\subsection{Local $F_0$}
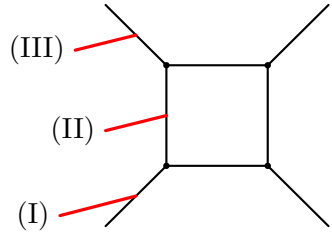
\label{sec:f0tables}
\begin{figure}[H]
	\centering
	\centering
	\renewcommand{\arraystretch}{1.5} 
	\begin{tabular}{c}
		\begin{tikzpicture}[
			x=2cm,y=2cm,
			line cap=round,line join=round,
			baseline=(current bounding box.center)
			]
			
			\tikzset{
				dual edge/.style={black, thick},
				dthree/.style={red, very thick},
			}
			
			\coordinate (K1) at (-1/3, 1/3);
			\coordinate (K2) at (-1/3,-1/3);
			\coordinate (K3) at ( 1/3,-1/3);
			\coordinate (K4) at ( 1/3, 1/3);
			\foreach \Q in {K1,K2,K3,K4}{\fill[black] (\Q) circle (1.2pt);}
			
			\draw[dual edge] (K1) -- (K2);
			\draw[dual edge] (K2) -- (K3);
			\draw[dual edge] (K3) -- (K4);
			\draw[dual edge] (K4) -- (K1);
			
			\coordinate (NW) at ($(K1)+(-0.40, 0.40)$);
			\draw[dual edge] (K1) -- (NW);
			\coordinate (SW) at ($(K2)+(-0.40,-0.40)$);
			\draw[dual edge] (K2) -- (SW);
			\draw[dual edge] (K3) -- ++( 0.40,-0.40);
			\draw[dual edge] (K4) -- ++( 0.40, 0.40);
			
			\coordinate (P) at ($(K2)!0.50!(K1)$);  
			\coordinate (Pend) at ($(P)+(-0.40,-0.1)$);
			\draw[dthree] (P) -- (Pend);   
			\node[left=0cm of Pend] () {(II)};
			
			\coordinate (Q) at ($(K1)!0.50!(NW)$);  
			\coordinate (Qend) at ($(Q)+(-0.40,-0.1)$);
			\draw[dthree] (Q) -- (Qend);   
			\node[left=0cm of Qend] () {(III)};
			
			\coordinate (R) at ($(K2)!0.50!(SW)$);  
			\coordinate (Rend) at ($(R)+(-0.50,-0.13)$);
			\draw[dthree] (R) -- (Rend);   
			\node[left=0cm of Rend] () {(I)};

		\end{tikzpicture}
		\\
		\begin{tabular}{@{}c@{}}
			\(T(X) = 1 + \underbrace{\left(-\frac{1+a}{q_1 q_2}+\frac{q_1 q_2  a (1+a) \mathfrak{q}}{(q_1 q_2 -a) (q_1 q_2 a-1)}+ \mathcal O (\mathfrak{q}^2 )\right)}_{t(a,\mathfrak q;q_1,q_2)} \frac{1}{X} + \frac{ a}{q_1^2 q_2^2 X^2} \),
			\\
			\(\left[\frac{\sqrt{a}}{q_1}(X+\frac{1}{X}) + \frac{\sqrt{\mathfrak q}}{q_1} \left(\frac{q_1^{D_X}}{\sqrt{\mathfrak q}} + \frac{\sqrt{\mathfrak q}}{q_1^{D_X}}\right) + t\left(a,\frac{\mathfrak q}{a q_1};q_1,1\right)\right]
			\left\langle Q_{-1}\left(\frac{\sqrt{aq_2} X}{q_1}\right) \right\rangle_{q_2=1} = 0\).
		\end{tabular}
	\end{tabular}
	\caption{Toric diagram for local $F_0$ (Fig. \ref{fig:toricp1p1}) with an AV brane, engineering a 5d $\mathcal N=1$ $\mathrm{U}(2)$ gauge theory with Chern-Simons level $k_\mathrm{CS}=0$ and the insertion of a $Q$-observable. Below are the corresponding $qq$-character \eqref{eq:qqF0} and the quantum mirror curve \eqref{eq:qcurvep1p1} with framing $f=-1$ and a rescaled parameter $\mathfrak q \to \frac{\mathfrak q}{aq_1q_2^{3/2}}$. The three different positions of the D3 brane correspond to three diffrent regions of the K\"ahler parameter space: I) $|X| < |\sqrt a|$, II) $|\sqrt{a}| < |X| < \frac{1}{|\sqrt{a}|}$, and III) $\frac{1}{|\sqrt a|} < |X|$.}
	\label{fig:F0}
\end{figure}

Local $F_0$ has two K\"ahler parameters corresponding to the two 2-cycles depicted by the horizontal and the vertical finite edges of the toric diagram. They translate to the 5d gauge coupling $\mathfrak q$ and the Coulomb moduli $a$ respectively. Insertion of the AV brane introduces a relative 2-cycle and a corresponding K\"ahler parameter $X$. We extract BPS invariants from the defect partition function $\left\langle Q_{-1}\left(\frac{\sqrt{aq_2} X}{q_1}\right) \right\rangle$. Below we make tables of invariants by expanding the partition function in different regimes. Rows and columns of the tables denote powers of $\mathfrak q$ and $X$, different tables correspond to different powers of $a$. Because of the rescaling of the argument of $Q_{-1}$ by $\sqrt{a}$, invariants appear at half-integral powers of $a$ iff powers of $X$ are odd.

The toric diagram and the quantum curve in Fig. \ref{fig:F0} are invariant under the inversion:
\begin{equation}
	(X, q_1) \mapsto (X^{-1}, q_1^{-1}),
\end{equation}
which relies on the non-trivial porperty 
\begin{equation}
	t\left(a,\frac{\mathfrak q}{aq_1};q_1,1\right) = \frac{1}{q_1^2} t\left(a,\frac{\mathfrak qq_1}{aq};\frac{1}{q_1},1\right)
\end{equation}
of the coefficient of $X^{-1}$ in the $qq$-character. Under this inversion, a D3 brane in position I is mapped to position III and vice versa. But a D3 brane in position II is mapped to itself. We expect to see this symmetry in the defect partition function and in turns in the BPS invariants. The above $\mathbb Z_2$-symmetry is evident in the Nekrasov-Shatashvili limit ($q_2 \to 1$) and needs to be extended to the refined case. We observe this symmetry in the refined case in terms of the invariants as follows:
\begin{equation}
	\text{for $i>0$:} \qquad \Lambda_{i,j,k}(\lambda, \mu) = -\Lambda_{-i,j,k}(\lambda^{-1}, \mu^{-1}). \label{invertedInv}
\end{equation}
This is manifest in the computed invariants below.

\paragraph{Case I: $|X| < |\sqrt{a}|$.}

Tables at low degrees in $a$ are sparse, we present the first few of them as series:
\begin{itemize}
	\item $a^{< -3/2}$: $\Lambda_{i,j,k < -\frac{3}{2}}(\lambda, \mu) = 0$.
	\item $a^{-3/2}$:
	\begin{equation}
		\sum_{i, j \in \mathbb Z} \Lambda_{i,j,-\frac{3}{2}}(\lambda, \mu) X^i \mathfrak q^j = \left(\frac{\mu ^4}{\lambda ^5} \mathfrak q + \frac{\mu ^5}{\lambda ^5} \mathfrak q^2 \right) X^3 + \mathcal O(X^5).
	\end{equation}
	
	\item $a^{-1}$:
	\begin{equation}\begin{split}
			&\; \sum_{i, j \in \mathbb Z} \Lambda_{i,j,-1}(\lambda, \mu) X^i \mathfrak q^j \\
			=&\; 
			\frac{\mu ^3}{\lambda ^3} \mathfrak q X^2 + \left[\left( \frac{\mu ^5}{\lambda ^7}+\frac{\mu ^7}{\lambda ^5} \right) \mathfrak q + \left( \frac{\mu ^6}{\lambda ^9}+\frac{\mu ^4}{\lambda ^9}+\frac{\mu ^8}{\lambda ^7}+\frac{3 \mu ^6}{\lambda ^7}+\frac{3 \mu ^8}{\lambda ^5}+\frac{\mu ^{10}}{\lambda ^3} \right) \mathfrak q^2 \right.
			\\
			&\; \left. + \left( \frac{\mu ^5}{\lambda ^{11}}+\frac{\mu ^3}{\lambda ^{11}}+\frac{\mu ^7}{\lambda ^9}+\frac{3 \mu ^5}{\lambda ^9}+\frac{\mu ^9}{\lambda ^7}+\frac{4 \mu ^7}{\lambda ^7}+\frac{\mu ^{11}}{\lambda ^5}+\frac{4 \mu ^9}{\lambda ^5}+\frac{3 \mu ^{11}}{\lambda ^3}+\frac{\mu ^{13}}{\lambda } \right) \mathfrak q^3 + \mathcal O(\mathfrak q^4)  \right] X^4
			\\
			&\; + \mathcal O(X^5).
	\end{split}\end{equation}

	\item $a^{-1/2}$:
	\begin{equation}
		\begin{split}
			&\; \sum_{i, j \in \mathbb Z} \Lambda_{i,j,-\frac{1}{2}}(\lambda, \mu) X^i \mathfrak q^j
			\\
			=&\; \frac{\mu ^2}{\lambda } X \mathfrak q + \left[\left(\frac{\mu ^4}{\lambda ^5}+\frac{\mu ^6}{\lambda ^3}\right) \mathfrak q + \left( \frac{\mu ^3}{\lambda ^7}+\frac{2 \mu ^5}{\lambda ^5}+\frac{2 \mu ^7}{\lambda ^3}+\frac{\mu ^9}{\lambda } \right) \mathfrak q^2 \right.
			\\
			&\; \left. \left( \frac{\mu ^2}{\lambda ^9}+\frac{2 \mu ^4}{\lambda ^7}+\frac{2 \mu ^6}{\lambda ^5}+\frac{2 \mu ^8}{\lambda ^3}+\lambda  \mu ^{12}+\frac{2 \mu ^{10}}{\lambda } \right) \mathfrak q^3 + \mathcal O(\mathfrak q^4) \right] X^3 + \mathcal O(X^5).
		\end{split}
	\end{equation}
	
\end{itemize}

Tables from $a^0$ to $a^{3/2}$ follows.
{\setlength{\tabcolsep}{2pt} 
	\renewcommand{\arraystretch}{1.1}
\begin{longtable}{>{\centering\arraybackslash}m{.5cm}||
			>{\centering\arraybackslash}m{.5cm}|
			>{\centering\arraybackslash}m{2cm}|
			>{\centering\arraybackslash}m{3cm}|
			>{\centering\arraybackslash}m{8cm}|
		}
		\hline
		& \multicolumn{4}{c|}{$a^0$} \\\hline
		& \(\mathfrak q^0\) & \(\mathfrak q^1\) & \(\mathfrak q^2\) & \(\mathfrak q^3\) 
		\\\toprule
		\endfirsthead
		
		& \multicolumn{4}{c|}{$a^0$} \\\hline
		& \(\mathfrak q^0\) & \(\mathfrak q^1\) & \(\mathfrak q^2\) & \(\mathfrak q^3\) 
		\\\midrule
		\endhead
		
		\midrule
		\endfoot
		
		\toprule
		\endlastfoot
		\( X^0 \) & \( 0 \) & \( \lambda  \mu +\frac{1}{\lambda  \mu } \) & \( 0 \) & \( 0 \) 
		\\\hline
		\( X^1 \) & \( 0 \) & \( 0 \) & \( 0 \) & \( 0 \) 
		\\\hline
		\( X^2 \) & \( 0 \) & \( \frac{\mu ^3}{\lambda ^3}+\frac{\mu ^5}{\lambda } \) & \( \frac{\mu ^2}{\lambda ^5}+\frac{\mu ^4}{\lambda ^3}+\lambda  \mu ^8+\frac{\mu ^6}{\lambda } \) & \( \frac{\mu }{\lambda ^7}+\frac{\mu ^3}{\lambda ^5}+\lambda ^3 \mu ^{11}+\frac{\mu ^5}{\lambda ^3}+\lambda  \mu ^9+\frac{\mu ^7}{\lambda } \) 
		\\\hline
		\( X^3 \) & \( 0 \) & \( 0 \) & \( 0 \) & \( 0 \) 
		\\
\end{longtable}
}

{\setlength{\tabcolsep}{2pt} 
	\renewcommand{\arraystretch}{1.1}
	\begin{longtable}{>{\centering\arraybackslash}m{.5cm}||
			>{\centering\arraybackslash}m{.5cm}|
			>{\centering\arraybackslash}m{2cm}|
			>{\centering\arraybackslash}m{3cm}|
			>{\centering\arraybackslash}m{8cm}|
		}
		\hline
		& \multicolumn{4}{c|}{$a^{1/2}$} \\\hline
		& \(\mathfrak q^0\) & \(\mathfrak q^1\) & \(\mathfrak q^2\) & \(\mathfrak q^3\) 
		\\\toprule
		\endfirsthead
		
		& \multicolumn{4}{c|}{$a^{1/2}$} \\\hline
		& \(\mathfrak q^0\) & \(\mathfrak q^1\) & \(\mathfrak q^2\) & \(\mathfrak q^3\) 
		\\\midrule
		\endhead
		
		\midrule
		\endfoot
		
		\toprule
		\endlastfoot
		\( \frac{1}{X} \) & \( -\frac{\lambda}{\mu} \) & \( 0 \) & \( 0 \) & \( 0 \) 
		\\\hline
		\( X^0 \) & \( 0 \) & \( 0 \) & \( 0 \) & \( 0 \) 
		\\\hline
		\( X^1 \) & \( \frac{\mu}{\lambda} \) & \( \frac{1}{\lambda ^3}+\lambda  \mu ^4+\frac{\mu ^2}{\lambda } \) & \( \frac{1}{\lambda ^5 \mu }+\lambda ^3 \mu ^7+\frac{\mu }{\lambda ^3}+\lambda  \mu ^5+\frac{\mu ^3}{\lambda } \) & \( \frac{1}{\lambda ^7 \mu ^2}+\lambda ^5 \mu ^{10}+\frac{1}{\lambda ^5}+\lambda ^3 \mu ^8+\frac{\mu ^2}{\lambda ^3}+\lambda  \mu ^6+\frac{\mu ^4}{\lambda } \) 
		\\\hline
		\( X^2 \) & \( 0 \) & \( 0 \) & \( 0 \) & \( 0 \) 
		\\
\end{longtable}
}

{\setlength{\tabcolsep}{2pt} 
	\renewcommand{\arraystretch}{1.1}
	\begin{longtable}{>{\centering\arraybackslash}m{.5cm}||
			>{\centering\arraybackslash}m{.5cm}|
			>{\centering\arraybackslash}m{1cm}|
			>{\centering\arraybackslash}m{3.5cm}|
			>{\centering\arraybackslash}m{8.5cm}|
		}
		\hline
		& \multicolumn{4}{c|}{$a^1$} \\\hline
		& \(\mathfrak q^0\) & \(\mathfrak q^1\) & \(\mathfrak q^2\) & \(\mathfrak q^3\) 
		\\\toprule
		\endfirsthead
		
		& \multicolumn{4}{c|}{$a^1$} \\\hline
		& \(\mathfrak q^0\) & \(\mathfrak q^1\) & \(\mathfrak q^2\) & \(\mathfrak q^3\) 
		\\\midrule
		\endhead
		
		\midrule
		\endfoot
		
		\toprule
		\endlastfoot
		\( X^0 \) & \( 0 \) & \( \lambda ^3 \mu ^3+\frac{1}{\lambda ^3 \mu ^3}+\lambda  \mu +\frac{1}{\lambda  \mu } \) & \( \lambda ^5 \mu ^6+\frac{1}{\lambda ^5 \mu ^6}+\lambda ^5 \mu ^4+\frac{1}{\lambda ^5 \mu ^4}+\lambda ^3 \mu ^4+\frac{1}{\lambda ^3 \mu ^4}+\lambda ^3 \mu ^2+\frac{1}{\lambda ^3 \mu ^2}+\lambda  \mu ^2+\frac{1}{\lambda  \mu ^2}+\lambda +\frac{1}{\lambda } \) & \( \lambda ^7 \mu ^9+\frac{1}{\lambda ^7 \mu ^9}+\lambda ^7 \mu ^7+\frac{1}{\lambda ^7 \mu ^7}+\lambda ^7 \mu ^5+\frac{1}{\lambda ^7 \mu ^5}+\lambda ^5 \mu ^7+\frac{1}{\lambda ^5 \mu ^7}+\lambda ^5 \mu ^5+\frac{1}{\lambda ^5 \mu ^5}+\lambda ^5 \mu ^3+\frac{1}{\lambda ^5 \mu ^3}+\lambda ^3 \mu ^5+\frac{1}{\lambda ^3 \mu ^5}+\lambda ^3 \mu ^3+\frac{1}{\lambda ^3 \mu ^3}+\lambda ^3 \mu +\frac{1}{\lambda ^3 \mu }+\lambda  \mu ^3+\frac{1}{\lambda  \mu ^3}+\lambda  \mu +\frac{\mu }{\lambda }+\frac{1}{\lambda  \mu }+\frac{\lambda }{\mu } \) 
		\\\hline
		\( X^1 \) & \( 0 \) & \( 0 \) & \( 0 \) & \( 0 \) 
		\\
\end{longtable}
}

{\setlength{\tabcolsep}{2pt} 
	\renewcommand{\arraystretch}{1.1}
	\begin{longtable}{>{\centering\arraybackslash}m{.5cm}||
			>{\centering\arraybackslash}m{.5cm}|
			>{\centering\arraybackslash}m{1cm}|
			>{\centering\arraybackslash}m{3.5cm}|
			>{\centering\arraybackslash}m{8.5cm}|
		}
		\hline
		& \multicolumn{4}{c|}{$a^{3/2}$} \\\hline
		& \(\mathfrak q^0\) & \(\mathfrak q^1\) & \(\mathfrak q^2\) & \(\mathfrak q^3\) 
		\\\toprule
		\endfirsthead
		
		& \multicolumn{4}{c|}{$a^{3/2}$} \\\hline
		& \(\mathfrak q^0\) & \(\mathfrak q^1\) & \(\mathfrak q^2\) & \(\mathfrak q^3\) 
		\\\midrule
		\endhead
		
		\midrule
		\endfoot
		
		\toprule
		\endlastfoot
		\( X^0 \) & \( 0 \) & \( 0 \) & \( 0 \) & \( 0 \) 
		\\
\end{longtable}
}

{\setlength{\tabcolsep}{2pt} 
	\renewcommand{\arraystretch}{1.1}
	\begin{longtable}{>{\centering\arraybackslash}m{.5cm}||
			>{\centering\arraybackslash}m{.5cm}|
			>{\centering\arraybackslash}m{1cm}|
			>{\centering\arraybackslash}m{3.5cm}|
			>{\centering\arraybackslash}m{8.5cm}|
		}
		\hline
		& \multicolumn{4}{c|}{$a^2$} \\\hline
		& \(\mathfrak q^0\) & \(\mathfrak q^1\) & \(\mathfrak q^2\) & \(\mathfrak q^3\) 
		\\\toprule
		\endfirsthead
		
		& \multicolumn{4}{c|}{$a^2$} \\\hline
		& \(\mathfrak q^0\) & \(\mathfrak q^1\) & \(\mathfrak q^2\) & \(\mathfrak q^3\) 
		\\\midrule
		\endhead
		
		\midrule
		\endfoot
		
		\toprule
		\endlastfoot
		\( X^0 \) & \( 0 \) & \( \lambda ^5 \mu ^5+\frac{1}{\lambda ^5 \mu ^5}+\lambda ^3 \mu ^3+\frac{1}{\lambda ^3 \mu ^3}+\lambda  \mu +\frac{1}{\lambda  \mu } \) & \( \lambda ^9 \mu ^8+\frac{1}{\lambda ^9 \mu ^8}+\lambda ^9 \mu ^6+\frac{1}{\lambda ^9 \mu ^6}+\lambda ^7 \mu ^{10}+\frac{1}{\lambda ^7 \mu ^{10}}+2 \lambda ^7 \mu ^8+\frac{2}{\lambda ^7 \mu ^8}+2 \lambda ^7 \mu ^6+\frac{2}{\lambda ^7 \mu ^6}+\lambda ^7 \mu ^4+\frac{1}{\lambda ^7 \mu ^4}+\lambda ^5 \mu ^8+\frac{1}{\lambda ^5 \mu ^8}+3 \lambda ^5 \mu ^6+\frac{3}{\lambda ^5 \mu ^6}+3 \lambda ^5 \mu ^4+\frac{3}{\lambda ^5 \mu ^4}+\lambda ^5 \mu ^2+\frac{1}{\lambda ^5 \mu ^2}+\lambda ^3 \mu ^6+\frac{1}{\lambda ^3 \mu ^6}+3 \lambda ^3 \mu ^4+\frac{3}{\lambda ^3 \mu ^4}+3 \lambda ^3 \mu ^2+\frac{3}{\lambda ^3 \mu ^2}+\lambda ^3+\frac{1}{\lambda ^3}+\lambda  \mu ^4+\frac{1}{\lambda  \mu ^4}+3 \lambda  \mu ^2+\frac{\mu ^2}{\lambda }+\frac{\lambda }{\mu ^2}+\frac{3}{\lambda  \mu ^2}+3 \lambda +\frac{3}{\lambda } \) & \( \lambda ^{13} \mu ^{11}+\frac{1}{\lambda ^{13} \mu ^{11}}+\lambda ^{13} \mu ^9+\frac{1}{\lambda ^{13} \mu ^9}+\lambda ^{13} \mu ^7+\frac{1}{\lambda ^{13} \mu ^7}+\lambda ^{11} \mu ^{13}+\frac{1}{\lambda ^{11} \mu ^{13}}+2 \lambda ^{11} \mu ^{11}+\frac{2}{\lambda ^{11} \mu ^{11}}+3 \lambda ^{11} \mu ^9+\frac{3}{\lambda ^{11} \mu ^9}+2 \lambda ^{11} \mu ^7+\frac{2}{\lambda ^{11} \mu ^7}+\lambda ^{11} \mu ^5+\frac{1}{\lambda ^{11} \mu ^5}+\lambda ^9 \mu ^{15}+\frac{1}{\lambda ^9 \mu ^{15}}+2 \lambda ^9 \mu ^{13}+\frac{2}{\lambda ^9 \mu ^{13}}+5 \lambda ^9 \mu ^{11}+\frac{5}{\lambda ^9 \mu ^{11}}+6 \lambda ^9 \mu ^9+\frac{6}{\lambda ^9 \mu ^9}+6 \lambda ^9 \mu ^7+\frac{6}{\lambda ^9 \mu ^7}+3 \lambda ^9 \mu ^5+\frac{3}{\lambda ^9 \mu ^5}+\lambda ^9 \mu ^3+\frac{1}{\lambda ^9 \mu ^3}+\lambda ^7 \mu ^{13}+\frac{1}{\lambda ^7 \mu ^{13}}+3 \lambda ^7 \mu ^{11}+\frac{3}{\lambda ^7 \mu ^{11}}+7 \lambda ^7 \mu ^9+\frac{7}{\lambda ^7 \mu ^9}+8 \lambda ^7 \mu ^7+\frac{8}{\lambda ^7 \mu ^7}+7 \lambda ^7 \mu ^5+\frac{7}{\lambda ^7 \mu ^5}+3 \lambda ^7 \mu ^3+\frac{3}{\lambda ^7 \mu ^3}+\lambda ^7 \mu +\frac{1}{\lambda ^7 \mu }+\lambda ^5 \mu ^{11}+\frac{1}{\lambda ^5 \mu ^{11}}+3 \lambda ^5 \mu ^9+\frac{3}{\lambda ^5 \mu ^9}+8 \lambda ^5 \mu ^7+\frac{8}{\lambda ^5 \mu ^7}+9 \lambda ^5 \mu ^5+\frac{9}{\lambda ^5 \mu ^5}+8 \lambda ^5 \mu ^3+\frac{8}{\lambda ^5 \mu ^3}+3 \lambda ^5 \mu +\frac{\mu }{\lambda ^5}+\frac{3}{\lambda ^5 \mu }+\frac{\lambda ^5}{\mu }+\lambda ^3 \mu ^9+\frac{1}{\lambda ^3 \mu ^9}+3 \lambda ^3 \mu ^7+\frac{3}{\lambda ^3 \mu ^7}+8 \lambda ^3 \mu ^5+\frac{8}{\lambda ^3 \mu ^5}+9 \lambda ^3 \mu ^3+\frac{\mu ^3}{\lambda ^3}+\frac{\lambda ^3}{\mu ^3}+\frac{9}{\lambda ^3 \mu ^3}+8 \lambda ^3 \mu +\frac{3 \mu }{\lambda ^3}+\frac{8}{\lambda ^3 \mu }+\frac{3 \lambda ^3}{\mu }+\lambda  \mu ^7+\frac{1}{\lambda  \mu ^7}+3 \lambda  \mu ^5+\frac{\mu ^5}{\lambda }+\frac{\lambda }{\mu ^5}+\frac{3}{\lambda  \mu ^5}+8 \lambda  \mu ^3+\frac{3 \mu ^3}{\lambda }+\frac{3 \lambda }{\mu ^3}+\frac{8}{\lambda  \mu ^3}+9 \lambda  \mu +\frac{8 \mu }{\lambda }+\frac{9}{\lambda  \mu }+\frac{8 \lambda }{\mu } \) 
		\\
\end{longtable}
}

\paragraph{Case II: $|\sqrt{a}| < |X| < \frac{1}{|\sqrt{a}|}$.}

For $a^0$: $\sum_{i,j, \in \mathbb Z} \Lambda_{i,j,0}(\lambda,\mu) X^i \mathfrak q^j = \frac{\mathfrak q}{\lambda\mu}$. Tables from $a^{1/2}$ to $a^2$ follow.

{\setlength{\tabcolsep}{2pt} 
	\renewcommand{\arraystretch}{1.1}
	\begin{longtable}{>{\centering\arraybackslash}m{.5cm}||
			>{\centering\arraybackslash}m{1cm}|
			>{\centering\arraybackslash}m{1cm}|
			>{\centering\arraybackslash}m{2cm}|
			>{\centering\arraybackslash}m{2cm}|
		}
		\hline
		& \multicolumn{4}{c|}{$a^{1/2}$} \\\hline
		& \(\mathfrak q^0\) & \(\mathfrak q^1\) & \(\mathfrak q^2\) & \(\mathfrak q^3\) 
		\\\toprule
		\endfirsthead
		
		& \multicolumn{4}{c|}{$a^{1/2}$} \\\hline
		& \(\mathfrak q^0\) & \(\mathfrak q^1\) & \(\mathfrak q^2\) & \(\mathfrak q^3\) 
		\\\midrule
		\endhead
		
		\midrule
		\endfoot
		
		\toprule
		\endlastfoot
		\( X^2 \) & \( 0 \) & \( 0 \) & \( 0 \) & \( 0 \) 
		\\\hline
		\( X^1 \) & \( \frac{\mu }{\lambda } \) & \( \frac{1}{\lambda ^3} \) & \( \frac{1}{\lambda ^5 \mu } \) & \( \frac{1}{\lambda ^7 \mu ^2} \) 
		\\\hline
		\( X^0 \) & \( 0 \) & \( 0 \) & \( 0 \) & \( 0 \) 
		\\\hline
		\( \frac{1}{X^1} \) & \( -\frac{\lambda }{\mu } \) & \( -\lambda ^3 \) & \( -\lambda ^5 \mu  \) & \( -\lambda ^7 \mu ^2 \) 
		\\\hline
		\( \frac{1}{X^2} \) & \( 0 \) & \( 0 \) & \( 0 \) & \( 0 \) 
	\end{longtable}
}

{\setlength{\tabcolsep}{2pt} 
	\renewcommand{\arraystretch}{1.1}
	\begin{longtable}{>{\centering\arraybackslash}m{.5cm}||
			>{\centering\arraybackslash}m{.5cm}|
			>{\centering\arraybackslash}m{1.5cm}|
			>{\centering\arraybackslash}m{3.5cm}|
			>{\centering\arraybackslash}m{6cm}|
		}
		\hline
		& \multicolumn{4}{c|}{$a^1$} \\\hline
		& \(\mathfrak q^0\) & \(\mathfrak q^1\) & \(\mathfrak q^2\) & \(\mathfrak q^3\) 
		\\\toprule
		\endfirsthead
		
		& \multicolumn{4}{c|}{$a^1$} \\\hline
		& \(\mathfrak q^0\) & \(\mathfrak q^1\) & \(\mathfrak q^2\) & \(\mathfrak q^3\) 
		\\\midrule
		\endhead
		
		\midrule
		\endfoot
		
		\toprule
		\endlastfoot
		\( X^3 \) & \( 0 \) & \( 0 \) & \( 0 \) & \( 0 \) 
		\\\hline
		\( X^2 \) & \( 0 \) & \( \frac{\mu }{\lambda ^5} \) & \( \frac{1}{\lambda ^9}+\frac{1}{\lambda ^7} \) & \( \frac{1}{\lambda ^{13} \mu }+\frac{1}{\lambda ^{11} \mu }+\frac{2}{\lambda ^9 \mu } \) 
		\\\hline
		\( X^1 \) & \( 0 \) & \( 0 \) & \( 0 \) & \( 0 \) 
		\\\hline
		\( X^0 \) & \( 0 \) & \( \frac{1}{\lambda ^3 \mu ^3}+\frac{1}{\lambda  \mu } \) & \( \frac{1}{\lambda ^5 \mu ^6}+\frac{1}{\lambda ^5 \mu ^4}+\frac{1}{\lambda ^3 \mu ^4}+\frac{1}{\lambda ^3 \mu ^2}+\frac{1}{\lambda  \mu ^2}+\lambda  \) & \( \frac{1}{\lambda ^7 \mu ^9}+\frac{1}{\lambda ^7 \mu ^7}+\frac{1}{\lambda ^7 \mu ^5}+\frac{1}{\lambda ^5 \mu ^7}+\frac{1}{\lambda ^5 \mu ^5}+\frac{1}{\lambda ^5 \mu ^3}+\frac{1}{\lambda ^3 \mu ^5}+\frac{1}{\lambda ^3 \mu ^3}+\lambda ^3 \mu +\frac{1}{\lambda  \mu ^3}+\frac{\lambda }{\mu }+\frac{1}{\lambda  \mu } \) 
		\\\hline
		\( \frac{1}{X} \) & \( 0 \) & \( 0 \) & \( 0 \) & \( 0 \) 
		\\\hline
		\( \frac{1}{X^2} \) & \( 0 \) & \( -\frac{\lambda ^5}{\mu } \) & \( -\lambda ^9-\lambda ^7 \) & \( -\lambda ^{13} \mu-\lambda ^{11} \mu -2 \lambda ^9 \mu  \) 
		\\\hline
		\( \frac{1}{X^3} \) & \( 0 \) & \( 0 \) & \( 0 \) & \( 0 \) 
	\end{longtable}
}

{\setlength{\tabcolsep}{2pt} 
	\renewcommand{\arraystretch}{1.1}
	\begin{longtable}{>{\centering\arraybackslash}m{.5cm}||
			>{\centering\arraybackslash}m{.5cm}|
			>{\centering\arraybackslash}m{2cm}|
			>{\centering\arraybackslash}m{4cm}|
			>{\centering\arraybackslash}m{7cm}|
		}
		\hline
		& \multicolumn{4}{c|}{$a^{3/2}$} \\\hline
		& \(\mathfrak q^0\) & \(\mathfrak q^1\) & \(\mathfrak q^2\) & \(\mathfrak q^3\) 
		\\\toprule
		\endfirsthead
		
		& \multicolumn{4}{c|}{$a^{3/2}$} \\\hline
		& \(\mathfrak q^0\) & \(\mathfrak q^1\) & \(\mathfrak q^2\) & \(\mathfrak q^3\) 
		\\\midrule
		\endhead
		
		\midrule
		\endfoot
		
		\toprule
		\endlastfoot
		\( X^4 \) & \( 0 \) & \( 0 \) & \( 0 \) & \( 0 \) 
		\\\hline
		\( X^3 \) & \( 0 \) & \( \frac{\mu ^2}{\lambda ^7} \) & \( \frac{\mu }{\lambda ^{13}}+\frac{\mu }{\lambda ^{11}}+\frac{2 \mu }{\lambda ^9} \) & \( \frac{1}{\lambda ^{19}}+\frac{1}{\lambda ^{17}}+\frac{3}{\lambda ^{15}}+\frac{3}{\lambda ^{13}}+\frac{3}{\lambda ^{11}} \) 
		\\\hline
		\( X^2 \) & \( 0 \) & \( 0 \) & \( 0 \) & \( 0 \) 
		\\\hline
		\( X^1 \) & \( 0 \) & \( \frac{1}{\lambda ^5 \mu ^2}+\frac{1}{\lambda ^3} \) & \( \frac{1}{\lambda ^9 \mu ^3}+\frac{1}{\lambda ^7 \mu ^5}+\frac{2}{\lambda ^7 \mu ^3}+\frac{1}{\lambda ^7 \mu }+\frac{1}{\lambda ^5 \mu ^3}+\frac{2}{\lambda ^5 \mu }+\frac{1}{\lambda ^3 \mu }+\frac{\mu }{\lambda } \) & \( \frac{1}{\lambda ^{13} \mu ^4}+\frac{1}{\lambda ^{11} \mu ^6}+\frac{2}{\lambda ^{11} \mu ^4}+\frac{1}{\lambda ^{11} \mu ^2}+\frac{1}{\lambda ^9 \mu ^8}+\frac{2}{\lambda ^9 \mu ^6}+\frac{4}{\lambda ^9 \mu ^4}+\frac{2}{\lambda ^9 \mu ^2}+\frac{1}{\lambda ^7 \mu ^6}+\frac{2}{\lambda ^7 \mu ^4}+\frac{4}{\lambda ^7 \mu ^2}+\frac{1}{\lambda ^5 \mu ^4}+\frac{2}{\lambda ^5 \mu ^2}+\frac{1}{\lambda ^5}+\frac{1}{\lambda ^3 \mu ^2}+\frac{2}{\lambda ^3}+\lambda  \mu ^2+\frac{1}{\lambda } \) 
		\\\hline
		\( X^0 \) & \( 0 \) & \( 0 \) & \( 0 \) & \( 0 \) 
		\\\hline
		\( \frac{1}{X} \) & \( 0 \) & \( -\lambda ^5 \mu ^2-\lambda ^3 \) & \( -\lambda ^9 \mu ^3-\lambda ^7 \mu ^5-2 \lambda ^7 \mu ^3-\lambda ^7 \mu -\lambda ^5 \mu ^3-2 \lambda ^5 \mu -\lambda ^3 \mu -\frac{\lambda }{\mu } \) & \( -\lambda ^{13} \mu ^4-\lambda ^{11} \mu ^6-2 \lambda ^{11} \mu ^4-\lambda ^{11} \mu ^2-\lambda ^9 \mu ^8-2 \lambda ^9 \mu ^6-4 \lambda ^9 \mu ^4-2 \lambda ^9 \mu ^2-\lambda ^7 \mu ^6-2 \lambda ^7 \mu ^4-4 \lambda ^7 \mu ^2-\lambda ^5 \mu ^4-2 \lambda ^5 \mu ^2-\lambda ^5-\lambda ^3 \mu ^2-2 \lambda ^3-\frac{1}{\lambda  \mu ^2}-\lambda  \) 
		\\\hline
		\( \frac{1}{X^2} \) & \( 0 \) & \( 0 \) & \( 0 \) & \( 0 \) 
		\\
\end{longtable}
}

{\setlength{\tabcolsep}{2pt} 
	\renewcommand{\arraystretch}{1.1}
	\begin{longtable}{>{\centering\arraybackslash}m{.5cm}||
			>{\centering\arraybackslash}m{.5cm}|
			>{\centering\arraybackslash}m{2cm}|
			>{\centering\arraybackslash}m{4cm}|
			>{\centering\arraybackslash}m{7cm}|
		}
		\hline
		& \multicolumn{4}{c|}{$a^2$} \\\hline
		& \(\mathfrak q^0\) & \(\mathfrak q^1\) & \(\mathfrak q^2\) & \(\mathfrak q^3\) 
		\\\toprule
		\endfirsthead
		
		& \multicolumn{4}{c|}{$a^2$} \\\hline
		& \(\mathfrak q^0\) & \(\mathfrak q^1\) & \(\mathfrak q^2\) & \(\mathfrak q^3\) 
		\\\midrule
		\endhead
		
		\midrule
		\endfoot
		
		\toprule
		\endlastfoot
		\( X^5 \) & \( 0 \) & \( 0 \) & \( 0 \) & \( 0 \) 
		\\\hline
		\( X^4 \) & \( 0 \) & \( \frac{\mu ^3}{\lambda ^9} \) & \( \frac{\mu ^2}{\lambda ^{17}}+\frac{\mu ^2}{\lambda ^{15}}+\frac{2 \mu ^2}{\lambda ^{13}}+\frac{2 \mu ^2}{\lambda ^{11}} \) & \( \frac{\mu }{\lambda ^{25}}+\frac{\mu }{\lambda ^{23}}+\frac{3 \mu }{\lambda ^{21}}+\frac{4 \mu }{\lambda ^{19}}+\frac{6 \mu }{\lambda ^{17}}+\frac{5 \mu }{\lambda ^{15}}+\frac{5 \mu }{\lambda ^{13}} \) 
		\\\hline
		\( X^3 \) & \( 0 \) & \( 0 \) & \( 0 \) & \( 0 \) 
		\\\hline
		\( X^2 \) & \( 0 \) & \( \frac{1}{\lambda ^7 \mu }+\frac{\mu }{\lambda ^5} \) & \( \frac{1}{\lambda ^{13} \mu ^2}+\frac{2}{\lambda ^{11} \mu ^2}+\frac{1}{\lambda ^{11}}+\frac{1}{\lambda ^9 \mu ^4}+\frac{3}{\lambda ^9 \mu ^2}+\frac{2}{\lambda ^9}+\frac{1}{\lambda ^7 \mu ^2}+\frac{3}{\lambda ^7}+\frac{1}{\lambda ^5}+\frac{\mu ^2}{\lambda ^3} \) & \( \frac{1}{\lambda ^{19} \mu ^3}+\frac{2}{\lambda ^{17} \mu ^3}+\frac{1}{\lambda ^{17} \mu }+\frac{1}{\lambda ^{15} \mu ^5}+\frac{5}{\lambda ^{15} \mu ^3}+\frac{2}{\lambda ^{15} \mu }+\frac{2}{\lambda ^{13} \mu ^5}+\frac{7}{\lambda ^{13} \mu ^3}+\frac{5}{\lambda ^{13} \mu }+\frac{1}{\lambda ^{11} \mu ^7}+\frac{3}{\lambda ^{11} \mu ^5}+\frac{8}{\lambda ^{11} \mu ^3}+\frac{7}{\lambda ^{11} \mu }+\frac{1}{\lambda ^9 \mu ^5}+\frac{3}{\lambda ^9 \mu ^3}+\frac{\mu }{\lambda ^9}+\frac{8}{\lambda ^9 \mu }+\frac{1}{\lambda ^7 \mu ^3}+\frac{2 \mu }{\lambda ^7}+\frac{3}{\lambda ^7 \mu }+\frac{3 \mu }{\lambda ^5}+\frac{1}{\lambda ^5 \mu }+\frac{\mu }{\lambda ^3}+\frac{\mu ^3}{\lambda } \) 
		\\\hline
		\( X^1 \) & \( 0 \) & \( 0 \) & \( 0 \) & \( 0 \) 
		\\
\end{longtable}
}

\paragraph{Case III: $|X| > \frac{1}{|\sqrt{a}|}$.} We checked that that invariants in this case at nonzero degrees in $X$ respect the symmetry \eqref{invertedInv}. Invariants at degree zero in $X$ are not related in this way so they need to be specified separately. These invariants simply vanish in the present case up to the orders checked.

\subsection{Local $F_1$} \label{app:f1}
\begin{figure}[H]
	\centering
	\centering
	\renewcommand{\arraystretch}{1.5} 
	%
	\end{tabular}
	\caption{Toric diagram for local $F_1$ (Fig. \ref{fig:localf1}) with an AV brane, engineering a 5d $\mathcal N=1$ $\mathrm{U}(2)$ gauge theory with Chern-Simons level $k_\mathrm{CS}=1$ and the insertion of a $Q$-observable. Below are the corresponding $qq$-character \eqref{eq:qqf1} and the quantum mirror curve \eqref{eq:qcurvef1} with framing $f=-1$ and a rescaled parameter $\mathfrak q \to \frac{-\mathfrak q}{aq_2\sqrt{q_1q_2}}$. The three different positions of the D3 brane correspond to three diffrent regions of the K\"ahler parameter space: I) $|X| < |\sqrt a|$, II) $|\sqrt{a}| < |X| < \frac{1}{|\sqrt{a}|}$, and III) $\frac{1}{|\sqrt a|} < |X|$.}
	\label{fig:F1}
\end{figure}

\paragraph{Case I: $|X|<|\sqrt{a}|$.}
\begin{itemize}
	\item $a^{-1}$: 
	\begin{equation}
		\sum_{i,j \in \mathbb Z} \Lambda_{i,j,-1}(\lambda,\mu) X^i \mathfrak q^j = \frac{\mu^5}{\lambda^6} \mathfrak q X^4 + \mathcal O(X^5).
	\end{equation}
	
	\item $a^{-1/2}$: 
	\begin{equation} 
		\sum_{i,j \in \mathbb Z} \Lambda_{i,j,-\frac{1}{2}}(\lambda,\mu) X^i \mathfrak q^j = \frac{\mu^4}{\lambda^4} \mathfrak q X^3 + \mathcal O(X^5).
	\end{equation}
	
	\item $a^0$:
	\begin{equation}\begin{split}
		&\; \sum_{i,j \in \mathbb Z} \Lambda_{i,j,0}(\lambda,\mu) X^i \mathfrak q^j 
		\\
		=&\; \frac{1}{\mu} \mathfrak q + \frac{\mu ^3}{\lambda ^2} \mathfrak q X^2 + \left[\left(\frac{\mu ^5}{\lambda ^6}+\frac{\mu ^7}{\lambda ^4}\right) \mathfrak q + \left(\frac{\mu ^4}{\lambda ^7}+\frac{2 \mu ^6}{\lambda ^5}+\frac{\mu ^8}{\lambda ^3}\right) \mathfrak q^2\right] X^4 + \mathcal O(X^5).
	\end{split}\end{equation}

	\item $a^{1/2}$:
	\begin{equation}\begin{split}
			&\; \sum_{i,j \in \mathbb Z} \Lambda_{i,j,\frac{1}{2}}(\lambda,\mu) X^i \mathfrak q^j 
			\\
			=&\; -\frac{\lambda}{\mu} X^{-1} + \left[\frac{\mu}{\lambda} + \left(\frac{1}{\lambda ^2}+\mu ^2\right) \mathfrak q \right] X + \left[\left(\frac{\mu ^4}{\lambda ^4}+\frac{\mu ^6}{\lambda ^2}\right) \mathfrak q + \left(\frac{\mu ^3}{\lambda ^5}+\frac{\mu ^5}{\lambda ^3}+\frac{\mu ^7}{\lambda }\right) \mathfrak q^2\right] X^3 
			\\
			&\; + \mathcal O(X^5).
	\end{split}\end{equation}
\end{itemize}

Tables from $a^1$ to $a^2$ follow.

{\setlength{\tabcolsep}{2pt} 
	\renewcommand{\arraystretch}{1.1}
	\begin{longtable}{>{\centering\arraybackslash}m{.5cm}||
			>{\centering\arraybackslash}m{.5cm}|
			>{\centering\arraybackslash}m{3cm}|
			>{\centering\arraybackslash}m{5cm}|
			>{\centering\arraybackslash}m{5cm}|
		}
		\hline
		& \multicolumn{4}{c|}{$a^1$} \\\hline
		& \(\mathfrak q^0\) & \(\mathfrak q^1\) & \(\mathfrak q^2\) & \(\mathfrak q^3\) 
		\\\toprule
		\endfirsthead
		
		& \multicolumn{4}{c|}{$a^1$} \\\hline
		& \(\mathfrak q^0\) & \(\mathfrak q^1\) & \(\mathfrak q^2\) & \(\mathfrak q^3\) 
		\\\midrule
		\endhead
		
		\midrule
		\endfoot
		
		\toprule
		\endlastfoot
	\( X^0 \) & \( 0 \) & \( \frac{1}{\lambda ^2 \mu ^3}+\lambda ^2 \mu +\frac{1}{\mu } \) & \( 0 \) & \( 0 \) 
	\\\hline
	\( X^1 \) & \( 0 \) & \( 0 \) & \( 0 \) & \( 0 \) 
	\\\hline
	\( X^2 \) & \( 0 \) & \( \frac{\mu }{\lambda ^4}+\frac{\mu ^3}{\lambda ^2}+\mu ^5 \) & \( \frac{1}{\lambda ^5}+\frac{\mu ^2}{\lambda ^3}+\lambda  \mu ^6+\frac{\mu ^4}{\lambda } \) & \( 0 \) 
	\\\hline
	\( X^3 \) & \( 0 \) & \( 0 \) & \( 0 \) & \( 0 \) 
	\\\hline
	\( X^4 \) & \( 0 \) & \( \frac{\mu ^5}{\lambda ^6}+\frac{\mu ^7}{\lambda ^4}+\frac{\mu ^9}{\lambda ^2} \) & \( \frac{\mu ^4}{\lambda ^9}+\frac{\mu ^2}{\lambda ^9}+\frac{\mu ^6}{\lambda ^7}+\frac{3 \mu ^4}{\lambda ^7}+\frac{\mu ^8}{\lambda ^5}+\frac{4 \mu ^6}{\lambda ^5}+\frac{\mu ^{10}}{\lambda ^3}+\frac{4 \mu ^8}{\lambda ^3}+\frac{\mu ^{12}}{\lambda }+\lambda  \mu ^{10}+\frac{2 \mu ^{10}}{\lambda } \) & \( \frac{\mu ^3}{\lambda ^{10}}+\frac{\mu }{\lambda ^{10}}+\frac{\mu ^5}{\lambda ^8}+\frac{2 \mu ^3}{\lambda ^8}+\frac{\mu ^7}{\lambda ^6}+\frac{3 \mu ^5}{\lambda ^6}+\frac{\mu ^9}{\lambda ^4}+\frac{3 \mu ^7}{\lambda ^4}+\lambda ^2 \mu ^{11}+\frac{\mu ^{11}}{\lambda ^2}+\frac{3 \mu ^9}{\lambda ^2}+\mu ^{13}+\mu ^{11} \) 
\end{longtable}
}

{\setlength{\tabcolsep}{2pt} 
	\renewcommand{\arraystretch}{1.1}
	\begin{longtable}{>{\centering\arraybackslash}m{.5cm}||
			>{\centering\arraybackslash}m{.5cm}|
			>{\centering\arraybackslash}m{3cm}|
			>{\centering\arraybackslash}m{5cm}|
			>{\centering\arraybackslash}m{5cm}|
		}
		\hline
		& \multicolumn{4}{c|}{$a^{3/2}$} \\\hline
		& \(\mathfrak q^0\) & \(\mathfrak q^1\) & \(\mathfrak q^2\) & \(\mathfrak q^3\) 
		\\\toprule
		\endfirsthead
		
		& \multicolumn{4}{c|}{$a^{3/2}$} \\\hline
		& \(\mathfrak q^0\) & \(\mathfrak q^1\) & \(\mathfrak q^2\) & \(\mathfrak q^3\) 
		\\\midrule
		\endhead
		
		\midrule
		\endfoot
		
		\toprule
		\endlastfoot
	\( X^0 \) & \( 0 \) & \( 0 \) & \( 0 \) & \( 0 \) 
	\\\hline
	\( X^1 \) & \( 0 \) & \( \frac{1}{\lambda ^4 \mu ^2}+\lambda ^2 \mu ^4+\frac{1}{\lambda ^2}+\mu ^2 \) & \( \frac{1}{\lambda ^5 \mu ^3}+\lambda ^3 \mu ^5+\frac{1}{\lambda ^3 \mu }+\lambda  \mu ^3+\frac{\mu }{\lambda } \) & \( 0 \) 
	\\\hline
	\( X^2 \) & \( 0 \) & \( 0 \) & \( 0 \) & \( 0 \) 
	\\\hline
	\( X^3 \) & \( 0 \) & \( \frac{\mu ^2}{\lambda ^6}+\frac{\mu ^4}{\lambda ^4}+\frac{\mu ^6}{\lambda ^2}+\mu ^8 \) & \( \frac{\mu }{\lambda ^9}+\frac{\mu ^3}{\lambda ^7}+\frac{2 \mu }{\lambda ^7}+\frac{\mu ^5}{\lambda ^5}+\frac{3 \mu ^3}{\lambda ^5}+\lambda ^3 \mu ^9+\frac{\mu ^7}{\lambda ^3}+\frac{3 \mu ^5}{\lambda ^3}+\lambda  \mu ^{11}+2 \lambda  \mu ^9+\frac{\mu ^9}{\lambda }+\frac{3 \mu ^7}{\lambda } \) & \( \frac{1}{\lambda ^{10}}+\frac{\mu ^2}{\lambda ^8}+\frac{1}{\lambda ^8}+\frac{\mu ^4}{\lambda ^6}+\frac{2 \mu ^2}{\lambda ^6}+\lambda ^4 \mu ^{10}+\frac{\mu ^6}{\lambda ^4}+\frac{2 \mu ^4}{\lambda ^4}+\lambda ^2 \mu ^{12}+\lambda ^2 \mu ^{10}+\frac{\mu ^8}{\lambda ^2}+\frac{2 \mu ^6}{\lambda ^2}+\mu ^{10}+2 \mu ^8 \) 
	\\\hline
	\( X^4 \) & \( 0 \) & \( 0 \) & \( 0 \) & \( 0 \) 
\end{longtable}
}

{\setlength{\tabcolsep}{2pt} 
	\renewcommand{\arraystretch}{1.1}
	\begin{longtable}{>{\centering\arraybackslash}m{.5cm}||
			>{\centering\arraybackslash}m{.5cm}|
			>{\centering\arraybackslash}m{2.5cm}|
			>{\centering\arraybackslash}m{4.5cm}|
			>{\centering\arraybackslash}m{6cm}|
		}
		\hline
		& \multicolumn{4}{c|}{$a^2$} \\\hline
		& \(\mathfrak q^0\) & \(\mathfrak q^1\) & \(\mathfrak q^2\) & \(\mathfrak q^3\) 
		\\\toprule
		\endfirsthead
		
		& \multicolumn{4}{c|}{$a^2$} \\\hline
		& \(\mathfrak q^0\) & \(\mathfrak q^1\) & \(\mathfrak q^2\) & \(\mathfrak q^3\) 
		\\\midrule
		\endhead
		
		\midrule
		\endfoot
		
		\toprule
		\endlastfoot
	\( X^0 \) & \( 0 \) & \( \frac{1}{\lambda ^4 \mu ^5}+\lambda ^4 \mu ^3+\frac{1}{\lambda ^2 \mu ^3}+\lambda ^2 \mu +\frac{1}{\mu } \) & \( \frac{1}{\lambda ^5 \mu ^8}+\frac{1}{\lambda ^5 \mu ^6}+\lambda ^5 \mu ^4+\lambda ^5 \mu ^2+\frac{1}{\lambda ^3 \mu ^6}+\frac{1}{\lambda ^3 \mu ^4}+\lambda ^3 \mu ^2+\lambda ^3+\frac{1}{\lambda  \mu ^4}+\frac{\lambda }{\mu ^2}+\frac{1}{\lambda  \mu ^2}+\lambda  \) & \( 0 \) 
	\\\hline
	\( X^1 \) & \( 0 \) & \( 0 \) & \( 0 \) & \( 0 \) 
	\\\hline
	\( X^2 \) & \( 0 \) & \( \frac{1}{\lambda ^6 \mu }+\frac{\mu }{\lambda ^4}+\lambda ^2 \mu ^7+\frac{\mu ^3}{\lambda ^2}+\mu ^5 \) & \( \frac{1}{\lambda ^9 \mu ^2}+\frac{2}{\lambda ^7 \mu ^2}+\frac{1}{\lambda ^7}+\lambda ^5 \mu ^8+\frac{\mu ^2}{\lambda ^5}+\frac{3}{\lambda ^5}+\lambda ^3 \mu ^{10}+2 \lambda ^3 \mu ^8+\frac{\mu ^4}{\lambda ^3}+\frac{3 \mu ^2}{\lambda ^3}+\lambda  \mu ^8+3 \lambda  \mu ^6+\frac{\mu ^6}{\lambda }+\frac{3 \mu ^4}{\lambda } \) & \( \frac{1}{\lambda ^{10} \mu ^3}+\frac{1}{\lambda ^8 \mu ^3}+\frac{1}{\lambda ^8 \mu }+\lambda ^6 \mu ^9+\frac{\mu }{\lambda ^6}+\frac{2}{\lambda ^6 \mu }+\lambda ^4 \mu ^{11}+\lambda ^4 \mu ^9+\frac{\mu ^3}{\lambda ^4}+\frac{2 \mu }{\lambda ^4}+\lambda ^2 \mu ^9+2 \lambda ^2 \mu ^7+\frac{\mu ^5}{\lambda ^2}+\frac{2 \mu ^3}{\lambda ^2}+\mu ^7+2 \mu ^5 \) 
	\\\hline
	\( X^3 \) & \( 0 \) & \( 0 \) & \( 0 \) & \( 0 \) 
	\\\hline
	\( X^4 \) & \( 0 \) & \( \frac{\mu ^3}{\lambda ^8}+\frac{\mu ^5}{\lambda ^6}+\frac{\mu ^7}{\lambda ^4}+\frac{\mu ^9}{\lambda ^2}+\mu ^{11} \) & \( \frac{\mu ^2}{\lambda ^{13}}+\frac{\mu ^4}{\lambda ^{11}}+\frac{2 \mu ^2}{\lambda ^{11}}+\frac{1}{\lambda ^{11}}+\frac{\mu ^6}{\lambda ^9}+\frac{3 \mu ^4}{\lambda ^9}+\frac{5 \mu ^2}{\lambda ^9}+\frac{\mu ^8}{\lambda ^7}+\frac{3 \mu ^6}{\lambda ^7}+\frac{7 \mu ^4}{\lambda ^7}+\lambda ^5 \mu ^{12}+\frac{\mu ^{10}}{\lambda ^5}+\frac{3 \mu ^8}{\lambda ^5}+\frac{8 \mu ^6}{\lambda ^5}+\lambda ^3 \mu ^{14}+2 \lambda ^3 \mu ^{12}+\frac{\mu ^{12}}{\lambda ^3}+\frac{3 \mu ^{10}}{\lambda ^3}+\frac{8 \mu ^8}{\lambda ^3}+\lambda  \mu ^{16}+2 \lambda  \mu ^{14}+\frac{\mu ^{14}}{\lambda }+4 \lambda  \mu ^{12}+\frac{3 \mu ^{12}}{\lambda }+\lambda  \mu ^{10}+\frac{7 \mu ^{10}}{\lambda } \) & \( \frac{\mu }{\lambda ^{16}}+\frac{\mu ^3}{\lambda ^{14}}+\frac{2 \mu }{\lambda ^{14}}+\frac{1}{\lambda ^{14} \mu }+\frac{\mu ^5}{\lambda ^{12}}+\frac{3 \mu ^3}{\lambda ^{12}}+\frac{1}{\lambda ^{12} \mu ^3}+\frac{6 \mu }{\lambda ^{12}}+\frac{2}{\lambda ^{12} \mu }+\frac{\mu ^7}{\lambda ^{10}}+\frac{3 \mu ^5}{\lambda ^{10}}+\frac{9 \mu ^3}{\lambda ^{10}}+\frac{8 \mu }{\lambda ^{10}}+\frac{1}{\lambda ^{10} \mu }+\lambda ^8 \mu ^{13}+\frac{\mu ^9}{\lambda ^8}+\frac{3 \mu ^7}{\lambda ^8}+\frac{10 \mu ^5}{\lambda ^8}+\frac{13 \mu ^3}{\lambda ^8}+\frac{\mu }{\lambda ^8}+\lambda ^6 \mu ^{15}+2 \lambda ^6 \mu ^{13}+\frac{\mu ^{11}}{\lambda ^6}+\frac{3 \mu ^9}{\lambda ^6}+\frac{10 \mu ^7}{\lambda ^6}+\frac{15 \mu ^5}{\lambda ^6}+\frac{\mu ^3}{\lambda ^6}+\lambda ^4 \mu ^{17}+3 \lambda ^4 \mu ^{15}+5 \lambda ^4 \mu ^{13}+\frac{\mu ^{13}}{\lambda ^4}+\frac{3 \mu ^{11}}{\lambda ^4}+\frac{10 \mu ^9}{\lambda ^4}+\frac{16 \mu ^7}{\lambda ^4}+\frac{\mu ^5}{\lambda ^4}+\lambda ^2 \mu ^{19}+2 \lambda ^2 \mu ^{17}+5 \lambda ^2 \mu ^{15}+\frac{\mu ^{15}}{\lambda ^2}+6 \lambda ^2 \mu ^{13}+\frac{3 \mu ^{13}}{\lambda ^2}+4 \lambda ^2 \mu ^{11}+\frac{10 \mu ^{11}}{\lambda ^2}+\frac{15 \mu ^9}{\lambda ^2}+\frac{\mu ^7}{\lambda ^2}+\mu ^{17}+3 \mu ^{15}+9 \mu ^{13}+12 \mu ^{11}+2 \mu ^9 \) 
\end{longtable}
}

\paragraph{Case II: $|\sqrt{a}|<|X|<\frac{1}{|\sqrt{a}|}$.}
\begin{itemize}
	\item $a^0$:
	\begin{equation}
		\sum_{i,j\in \mathbb Z} \Lambda_{i,j,0}(\lambda, \mu) X^i \mathfrak q^j = \frac{\mathfrak q}{\mu}.
	\end{equation}
	
	\item $a^{1/2}$:
	\begin{equation}
		\sum_{i,j\in \mathbb Z} \Lambda_{i,j,\frac{1}{2}}(\lambda, \mu) X^i \mathfrak q^j = -\frac{\lambda}{\mu} X^{-1} + \left(\frac{\mu}{\lambda} + \frac{\mathfrak q}{\lambda^2}\right)X.
	\end{equation}
	
	\item $a^1$:
	\begin{equation}
		\sum_{i,j\in \mathbb Z} \Lambda_{i,j,1}(\lambda, \mu) X^i \mathfrak q^j = \left(\frac{1}{\lambda ^2 \mu ^3}+\frac{1}{\mu }\right) \mathfrak q + \left(\frac{\mu }{\lambda ^4} + \frac{1}{\lambda ^5}\right) X^2.
	\end{equation}
	
	\item $a^{3/2}$:
	\begin{equation}\begin{split}
			\sum_{i,j\in \mathbb Z} \Lambda_{i,j,\frac{3}{2}}(\lambda, \mu) X^i \mathfrak q^j 
			=&\; -\frac{\lambda^4}{X} + \left[\left(\frac{1}{\lambda ^4 \mu ^2}+\frac{1}{\lambda ^2}\right)\mathfrak q + \left(\frac{1}{\lambda ^5 \mu ^3}+\frac{1}{\lambda ^3 \mu }\right) \mathfrak q^2\right] X
			\\
			&\; +\left[\frac{\mu ^2}{\lambda ^6} \mathfrak q + \left(\frac{\mu }{\lambda ^9}+\frac{\mu }{\lambda ^7}\right) \mathfrak q^2 + \frac{\mathfrak q^3}{\lambda ^{10}} \right]X^3.
	\end{split}\end{equation}
\end{itemize}

Tables for $a^2$ and $a^{5/2}$ follow.

{\setlength{\tabcolsep}{2pt} 
	\renewcommand{\arraystretch}{1.1}
\begin{longtable}{>{\centering\arraybackslash}m{1cm}||
			>{\centering\arraybackslash}m{.5cm}|
			>{\centering\arraybackslash}m{2.5cm}|
			>{\centering\arraybackslash}m{4.25cm}|
			>{\centering\arraybackslash}m{4.25cm}|
			>{\centering\arraybackslash}m{1.5cm}|
		}
		\hline
		& \multicolumn{5}{c|}{$a^2$} \\\hline
		& \(\mathfrak q^0\) & \(\mathfrak q^1\) & \(\mathfrak q^2\) & \(\mathfrak q^3\) & \(\mathfrak q^4\)
		\\\toprule
		\endfirsthead
		
		& \multicolumn{5}{c|}{$a^2$} \\\hline
		& \(\mathfrak q^0\) & \(\mathfrak q^1\) & \(\mathfrak q^2\) & \(\mathfrak q^3\) & \(\mathfrak q^4\)
		\\\midrule
		\endhead
		
		\midrule
		\endfoot
		
		\toprule
		\endlastfoot
		\( X^{>4} \) & \( 0 \) & \( 0 \) & \( 0 \) & \( 0 \) & \( 0 \) 
		\\\hline
		\( X^4 \) & \( 0 \) & \( \frac{\mu ^3}{\lambda ^8} \) & \( \frac{\mu ^2}{\lambda ^{13}}+\frac{\mu ^2}{\lambda ^{11}}+\frac{2 \mu ^2}{\lambda ^9} \) & \( \frac{\mu }{\lambda ^{16}}+\frac{\mu }{\lambda ^{14}}+\frac{2 \mu }{\lambda ^{12}}+\frac{\mu }{\lambda ^{10}} \) & \( \frac{1}{\lambda ^{17}}+\frac{1}{\lambda ^{13}} \) 
		\\\hline
		\( X^3 \) & \( 0 \) & \( 0 \) & \( 0 \) & \( 0 \) & \( 0 \) 
		\\\hline
		\( X^2 \) & \( 0 \) & \( \frac{1}{\lambda ^6 \mu }+\frac{\mu }{\lambda ^4} \) & \( \frac{1}{\lambda ^9 \mu ^2}+\frac{2}{\lambda ^7 \mu ^2}+\frac{1}{\lambda ^7}+\frac{2}{\lambda ^5} \) & \( \frac{1}{\lambda ^{10} \mu ^3}+\frac{1}{\lambda ^8 \mu ^3}+\frac{1}{\lambda ^8 \mu }+\frac{1}{\lambda ^6 \mu } \) & \( 0 \) 
		\\\hline
		\( X^1 \) & \( 0 \) & \( 0 \) & \( 0 \) & \( 0 \) & \( 0 \) 
		\\\hline
		\( X^0 \) & \( 0 \) & \( \frac{1}{\lambda ^4 \mu ^5}+\frac{1}{\lambda ^2 \mu ^3}+\frac{1}{\mu } \) & \( \frac{1}{\lambda ^5 \mu ^8}+\frac{1}{\lambda ^5 \mu ^6}+\frac{1}{\lambda ^3 \mu ^6}+\frac{1}{\lambda ^3 \mu ^4}+\lambda ^3+\frac{1}{\lambda  \mu ^4}+\frac{\lambda }{\mu ^2}+\frac{1}{\lambda  \mu ^2} \) & \( 0 \) & \( 0 \) 
		\\\hline
		\( \frac{1}{X} \) & \( 0 \) & \( 0 \) & \( 0 \) & \( 0 \) & \( 0 \) 
		\\\hline
		\( \frac{1}{X^2} \) & \( 0 \) & \( -\frac{\lambda ^6}{\mu } \) & \( 0 \) & \( 0 \) & \( 0 \) 
		\\
\end{longtable}
}

{\setlength{\tabcolsep}{2pt} 
	\renewcommand{\arraystretch}{1.1}
	\begin{longtable}{>{\centering\arraybackslash}m{1cm}||
			>{\centering\arraybackslash}m{.5cm}|
			>{\centering\arraybackslash}m{2cm}|
			>{\centering\arraybackslash}m{3.5cm}|
			>{\centering\arraybackslash}m{3.5cm}|
			>{\centering\arraybackslash}m{3.5cm}|
		}
		\hline
		& \multicolumn{5}{c|}{$a^{5/2}$} \\\hline
		& \(\mathfrak q^0\) & \(\mathfrak q^1\) & \(\mathfrak q^2\) & \(\mathfrak q^3\) & \(\mathfrak q^4\)
		\\\toprule
		\endfirsthead
		
		& \multicolumn{5}{c|}{$a^{5/2}$} \\\hline
		& \(\mathfrak q^0\) & \(\mathfrak q^1\) & \(\mathfrak q^2\) & \(\mathfrak q^3\) & \(\mathfrak q^4\)
		\\\midrule
		\endhead
		
		\midrule
		\endfoot
		
		\toprule
		\endlastfoot
		\( X^{>5} \) & \( 0 \) & \( 0 \) & \( 0 \) & \( 0 \) & \( 0 \) 
		\\\hline
		\( X^5 \) & \( 0 \) & \( \frac{\mu ^4}{\lambda ^{10}} \) & \( \frac{\mu ^3}{\lambda ^{17}}+\frac{\mu ^3}{\lambda ^{15}}+\frac{2 \mu ^3}{\lambda ^{13}}+\frac{2 \mu ^3}{\lambda ^{11}} \) & \( \frac{\mu ^2}{\lambda ^{22}}+\frac{\mu ^2}{\lambda ^{20}}+\frac{3 \mu ^2}{\lambda ^{18}}+\frac{3 \mu ^2}{\lambda ^{16}}+\frac{4 \mu ^2}{\lambda ^{14}}+\frac{2 \mu ^2}{\lambda ^{12}} \) & \( \frac{\mu }{\lambda ^{25}}+\frac{\mu }{\lambda ^{23}}+\frac{2 \mu }{\lambda ^{21}}+\frac{3 \mu }{\lambda ^{19}}+\frac{3 \mu }{\lambda ^{17}}+\frac{3 \mu }{\lambda ^{15}}+\frac{\mu }{\lambda ^{13}} \) 
		\\\hline
		\( X^4 \) & \( 0 \) & \( 0 \) & \( 0 \) & \( 0 \) & \( 0 \) 
		\\\hline
		\( X^3 \) & \( 0 \) & \( \frac{1}{\lambda ^8}+\frac{\mu ^2}{\lambda ^6} \) & \( \frac{1}{\lambda ^{13} \mu }+\frac{\mu }{\lambda ^{11}}+\frac{2}{\lambda ^{11} \mu }+\frac{2 \mu }{\lambda ^9}+\frac{3}{\lambda ^9 \mu }+\frac{3 \mu }{\lambda ^7} \) & \( \frac{1}{\lambda ^{16} \mu ^2}+\frac{2}{\lambda ^{14} \mu ^2}+\frac{1}{\lambda ^{14}}+\frac{4}{\lambda ^{12} \mu ^2}+\frac{2}{\lambda ^{12}}+\frac{3}{\lambda ^{10} \mu ^2}+\frac{4}{\lambda ^{10}}+\frac{3}{\lambda ^8} \) & \( \frac{1}{\lambda ^{17} \mu ^3}+\frac{1}{\lambda ^{15} \mu ^3}+\frac{1}{\lambda ^{15} \mu }+\frac{2}{\lambda ^{13} \mu ^3}+\frac{1}{\lambda ^{13} \mu }+\frac{1}{\lambda ^{11} \mu ^3}+\frac{2}{\lambda ^{11} \mu }+\frac{1}{\lambda ^9 \mu } \) 
		\\\hline
		\( X^2 \) & \( 0 \) & \( 0 \) & \( 0 \) & \( 0 \) & \( 0 \) 
		\\
\end{longtable}
}

\paragraph{Case III: $|\sqrt{a}|<|X|$.}
For $a^{-3/2}$:
\begin{equation}
	\sum_{i,j \in \mathbb Z} \Lambda_{i,j,-\frac{3}{2}}(\lambda,\mu) X^i \mathfrak q^j = \left[-\frac{\lambda ^6}{\mu ^4} \mathfrak q - \left(\frac{\lambda ^9}{\mu ^5}+\frac{\lambda ^7}{\mu ^5}\right) \mathfrak q^2 -\frac{\lambda ^{10}}{\mu ^6} \mathfrak q^3 + \mathcal O(\mathfrak q^4) \right] X^{-3}
\end{equation}

Tables from $a^{-1}$ to $a^2$ follow.
{\setlength{\tabcolsep}{2pt} 
	\renewcommand{\arraystretch}{1.1}
	\begin{longtable}{>{\centering\arraybackslash}m{1.5cm}||
			>{\centering\arraybackslash}m{.5cm}|
			>{\centering\arraybackslash}m{1.5cm}|
			>{\centering\arraybackslash}m{4.5cm}|
			>{\centering\arraybackslash}m{6cm}|
		}
		\hline
		& \multicolumn{4}{c|}{$a^{-1}$} \\\hline
		& \(\mathfrak q^0\) & \(\mathfrak q^1\) & \(\mathfrak q^2\) & \(\mathfrak q^3\) 
		\\\toprule
		\endfirsthead
		
		& \multicolumn{4}{c|}{$a^{-1}$} \\\hline
		& \(\mathfrak q^0\) & \(\mathfrak q^1\) & \(\mathfrak q^2\) & \(\mathfrak q^3\) 
		\\\midrule
		\endhead
		
		\midrule
		\endfoot
		
		\toprule
		\endlastfoot
	\( X^0, \frac{1}{X} \) & \( 0 \) & \( 0 \) & \( 0 \) & \( 0 \) 
	\\\hline
	\( \frac{1}{X^2} \) & \( 0 \) & \( -\frac{\lambda ^4}{\mu ^3} \) & \( -\frac{\lambda ^5}{\mu ^4} \) & \( 0 \) 
	\\\hline
	\( \frac{1}{X^3} \) & \( 0 \) & \( 0 \) & \( 0 \) & \( 0 \) 
	\\\hline
	\( \frac{1}{X^4} \) & \( 0 \) & \( -\frac{\lambda ^8}{\mu ^5}-\frac{\lambda ^6}{\mu ^7} \) & \( -\frac{\lambda ^{13}}{\mu ^6}-\frac{\lambda ^{11}}{\mu ^8}-\frac{2 \lambda ^{11}}{\mu ^6}-\frac{2 \lambda ^9}{\mu ^8}-\frac{3 \lambda ^9}{\mu ^6}-\frac{3 \lambda ^7}{\mu ^8} \) & \( -\frac{\lambda ^{16}}{\mu ^7}-\frac{\lambda ^{14}}{\mu ^9}-\frac{2 \lambda ^{14}}{\mu ^7}-\frac{2 \lambda ^{12}}{\mu ^9}-\frac{4 \lambda ^{12}}{\mu ^7}-\frac{4 \lambda ^{10}}{\mu ^9}-\frac{3 \lambda ^{10}}{\mu ^7}-\frac{3 \lambda ^8}{\mu ^9} \) 
\end{longtable}
}

{\setlength{\tabcolsep}{2pt} 
	\renewcommand{\arraystretch}{1.1}
	\begin{longtable}{>{\centering\arraybackslash}m{.5cm}||
			>{\centering\arraybackslash}m{.5cm}|
			>{\centering\arraybackslash}m{2cm}|
			>{\centering\arraybackslash}m{4cm}|
			>{\centering\arraybackslash}m{4cm}|
		}
		\hline
		& \multicolumn{4}{c|}{$a^{-1/2}$} \\\hline
		& \(\mathfrak q^0\) & \(\mathfrak q^1\) & \(\mathfrak q^2\) & \(\mathfrak q^3\) 
		\\\toprule
		\endfirsthead
		
		& \multicolumn{4}{c|}{$a^{-1/2}$} \\\hline
		& \(\mathfrak q^0\) & \(\mathfrak q^1\) & \(\mathfrak q^2\) & \(\mathfrak q^3\) 
		\\\midrule
		\endhead
		
		\midrule
		\endfoot
		
		\toprule
		\endlastfoot
	\( X^0 \) & \( 0 \) & \( 0 \) & \( 0 \) & \( 0 \) 
	\\\hline
	\( \frac{1}{X} \) & \( 0 \) & \( -\frac{\lambda ^2}{\mu ^2} \) & \( 0 \) & \( 0 \) 
	\\\hline
	\( \frac{1}{X^2} \) & \( 0 \) & \( 0 \) & \( 0 \) & \( 0 \) 
	\\\hline
	\( \frac{1}{X^3} \) & \( 0 \) & \( -\frac{\lambda ^6}{\mu ^4}-\frac{\lambda ^4}{\mu ^6} \) & \( -\frac{\lambda ^9}{\mu ^5}-\frac{\lambda ^7}{\mu ^7}-\frac{2 \lambda ^7}{\mu ^5}-\frac{2 \lambda ^5}{\mu ^7} \) & \( -\frac{\lambda ^{10}}{\mu ^6}-\frac{\lambda ^8}{\mu ^8}-\frac{\lambda ^8}{\mu ^6}-\frac{\lambda ^6}{\mu ^8} \) 
	\\\hline
	\( \frac{1}{X^4} \) & \( 0 \) & \( 0 \) & \( 0 \) & \( 0 \) 
\end{longtable}
}

{\setlength{\tabcolsep}{2pt} 
	\renewcommand{\arraystretch}{1.1}
	\begin{longtable}{>{\centering\arraybackslash}m{1cm}||
			>{\centering\arraybackslash}m{.5cm}|
			>{\centering\arraybackslash}m{1.5cm}|
			>{\centering\arraybackslash}m{4.5cm}|
			>{\centering\arraybackslash}m{6.5cm}|
		}
		\hline
		& \multicolumn{4}{c|}{$a^0$} \\\hline
		& \(\mathfrak q^0\) & \(\mathfrak q^1\) & \(\mathfrak q^2\) & \(\mathfrak q^3\) 
		\\\toprule
		\endfirsthead
		
		& \multicolumn{4}{c|}{$a^0$} \\\hline
		& \(\mathfrak q^0\) & \(\mathfrak q^1\) & \(\mathfrak q^2\) & \(\mathfrak q^3\) 
		\\\midrule
		\endhead
		
		\midrule
		\endfoot
		
		\toprule
		\endlastfoot
	\( X^0, \frac{1}{X} \) & \( 0 \) & \( 0 \) & \( 0 \) & \( 0 \) 
	\\\hline
	\( \frac{1}{X^2} \) & \( 0 \) & \( -\frac{\lambda ^4}{\mu ^3}-\frac{\lambda ^2}{\mu ^5} \) & \( -\frac{\lambda ^5}{\mu ^4}-\frac{\lambda ^3}{\mu ^6} \) & \( 0 \) 
	\\\hline
	\( \frac{1}{X^3} \) & \( 0 \) & \( 0 \) & \( 0 \) & \( 0 \) 
	\\\hline
	\( \frac{1}{X^4} \) & \( 0 \) & \( -\frac{\lambda ^8}{\mu ^5}-\frac{\lambda ^6}{\mu ^7}-\frac{\lambda ^4}{\mu ^9} \) & \( -\frac{\lambda ^{13}}{\mu ^6}-\frac{\lambda ^{11}}{\mu ^8}-\frac{2 \lambda ^{11}}{\mu ^6}-\frac{\lambda ^{11}}{\mu ^4}-\frac{\lambda ^9}{\mu ^{10}}-\frac{3 \lambda ^9}{\mu ^8}-\frac{5 \lambda ^9}{\mu ^6}-\frac{2 \lambda ^7}{\mu ^{10}}-\frac{6 \lambda ^7}{\mu ^8}-\frac{5 \lambda ^5}{\mu ^{10}}-\frac{\lambda ^3}{\mu ^{12}} \) & \( -\frac{\lambda ^{16}}{\mu ^7}-\frac{\lambda ^{16}}{\mu ^5}-\frac{\lambda ^{14}}{\mu ^9}-\frac{3 \lambda ^{14}}{\mu ^7}-\frac{2 \lambda ^{14}}{\mu ^5}-\frac{\lambda ^{12}}{\mu ^{11}}-\frac{4 \lambda ^{12}}{\mu ^9}-\frac{7 \lambda ^{12}}{\mu ^7}-\frac{3 \lambda ^{12}}{\mu ^5}-\frac{3 \lambda ^{10}}{\mu ^{11}}-\frac{9 \lambda ^{10}}{\mu ^9}-\frac{7 \lambda ^{10}}{\mu ^7}-\frac{\lambda ^8}{\mu ^{13}}-\frac{7 \lambda ^8}{\mu ^{11}}-\frac{10 \lambda ^8}{\mu ^9}-\frac{2 \lambda ^6}{\mu ^{13}}-\frac{7 \lambda ^6}{\mu ^{11}}-\frac{3 \lambda ^4}{\mu ^{13}} \) 
\end{longtable}
}

{\setlength{\tabcolsep}{2pt} 
	\renewcommand{\arraystretch}{1.1}
	\begin{longtable}{>{\centering\arraybackslash}m{.5cm}||
			>{\centering\arraybackslash}m{1cm}|
			>{\centering\arraybackslash}m{2.5cm}|
			>{\centering\arraybackslash}m{4cm}|
			>{\centering\arraybackslash}m{6cm}|
		}
		\hline
		& \multicolumn{4}{c|}{$a^{1/2}$} \\\hline
		& \(\mathfrak q^0\) & \(\mathfrak q^1\) & \(\mathfrak q^2\) & \(\mathfrak q^3\) 
		\\\toprule
		\endfirsthead
		
		& \multicolumn{4}{c|}{$a^{1/2}$} \\\hline
		& \(\mathfrak q^0\) & \(\mathfrak q^1\) & \(\mathfrak q^2\) & \(\mathfrak q^3\) 
		\\\midrule
		\endhead
		
		\midrule
		\endfoot
		
		\toprule
		\endlastfoot
		\( X^1 \) & \( \frac{\mu}{\lambda} \) & \( 0 \) & \( 0 \) & \( 0 \) 
		\\\hline
	\( X^0 \) & \( 0 \) & \( 0 \) & \( 0 \) & \( 0 \) 
	\\\hline
	\( \frac{1}{X} \) & \( -\frac{\lambda }{\mu } \) & \( -\frac{\lambda ^2}{\mu ^2}-\frac{1}{\mu ^4} \) & \( 0 \) & \( 0 \) 
	\\\hline
	\( \frac{1}{X^2} \) & \( 0 \) & \( 0 \) & \( 0 \) & \( 0 \) 
	\\\hline
	\( \frac{1}{X^3} \) & \( 0 \) & \( -\frac{\lambda ^6}{\mu ^4}-\frac{\lambda ^4}{\mu ^6}-\frac{\lambda ^2}{\mu ^8} \) & \( -\frac{\lambda ^9}{\mu ^5}-\frac{\lambda ^9}{\mu ^3}-\frac{\lambda ^7}{\mu ^7}-\frac{3 \lambda ^7}{\mu ^5}-\frac{\lambda ^5}{\mu ^9}-\frac{4 \lambda ^5}{\mu ^7}-\frac{3 \lambda ^3}{\mu ^9}-\frac{\lambda }{\mu ^{11}} \) & \( -\frac{\lambda ^{12}}{\mu ^4}-\frac{2 \lambda ^{10}}{\mu ^6}-\frac{2 \lambda ^{10}}{\mu ^4}-\frac{2 \lambda ^8}{\mu ^8}-\frac{3 \lambda ^8}{\mu ^6}-\frac{2 \lambda ^6}{\mu ^{10}}-\frac{4 \lambda ^6}{\mu ^8}-\frac{\lambda ^4}{\mu ^{12}}-\frac{3 \lambda ^4}{\mu ^{10}}-\frac{2 \lambda ^2}{\mu ^{12}} \) 
	\\\hline
	\( \frac{1}{X^4} \) & \( 0 \) & \( 0 \) & \( 0 \) & \( 0 \) 
\end{longtable}
}

{\setlength{\tabcolsep}{2pt} 
	\renewcommand{\arraystretch}{1.1}
	\begin{longtable}{>{\centering\arraybackslash}m{1cm}||
			>{\centering\arraybackslash}m{.5cm}|
			>{\centering\arraybackslash}m{2cm}|
			>{\centering\arraybackslash}m{4cm}|
			>{\centering\arraybackslash}m{6.5cm}|
		}
		\hline
		& \multicolumn{4}{c|}{$a^1$} \\\hline
		& \(\mathfrak q^0\) & \(\mathfrak q^1\) & \(\mathfrak q^2\) & \(\mathfrak q^3\) 
		\\\toprule
		\endfirsthead
		
		& \multicolumn{4}{c|}{$a^1$} \\\hline
		& \(\mathfrak q^0\) & \(\mathfrak q^1\) & \(\mathfrak q^2\) & \(\mathfrak q^3\) 
		\\\midrule
		\endhead
		
		\midrule
		\endfoot
		
		\toprule
		\endlastfoot
	\( X^0, \frac{1}{X} \) & \( 0 \) & \( 0 \) & \( 0 \) & \( 0 \) 
	\\\hline
	\( \frac{1}{X^2} \) & \( 0 \) & \( -\frac{\lambda ^4}{\mu ^3}-\frac{\lambda ^2}{\mu ^5}-\frac{1}{\mu ^7} \) & \( -\frac{\lambda ^7}{\mu ^2}-\frac{2 \lambda ^5}{\mu ^4}-\frac{2 \lambda ^3}{\mu ^6}-\frac{1}{\lambda  \mu ^{10}}-\frac{2 \lambda }{\mu ^8} \) & \( -\frac{\lambda ^8}{\mu ^3}-\frac{\lambda ^6}{\mu ^5}-\frac{\lambda ^4}{\mu ^7}-\frac{\lambda ^2}{\mu ^9}-\frac{1}{\mu ^{11}} \) 
	\\\hline
	\( \frac{1}{X^3} \) & \( 0 \) & \( 0 \) & \( 0 \) & \( 0 \) 
	\\
\end{longtable}
}

{\setlength{\tabcolsep}{2pt} 
	\renewcommand{\arraystretch}{1.1}
	\begin{longtable}{>{\centering\arraybackslash}m{.5cm}||
			>{\centering\arraybackslash}m{.5cm}|
			>{\centering\arraybackslash}m{2.5cm}|
			>{\centering\arraybackslash}m{3.5cm}|
			>{\centering\arraybackslash}m{7cm}|
		}
		\hline
		& \multicolumn{4}{c|}{$a^{3/2}$} \\\hline
		& \(\mathfrak q^0\) & \(\mathfrak q^1\) & \(\mathfrak q^2\) & \(\mathfrak q^3\) 
		\\\toprule
		\endfirsthead
		
		& \multicolumn{4}{c|}{$a^{3/2}$} \\\hline
		& \(\mathfrak q^0\) & \(\mathfrak q^1\) & \(\mathfrak q^2\) & \(\mathfrak q^3\) 
		\\\midrule
		\endhead
		
		\midrule
		\endfoot
		
		\toprule
		\endlastfoot
	\( X^0 \) & \( 0 \) & \( 0 \) & \( 0 \) & \( 0 \) 
	\\\hline
	\( \frac{1}{X} \) & \( 0 \) & \( -\lambda ^4-\frac{1}{\lambda ^2 \mu ^6}-\frac{\lambda ^2}{\mu ^2}-\frac{1}{\mu ^4} \) & \( -\frac{\lambda ^5}{\mu }-\frac{1}{\lambda ^3 \mu ^9}-\frac{\lambda ^3}{\mu ^3}-\frac{1}{\lambda  \mu ^7}-\frac{\lambda }{\mu ^5} \) & \( 0 \) 
	\\\hline
	\( \frac{1}{X^2} \) & \( 0 \) & \( 0 \) & \( 0 \) & \( 0 \) 
	\\\hline
	\( \frac{1}{X^3} \) & \( 0 \) & \( -\frac{\lambda ^6}{\mu ^4}-\frac{\lambda ^4}{\mu ^6}-\frac{\lambda ^2}{\mu ^8}-\frac{1}{\mu ^{10}} \) & \( -\frac{\lambda ^{11}}{\mu ^3}-\frac{\lambda ^{11}}{\mu }-\frac{2 \lambda ^9}{\mu ^5}-\frac{3 \lambda ^9}{\mu ^3}-\frac{2 \lambda ^7}{\mu ^7}-\frac{5 \lambda ^7}{\mu ^5}-\frac{2 \lambda ^5}{\mu ^9}-\frac{6 \lambda ^5}{\mu ^7}-\frac{1}{\lambda ^3 \mu ^{13}}-\frac{2 \lambda ^3}{\mu ^{11}}-\frac{6 \lambda ^3}{\mu ^9}-\frac{1}{\lambda  \mu ^{15}}-\frac{\lambda }{\mu ^{13}}-\frac{2}{\lambda  \mu ^{13}}-\frac{5 \lambda }{\mu ^{11}} \) & \( -\frac{\lambda ^{14}}{\mu ^4}-\frac{2 \lambda ^{14}}{\mu ^2}-\frac{\lambda ^{12}}{\mu ^6}-\frac{5 \lambda ^{12}}{\mu ^4}-\frac{3 \lambda ^{12}}{\mu ^2}-\lambda ^{12}-\frac{\lambda ^{10}}{\mu ^8}-\frac{7 \lambda ^{10}}{\mu ^6}-\frac{8 \lambda ^{10}}{\mu ^4}-\frac{\lambda ^{10}}{\mu ^2}-\frac{\lambda ^8}{\mu ^{10}}-\frac{7 \lambda ^8}{\mu ^8}-\frac{11 \lambda ^8}{\mu ^6}-\frac{\lambda ^8}{\mu ^4}-\frac{\lambda ^6}{\mu ^{12}}-\frac{7 \lambda ^6}{\mu ^{10}}-\frac{12 \lambda ^6}{\mu ^8}-\frac{\lambda ^6}{\mu ^6}-\frac{1}{\lambda ^4 \mu ^{16}}-\frac{\lambda ^4}{\mu ^{14}}-\frac{7 \lambda ^4}{\mu ^{12}}-\frac{12 \lambda ^4}{\mu ^{10}}-\frac{\lambda ^4}{\mu ^8}-\frac{1}{\lambda ^2 \mu ^{18}}-\frac{\lambda ^2}{\mu ^{16}}-\frac{1}{\lambda ^2 \mu ^{16}}-\frac{5 \lambda ^2}{\mu ^{14}}-\frac{2}{\lambda ^2 \mu ^{14}}-\frac{11 \lambda ^2}{\mu ^{12}}-\frac{\lambda ^2}{\mu ^{10}}-\frac{3}{\mu ^{16}}-\frac{7}{\mu ^{14}}-\frac{1}{\mu ^{12}} \) 
	\\
\end{longtable}
}

{\setlength{\tabcolsep}{2pt} 
	\renewcommand{\arraystretch}{1.1}
	\begin{longtable}{>{\centering\arraybackslash}m{.5cm}||
			>{\centering\arraybackslash}m{.5cm}|
			>{\centering\arraybackslash}m{1.5cm}|
			>{\centering\arraybackslash}m{4cm}|
			>{\centering\arraybackslash}m{7.5cm}|
		}
		\hline
		& \multicolumn{4}{c|}{$a^2$} \\\hline
		& \(\mathfrak q^0\) & \(\mathfrak q^1\) & \(\mathfrak q^2\) & \(\mathfrak q^3\) 
		\\\toprule
		\endfirsthead
		
		& \multicolumn{4}{c|}{$a^2$} \\\hline
		& \(\mathfrak q^0\) & \(\mathfrak q^1\) & \(\mathfrak q^2\) & \(\mathfrak q^3\) 
		\\\midrule
		\endhead
		
		\midrule
		\endfoot
		
		\toprule
		\endlastfoot
	\( X^0 \) & \( 0 \) & \( 0 \) & \( 0 \) & \( 0 \) 
	\\\hline
	\( \frac{1}{X} \) & \( 0 \) & \( 0 \) & \( 0 \) & \( 0 \) 
	\\\hline
	\( \frac{1}{X^2} \) & \( 0 \) & \( -\frac{\lambda ^6}{\mu }-\frac{\lambda ^4}{\mu ^3}-\frac{1}{\lambda ^2 \mu ^9}-\frac{\lambda ^2}{\mu ^5}-\frac{1}{\mu ^7} \) & \( -\frac{\lambda ^9}{\mu ^2}-\lambda ^9-\frac{\lambda ^7}{\mu ^4}-\frac{3 \lambda ^7}{\mu ^2}-\frac{1}{\lambda ^5 \mu ^{12}}-\frac{\lambda ^5}{\mu ^6}-\frac{4 \lambda ^5}{\mu ^4}-\frac{1}{\lambda ^3 \mu ^{14}}-\frac{2}{\lambda ^3 \mu ^{12}}-\frac{\lambda ^3}{\mu ^8}-\frac{4 \lambda ^3}{\mu ^6}-\frac{1}{\lambda  \mu ^{12}}-\frac{\lambda }{\mu ^{10}}-\frac{4}{\lambda  \mu ^{10}}-\frac{4 \lambda }{\mu ^8} \) & \( -\frac{\lambda ^{12}}{\mu }-\frac{2 \lambda ^{10}}{\mu ^3}-\lambda ^{10} \mu -\frac{2 \lambda ^{10}}{\mu }-\frac{2 \lambda ^8}{\mu ^5}-\frac{4 \lambda ^8}{\mu ^3}-\frac{\lambda ^8}{\mu }-\frac{1}{\lambda ^6 \mu ^{15}}-\frac{2 \lambda ^6}{\mu ^7}-\frac{5 \lambda ^6}{\mu ^5}-\frac{\lambda ^6}{\mu ^3}-\frac{1}{\lambda ^4 \mu ^{17}}-\frac{1}{\lambda ^4 \mu ^{15}}-\frac{1}{\lambda ^4 \mu ^{13}}-\frac{2 \lambda ^4}{\mu ^9}-\frac{5 \lambda ^4}{\mu ^7}-\frac{\lambda ^4}{\mu ^5}-\frac{2}{\lambda ^2 \mu ^{15}}-\frac{4}{\lambda ^2 \mu ^{13}}-\frac{2 \lambda ^2}{\mu ^{11}}-\frac{1}{\lambda ^2 \mu ^{11}}-\frac{5 \lambda ^2}{\mu ^9}-\frac{\lambda ^2}{\mu ^7}-\frac{2}{\mu ^{13}}-\frac{5}{\mu ^{11}}-\frac{1}{\mu ^9} \) 
	\\\hline
	\( \frac{1}{X^3} \) & \( 0 \) & \( 0 \) & \( 0 \) & \( 0 \) 
	\\
\end{longtable}
}

	\bibliographystyle{utphys}
	\bibliography{kgr}
	
\end{document}